\documentclass[aps,preprint,showpacs,preprintnumbers,amsmath,amssymb,showkeys,superscriptaddress,titlepage,secnumarabic]{revtex4-2}
\usepackage{graphicx}
\usepackage{dcolumn}
\usepackage{bm}
\usepackage{textcomp}
\usepackage{makecell}
\usepackage{hyperref}

\numberwithin{figure}{section}

\begin{document}

\title{The $^{8}$Be nucleus and the Hoyle state in dissociation of relativistic nuclei}
\author{D.A.~Artemenkov}
\affiliation{Joint Institute for Nuclear Research (JINR), Dubna, Russia}

\author{A.A.~Zaitsev}
\affiliation{Joint Institute for Nuclear Research (JINR), Dubna, Russia}
\affiliation{P.N. Lebedev Physical Institute of the Russian Academy of Sciences (LPI), Moscow, Russia}

\author{P.I.~Zarubin}
\affiliation{Joint Institute for Nuclear Research (JINR), Dubna, Russia}
\affiliation{P.N. Lebedev Physical Institute of the Russian Academy of Sciences (LPI), Moscow, Russia}

\begin{abstract}
 The possibility of recording fragmentation events of relativistic nuclei in a nuclear emulsion, discovered back in the pioneering era of cosmic ray physics, opens up the prospect of using this method to study extremely cold ensembles of H and He nuclei in the interests of developing the physics of nuclear clustering and, possibly, expanding the scenarios of nuclear astrophysics. The results of the BECQUEREL experiment at JINR, obtained on unstable states in the relativistic dissociation of nuclei in a nuclear emulsion providing complete detection of fragments with record resolution are presented. According to the invariant masses calculated from the emission angles in the fragmentation cone, the decays of $^8$Be(0$^+$), $^8$Be(2$^+$), $^9$Be(1.7), $^9$B, $^6$Be, $^{12}$C(0$^+_2$) or the Hoyle state and $^{12}$C(3$^-$) have been identified. The contribution of $^8$Be(0$^+$), $^9$B and $^{12}$C(0$^+_2$) increases rapidly with the $\alpha$-particle multiplicity. Their structure and the diversity of parent nuclei suggest the fusion of the latter. The usage of automated microscopy for an analysis of exposures at the JINR NICA accelerator complex becomes a modern basis to apply the nuclear emulsion method.
\end{abstract}
\maketitle

\section{Introduction}
\subsection{Preface}
The presence of spin-paired quartets of protons and neutrons in the structure of light nuclei manifests itself in the intense production of $\alpha$-particles in a wide variety of nuclear reactions and decays. Playing a key role in the structure of light nuclei and nucleosynthesis, alpha clustering of nucleons remains the focus of research with nuclear beams at the energies up to several tens of MeV per nucleon. The history and state of research in this area have been recently reviewed \cite{1}. Extensive and detailed information on the ground and excited states of nuclei have been compiled, serving as the foundation for new research enabled one by increasing experimental capabilities and theoretical advances \cite{2}. At the same time, the region above 1 GeV per nucleon, seemingly remotes from this topic, offers unique opportunities to expand the understanding of cluster physics. In this field, clustering accompanies the sub-nucleon physics including isobars, hyper-nuclei, and partons in nuclear matter. Nucleon clustering is also one of the fundamental factors in the study of cosmic rays and applications.

This review discusses the motivation and main results on $\alpha$-particle fragmentation of relativistic nuclei in nuclear emulsions (NEs), obtained in the last decade in the BECQUEREL experiment performed at the JINR Laboratory of High Energy Physics. The review is based on the unifying publications \cite{3,4,5}, which in their turn contain references to earlier publications, as well as a summary of recent results from the reports at the conferences ``Nucleus-2024'' (Dubna) \cite{6} and ``Nucleus-2025'' \cite{7} (St. Petersburg). The \href{https://becquerel.jinr.ru/}{BECQUEREL} experiment has been based on the experience of using NE accumulated at the V.I. Veksler and A.M. Baldin Laboratory of High Energy Physics at JINR since its foundation in 1953 (recent notes \cite{3}). Its practical base is the chemical and microscopy laboratories of the Thick-Film Nuclear Emulsions Sector at VBLHEP, where this work has not been interrupted since then. The use of the NE method still requires careful and sufficiently long-term work of highly qualified researchers and laboratory technicians on viewing and precision measuring microscopes.

The current focus is the ground state $^8$Be(0$^+$) of the unbound isotope $^8$Be which has a pronounced 2$\alpha$-particle structure. The $^8$Be(0$^+$) is followed by its 3$\alpha$-analogue – the second excited state $^{12}$C(0$^+_2$) of the isotope $^{12}$C (and the first unbound) known as the Hoyle state (review \cite{8}). Being an indispensable product of the $^{12}$C(0$^+_2$) $\alpha$-decay with the energy of only 285 keV, the $^8$Be(0$^+$) then decays into a 2$\alpha$-particle pair with the energy of 92 keV. Such low energy values determine their widths of 5.6 and 9.3 eV. They correspond to lifetimes in excess of 10$^{-17}$ seconds, unusually long on the nuclear scale. This combination determines the very opportunity and rate of $^{12}$C synthesis in the final stages of the evolution of red giants in the reverse order to the decays of $^8$Be(0$^+$) and $^{12}$C(0$^+_2$). 

When an $\alpha$-particle is captured by the momentarily emerging $^8$Be(0$^+$) in the $^{12}$C(0$^+_2$) resonance, the latter can, with a probability of 1/2500, undergo a transition to the ground state of $^{12}$C(0$^+_1$) via 2$\gamma$-decay 0$^+$ $\to$ 2$^+$ $\to$ 0$^+$ or production of $e^+e^-$($\pi$) pairs. The importance of the $^{12}$C isotope for organic life inspires a particular interest in studying the structures of $^8$Be(0$^+$) and $^{12}$C(0$^+_2$). They are complemented by the ground 2$\alpha p$-state $^9$B(3/2$^-$) of the even less stable $^9$B isotope, which has a width of 0.54 keV and decays into $^8$Be(0$^+$)$p$ with the energy of 185 keV. Following the isotopes He and H, the above trio serves as test objects in the theory of nuclear structure and nuclear forces at the lower limits of nuclear density and temperature beyond which the physics of electron-ion plasma begins. 

The search for more complex states of real $\alpha$-particles and nucleons near the corresponding coupling thresholds begins with $^8$Be(0$^+$), $^9$B and $^{12}$C(0$^+_2$). The findings of the relevant laboratory studies provide the basis to develop scenarios for nuclear astrophysics. The primary question in the context is as follows: is the formation of $^{12}$C(0$^+_2$) possible in the relativistic fragmentation of nuclei? Microscopic reproduction of stellar synthesis of the $^{12}$C isotope could point to the inverse process of $\alpha$ + $^8$Be(0$^+$) $\to$ $^{12}$C(0$^+_2$). The mechanism may be a secondary interaction between emerging $\alpha$-particles. Then, an increase in their multiplicity suggests an enhancement of $^8$Be(0$^+$) and $^{12}$C(0$^+_2$). The condition of this production is a sufficiently small value of the invariant Lorentz factor of relative motion minus 1. These values are determined by the products of the $4$-momenta of particle pairs normalized by their masses, or by the products of the $4$-velocities according to A.M. Baldin as discussed below \cite{9,10}. 

In NE exposed in the stratosphere in the late 1940s, tracks of nuclei of energy of several GeV per nucleon, as well as the stars they created, were discovered in the composition of cosmic rays. Peripheral collisions were observed where the dominant share of the initial energy and charge was transferred to relativistic fragments. The hidden clustering of the ground states of the detected nuclei amazingly appeared in the group generation of He and H nuclei. These pioneering observations already hinted at the opportunity of a comprehensive study of the inherently non-relativistic ensembles of the lightest nuclei, i.e. having relative velocities much less than the speed of light. Considering the final states of relativistic fragmentation on the invariant basis could shed light on important issues in the few body nuclear physics which refers to a seemingly different field.

\begin{figure}
\includegraphics[width=0.6\textwidth]{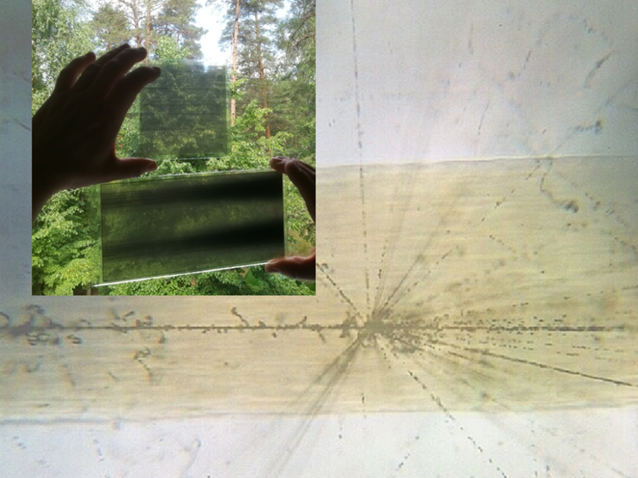}
\caption{\label{fig:1.1}Illustration of the NE resolution. The background is a photograph of the interaction of the $^{32}$S nucleus at 200 GeV per nucleon, superimposed at the same resolution on a 60 $\mu$m thick hair.}
\end{figure}

Longitudinal exposures of NE layers to the beams of nuclei of the known type and energy in high-energy accelerators allows this idea to be given the scope necessary to study the nuclear structure itself. In this case the directions of the tracks can be determined completely and with the best resolution of 0.5 $\mu$m (microns). The insert in Figure~\ref{fig:1.1} shows the developed layers of nuclear energy with an initial thickness. The top plate demonstrates the uniform NE exposure in the defocused beam of $^{84}$Kr nuclei at 950 MeV per nucleon at the SIS-18 synchrotron in GSI. The bottom plate shows the longitudinal NE exposure with a shift between two beam spills of $^{32}$S nuclei at 200 GeV per nucleon at the SPS synchrotron at CERN: the cascade of produced particles is developed after approximately 10 cm. The background is a photograph of the interaction of the $^{32}$S nucleus, combined at the same resolution with one hair 60-$\mu$m thick. It can be argued that NE provides the best spatial projection of the event, which occurred on the scale of the micro world.

\begin{figure}
\includegraphics[width=0.9\textwidth]{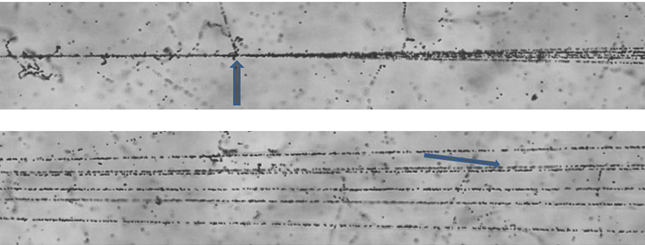}
\caption{\label{fig:1.2}Macro photograph of the dissociation of a $^{24}$Mg nucleus at 3.65 GeV per nucleon in NE into six He nuclei.}
\end{figure}

On the BECQUEREL experiment \href{https://becquerel.jinr.ru/}{website}, one can view \href{https://becquerel.jinr.ru/movies/movies.html}{video} recordings of relativistic dissociation events of the nuclei studied, including those discussed below, filmed by using digital cameras on microscopes. As an example, Figure~\ref{fig:1.2} shows a projection (\href{https://becquerel.jinr.ru/movies/movavi/mg24/mg24.1/mg24-l.avi}{video}) of a macro photograph of the dissociation event of the $^{24}$Mg nucleus at 4.5 GeV/$c$ in a nuclear beam onto six He nuclei. The top image illustrates the trace of the incoming Mg nucleus with the traces of $\delta$-electrons (left), the interaction vertex (arrow) with a single trace of a target fragment, and a sharp decrease in ionization due to forming the ensemble of relativistic fragments. In the bottom image, when tracking to 1 mm, the He traces are clearly separated. The narrow pair of tracks (arrow) corresponds to the decay of $^8$Be(0$^+$) at the vertex. This NE exposure was performed in the mid-80s at the JINR Synchrophasotron, using a laser ion source.

Identification of H and He isotope tracks in NE layers is possible by using $P$$\beta$$c$ values, where P is the total momentum and $\beta$c is the particle velocity, determined via the average angles of multiple Coulomb scattering. Due to the fragment momentum "quantization", their mass numbers are determined as $P$$\beta$$c$/($P_0$$\beta_0$$c$), where $P_0$ is the momentum per nucleon of the parent nucleus, and $\beta_0$ is its velocity. The accuracy of $P$$\beta$$c$, 20-30\%, is achieved by measuring planar displacements at least 100 points along a 2-5 cm track $\langle D \rangle$. The presented event is, thus, identified as $^{24}$Mg $\to$ 5$^4$He + $^3$He (lower trace). This technique was developed theoretically and experimentally in the early cosmic ray studies mentioned below to estimate $P$$\beta$$c$ of particles entering or produced in NE up to the values of 50 GeV/$c$. It has found an application in NE exposures at accelerators where $P_0$ values are fixed.

The final states are best interpreted in glancing collisions leading to coherent dissociation on heavy target nuclei without forming the target fragments or produced mesons (or ``white'' stars). The relativistic fragments are concentrated in a narrow angular cone, limited by the Fermi motion of nucleons in the parent nucleus of the order of [200 MeV/$c$]/$P_0$, where $P_0$ is the initial momentum per nucleon. This circumstance makes the simultaneous three-dimensional tracking possible in a single NE layer over the lengths of the order of 1 mm, sufficient for angular measurements with an accuracy of the order of 10$^{-3}$ rad. Angular correlations of fragments can be projected onto the invariant masses of their ensembles, determined in the approximation that $P_0$ (or the initial velocity) are conserved by the fragments \cite{3}.

In the most general case, the internal energy of a several particle ensemble $Q$ can be defined as the difference between the invariant mass of the system $M^*$ and the mass of the primary nucleus, or the sum of the particle masses $M$, i.e., $Q$ = $M^*$ - $M$. The $M^*$ is determined as the sum of all scalar products $M^{*2}$ = $\Sigma$($P_i\cdot P_k$) of the particle $4$-momenta $P_{i,k}$. The subtraction of $M$ is performed for convenience data presentation, and $Q$ is also called the invariant mass. Reconstructing $Q$ allows one to identify decays of unstable nuclei.

The decays of $^8$Be, $^9$B, and $^{12}$C(0$^+_2$) occurring at extremely low energy $Q$ should manifest themselves as pairs and triplets of He and H fragments with the smallest spread out angle. In the relativistic case, they can be identified at the origins of distributions over corresponding invariant masses Q calculated in the approximation of conservation of an initial momentum per nucleon of a parent nucleus. In these cases, the assumption that the He fragment corresponds to an $\alpha$-particle, and the H fragment to a proton, is justified. This is the framework of the study discussed below.

\subsection{Historical notes}
The basis of the NE method was formed between the second half of the 1940s and the first half of the 1950s while high-altitude cosmic ray studies which provided the foundation for elementary particle physics. These achievements became possible due to the development of NE possessing sensitivity down to minimally ionizing (or relativistic) particles \cite{11,12}. Fundamental discoveries based on observations in NE exposures are presented in the book by S. Powell, D. Perkins, and W. Fowler \cite{13,14}. Its concluding chapter is dedicated to the discovery of relativistic nuclei. Even then, the NE method with the 0.5 $\mu$m resolution, that is still unrivaled, provided observations of relativistic particle tracks from singly charged to heavy nuclei together with target fragment tracks. 

\begin{figure}
\includegraphics[width=0.9\textwidth]{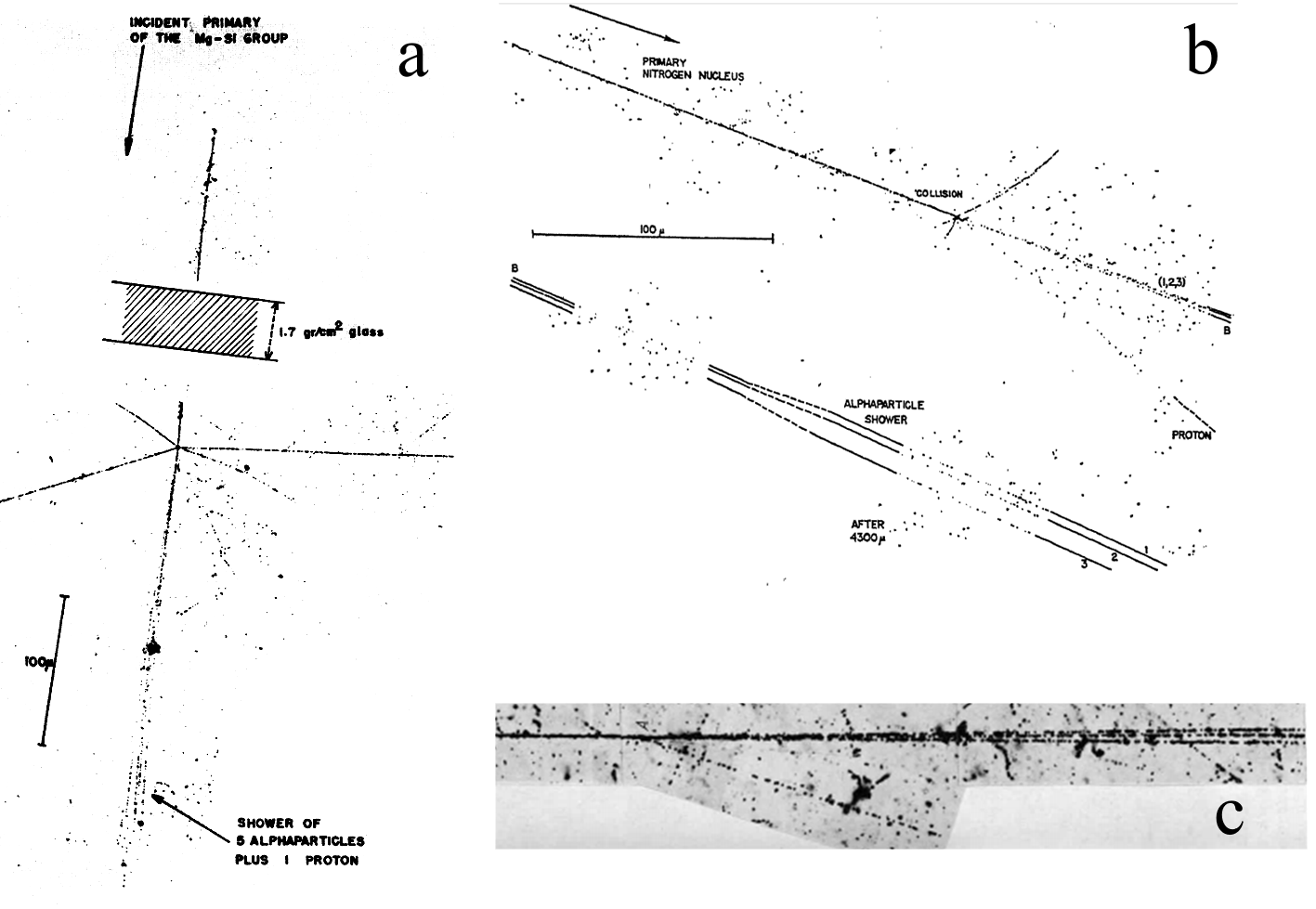}
\caption{\label{fig:1.3}(a) Fragmentation of Mg-Si nucleus group of cosmic origin at several GeV per nucleon \cite{15}. (b) Fragmentation of a cosmic origin nucleus $^{14}$N \cite{16}. (c) Fragmentation of the cosmic origin nucleus $^{12}$C into a triple of $\alpha$-particles accompanied by a recoil proton \cite{14}.}
\end{figure}

Tracks of relativistic nuclei of cosmic origin and the stars they formed, were discovered in the late 1940s by Bradt and Peters in NE stacks exposed in the stratosphere \cite{15}. They immediately proposed a model of geometric overlap of colliding nuclei which described the average ranges to inelastic interactions in the NE substance. Even then, amazingly complete observations of peripheral interactions of nuclei with the energy of several GeV per nucleon were made forming particle ensembles in a narrow cone of relativistic fragmentation. Figures~\ref{fig:1.3}(a) and ~\ref{fig:1.3}(b) reproduce the first events recorded by hand-drawing projections from microscopes. They immediately revealed the most probable channels and kinematic features of dissociation. Photography soon became possible (Figure~\ref{fig:1.3}(c)). This class of events has become the object of systematic study in the BECQUEREL experiment.

A fundamental contribution to the development of the nuclear neutron emission method during the construction of the Synchrophasotron in 1950-1957 (JINR) was made by collaboration with the Lebedev Physical Institute of the USSR Academy of Sciences (FIAN, Moscow), the parent institution of the JINR High Energy Laboratory, and the Institute of Theoretical and Experimental Physics (Laboratory No. 3, Moscow). Exposures in the late 1940s of NE G5 by the British company Ilford to photons generated by the electron synchrotron C-25 of the Lebedev Physical Institute resulted in the discovery of the phenomenon of photoproduction of charged mesons (or ``the nuclear properties of the light'' as defined at that time).

The proprietary technology for producing substrate-free 500 $\mu$m NE layers possessing relativistic sensitivity (the BR-2 type) was developed in the 1950s at the Scientific Research Film and Photo Institute (NIKFI, Moscow), where they were produced until the mid-2000s \cite{17,18}. The BECQUEREL project website has preserved publications of that time which are of more than just historical significance.

\begin{figure}
\includegraphics[width=0.9\textwidth]{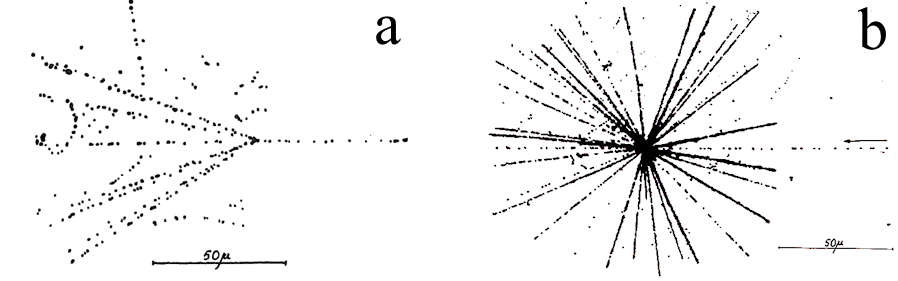}
\caption{\label{fig:1.4}(a) The event of coherent dissociation of a proton at 10 GeV with the formation of eight relativistic particles \cite{20}. (b) Event of complete destruction of a heavy nucleus by a proton at 10 GeV.}
\end{figure}

Stacks of BR-2 NE exposed in 1957 to protons at the JINR Synchrophasotron, became evidence of the record-breaking energy of 10 GeV achieved there. They served as the material for the first overview studies. The search for events and track measurements in NE using microscopes was mastered by scientists from the member countries of the newly established JINR and the laboratory assistants they supervised. Collaboration with the Lebedev Physical Institute ensured compliance with the world level of research at that time and established scientific ties which have not been interrupted since then. Among the physical results of that period, the observations of events of coherent dissociation of protons at 10 GeV, not accompanied by fragments of the target with multiple production of mesons (Figure 1.4(a)) and then 70 GeV (IHEP), as well as events of complete destruction of heavy nuclei (Figure 1.4(b)) have not lost their relevance \cite{19,20,21}. They later provided an important analogy in the study of coherent dissociation of relativistic nuclei.

In the early 1970s, on the initiative of A.M. Baldin, research was proposed at the intersection of nuclear and elementary particle physics called relativistic nuclear physics. The priority was to establish the limits of applicability of the proton-neutron model of the atomic nucleus with the prospect of developing concepts of nuclear matter at the quark-parton level. The starting points were IHEP data demonstrating the scale-invariant behavior of the inclusive spectra of produced hadrons. On the other hand, research of deep inelastic electron scattering at Stanford (USA) revealed a point-like (or parton) structure of nucleons. This research was supported by the validity of the Matveev-Muradyan-Tavkhelidze quark counting rules for elastic hadron scattering with large momentum transfers \cite{22}.

The new direction is motivated the acceleration of light nuclei at the JINR Synchrophasotron, the creation of a slow extraction system and beam transport channels. Confirming A.M. Baldin's hypothesis that meson production reaches the asymptotic regime of limiting fragmentation for the entire range of target nuclei, the study of the cumulative effect resulted in introducing the quark-parton structure function of nuclei. Beneficial in its broad implications, A.M. Baldin's initiative discovered new opportunities for survey observations in NE and bubble chambers, electronic experiments on deuteron structure, the search for hyper-nuclei and nuclear $\Delta$-isobars in a streamer chamber, and radiobiological research. The exposures to $^{12}$C, $^{16}$O, $^{24}$Mg, $^{28}$Si and $^{6,7}$Li nuclei provided the physical motivation for the BECQUEREL project on relativistic radioactive nuclei. In the 1970s, NE was exposed to nuclei at Bevalac LBL, and in the 1990s at AGS (BNL) and SPS (CERN).

\begin{figure}
\includegraphics[width=0.9\textwidth]{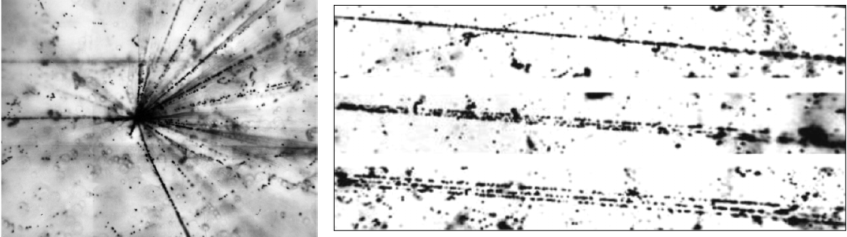}
\caption{\label{fig:1.5}NE exposure at the JINR Synchrophasotron to $^{12}$C nuclei at 3.65 GeV per nucleon (1974): complete destruction (left) and coherent dissociation into the $\alpha$-particle triple.}
\end{figure}
 
The development of relativistic nuclear physics was supported by communities with extensive experience in the field of NE application. Proton synchrotrons with injection systems modernized for ion acceleration gave opportunities for systematically studying the interactions of a wide variety of nuclei at specific energy. The developed stacks of layers were shared with other research centers around the world to be analyzed. It was in the spirit of the tradition of emulsion collaborations that appeared during the pioneering period of cosmic ray research. Irradiation of nuclear irradiation with $^{12}$C, $^{16}$O, $^{24}$Mg, and $^{28}$Si nuclei, as well as $^{6,7}$Li, provided the physical motivation for the BECQUEREL project on relativistic radioactive nuclei.

Searches for collisions of incident nuclei with NE nuclei were carried out on primary tracks without sampling. These studies provided a comprehensive picture of possible interaction types, from coherent dissociation of nuclei, unaccompanied by target fragments, to their complete disruption with secondary tracks spread out across the entire solid angle (Figure~\ref{fig:1.5}). This diversity of the observations became a kind of fork in the road when choosing how to formulate physical problems (or questions posed to Nature in the words by V.I. Veksler). Research on central interactions shifted entirely toward specialized spectrometers, because of which the pioneering role of the NE method was overshadowed. At the same time, the data on relativistic fragments generated in peripheral interactions retain their uniqueness, and the preserved NE layers allow one to carry out the targeted analysis in this direction.

One of the features revealed at several GeV per nucleon was the presence of $\alpha$-particle pairs with opening angles of the order 10$^{-3}$ rad in the relativistic fragmentation cone (e.g., \cite{23,24}). Corresponding to decays of the extremely short-lived $^8$Be nucleus, these results indicated the opportunity of studying $\alpha$ clustering within a relativistic approach starting with the minimum relative energy. The opportunity of the uniform comparison of the final states of fragments is indicated by the invariant approach of A.M. Baldin, who provided the general methodological basis to classify the processes of multiple particle formation \cite{9,10}. In the present study, this approach is applied to the case of extremely small $4$-velocity differences.

Until now, the complete detection of ensembles of the lightest relativistic fragments has been demonstrated only in NE. However, this method does not provide for momentum analysis. This limitation is compensated by the data from magnetic spectrometers (e.g., \cite{25}). In particular, the small momentum spread out of $\alpha$-particles justifies the approximation to conserve the initial momentum per nucleon. The accepted approximations can be verified by measuring 11,000 interactions of $^{16}$O nuclei at 2.4 GeV per nucleon using the JINR 1-meter hydrogen bubble chamber of (VPK-100) placed in the magnetic field \cite{26}. In particular, they have confirmed the dominance of $^4$He among fragments, especially in the narrow $\alpha$-pairs of $^8$Be \cite{27}.

\begin{figure}
\includegraphics[width=0.9\textwidth]{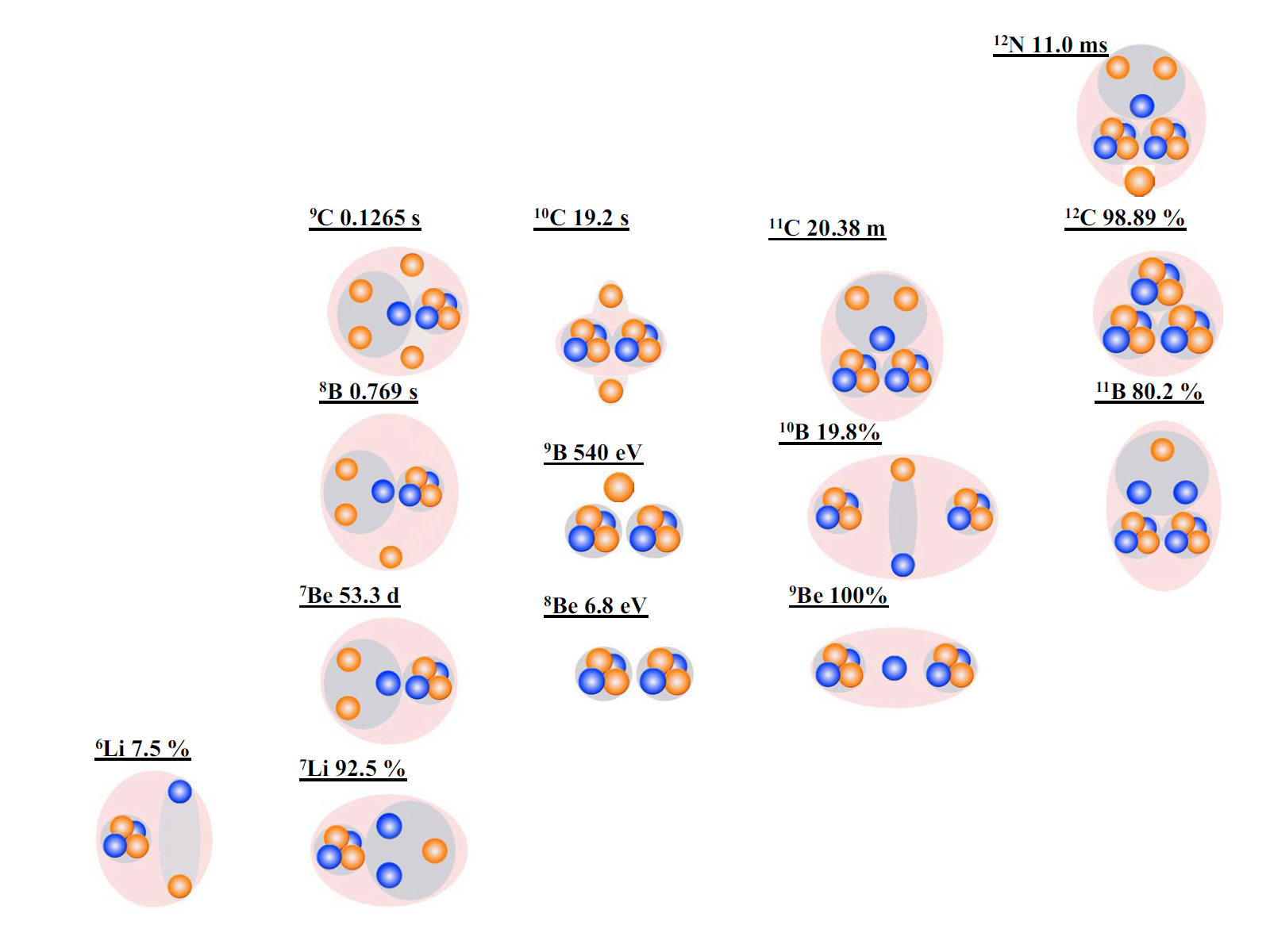}
\caption{\label{fig:1.6}Diagram of cluster degrees of freedom in stable and neutron-deficient nuclei; abundances or lifetimes of isotopes, their spins and parities are indicated: orange circles correspond to protons and the blue ones - to neutrons; clusters are marked as dark background \cite{3}.}
\end{figure}

The advantage of the relativistic approach is as follows: at small energy-momentum transfers in the final states of fragments near thresholds, the structure of the nuclei under study should be reflected most completely. On magnetic spectrometers, the above information is limited to the momentum distributions of relativistic fragments with charges close to the initial nucleus (e.g., \cite{28,29,30}). Detection of one or two accompanying protons or neutrons is feasible in the range of hundreds of MeV per nucleon (e.g., \cite{31,32,33}). Beams of relativistic radioactive nuclei make it possible to qualitatively expand research. They can be formed by means of magneto-optical separation of fragmentation products or charge exchange of accelerated nuclei. However, this approach loses fundamentally important channels containing only He and H fragments, including those from the decays of $^8$Be(0$^+$) and $^9$B. For example, dissociation events of the $^{11}$C isotope in NE containing only He and H fragments account for 80\% of events with conservation of the primary charge in the fragmentation cone. As part of fragmentation, the reconstruction of $^8$Be(0$^+$) and $^9$B is necessary in the search for unstable states which decay with their participation. This direction of research at the JINR Synchrophasotron began with the reactions $\alpha$ $\to$ t, $^7$Li $\to$ $^7$Be and $^6$Li $\to$ $^6$He. The extracted beam transport channels constructed in the 1980s were used \cite{34}.

The extraction of the beam from the JINR Nuclotron in 2002 made it possible to continue NE exposures to light nuclei, including the radioactive ones, within the framework of the BECQUEREL experiments (Figure~\ref{fig:1.6}) \cite{3,35}. Known and previously unobserved features of the clustering the isotopes: $^{7,9}$Be, $^{8,10,11}$B, $^{10,11}$C, and $^{12,14}$N were identified, manifesting themselves in the probabilities of dissociation channels. In the case of $^{10}$B, $^{10}$C, and $^{11}$C, the relativistic decays $^9$B $\to$ $^8$Be(0$^+$)p were found.

The production of NE layers lasted in Moscow for four decades and was completed in the 2000s. Since the 2010s, the BECQUEREL experiments have stimulated their production at the Slavich Company (Pereslavl-Zalessky). The NE quality was demonstrated in measurements of short-range particles (reviews \cite{36,37,38}). In $\beta$-decays of stopped $^8$He nuclei, the decays of $^8$Be(2$^+$) were reconstructed and the $^8$He atom drift was discovered. The $^{12}$C and $^{10}$B were studied by 14.1 MeV neutrons and thermal neutrons, respectively. The 3$\alpha$-splitting of $^{12}$C by relativistic muons and hadrons was studied to estimate it as a source of natural $^4$He production. The ternary fission of $^{252}$Cf and the U $\alpha$-activity were investigated. The NE samples were calibrated without light shielding in vacuum with Kr and Xe ions at 1.2 and 3 MeV per nucleon.

Along with its scientific novelty, the resumption of NE use in the BECQUEREL experiment proved to be critical for preserving the technology and familiarizing young researchers with the analysis methods. Using digital cameras mounted on microscopes uniquely detailed and high-resolution \href{https://becquerel.jinr.ru/movies/movies.html}{videos} of nuclear interactions were collected at the BECQUEREL experiment site. In practice, it was demonstrated that the NE method thought to be consigned to history has inspiring perspectives and deserves updating based on a state-of-art image analysis.

\subsection{Introductory remarks on nuclear clustering}
Clustering of the nuclear structure manifests itself in the formation of the lightest nuclei, whose common feature is the absence of bound excitations. They are primarily represented by $\alpha$-particles, and in the light odd nuclei – by deuterons ($d$), tritons ($t$), and $^3$He nuclei ($h$, ``helions''). In $^8$Be(0$^+$) $\alpha$-clustering is most clearly expressed. Its lifetime (of the order of the $\pi^0$-meson lifetime) exceeding the duration of the generating reactions by several orders of magnitude, gives the argument to consider $^8$Be(0$^+$) to be not a resonance, but rather a fragment, though – the exotic one.

The assumption that ``the $^8$Be(0$^+$) fragment is a component of the parent nuclei'' is doubtful due to its size and deformation. This would have resulted in decreasing of the $^8$Be(0$^+$) yield in fragmentation with increasing $\alpha$-particle multiplicity. Another scenario is the dynamic deformation of the shell structure of the parent nucleus, with the formation of $\alpha$-particle as the first magic nucleus. An alternative mechanism could be the fusion of the forming $\alpha$-particles. If the number of $\alpha$-particles grew, the contribution of $^8$Be(0$^+$) would have increased. Thus, the spatial extent of $^8$Be(0$^+$) makes it, on the one hand, a probe of reaction dynamics, and on the other hand, – a potential basis to generate more complex states by capturing $\alpha$-particles and neutrons.

Since the discovery of radioactivity the experimental and theoretical studies of nucleon clustering in the structure and interactions of atomic nuclei have remained a dynamically developing area of microscopic physics (reviews \cite{1,8,39,40,41,42,43,44,45,46,47,48,49,50,51,52}). In the focus of research there is an opportunity of existence of short-lived condensate or molecular-like states based on $\alpha$-particles. The spatial extent of these states makes them, on the one hand, probes of reaction dynamics, and on the other hand, – a potential basis for the generation of more complex states by capturing $\alpha$-particles and neutrons. Their decays can be reconstructed in the experiments with nuclear beams, including the radioactive ones, primarily at the energy from several to tens of MeV per nucleon. Upon reaching the energy of several GeV per nucleon, the kinematic regions of fragmentation of colliding nuclei are clearly separated, and the collision duration is minimal. These factors can facilitate the interpretation of the final states of fragments. The BECQUEREL experiment is aimed at developing this approach further.

The current interest in unstable $\alpha$-particle states is motivated by the concept of the $\alpha$-particle Bose-Einstein condensate ($\alpha$BEC). It was proposed in the early 2000s in analogy with the quantum gases of atomic physics discovered in the 1990s (review \cite{47}). $\alpha$BEC can manifest itself as excitations of $\alpha$-multiple nuclei just above the corresponding thresholds. Coexisting with fermionic excitations, they are considered to be based on a boson-type mean field formed by the gas of nearly ideal bosons in the S-state at the average density four times lower than the normal one. $^8$Be(0$^+$) and $^{12}$C(0$^+_2$) are described as 2- and 3$\alpha$BEC states. The 0$^+_6$ state of the $^{16}$O nucleus at 15.1 MeV (660 keV above the 4$\alpha$ threshold) is considered to be the 4$\alpha$ analogue of $^{12}$C(0$^+_2$), decaying into $^{12}$C(0$^+_2$)$\alpha$ or 2$^8$Be(0$^+$). Experimental approaches to searching for $\alpha$BEC in the fragmentation of light nuclei have been proposed, including the one presented here (review \cite{40}). New opportunities to search for $\alpha$BEC phenomena have appeared with increasing energy and mass numbers of the parent nuclei. It is highly significant to demonstrate the universality of $\alpha$BEC candidates based on relativistic invariance.

States $\alpha$BEC can be formed by $\alpha$-particles independent of the parent nucleus, representing a special state of nuclear matter with extremely low density and temperature on the nuclear scale. Indeed, the sizes and lifetimes of $^8$Be(0$^+$) and $^{12}$C(0$^+_2$) assume sequential fusions: 2$\alpha$ $\to$ $^8$Be(0$^+$), $^8$Be(0$^+$)$\alpha$ $\to$ $^{12}$C(0$^+_2$), $^{12}$C(0$^+_2$)$\alpha$ $\to$ $^{16}$O(0$^+_6$), 2$^8$Be(0$^+$) $\to$ $^{16}$O(0$^+_6$). The $^8$Be(0$^+$) signal can be enhanced in the events with high $\alpha$-particle multiplicity, accessible in the relativistic fragmentation of heavy nuclei. Therefore, the question of the dynamics of $^8$Be(0$^+$) formation is the starting point of the study. The exotic 3$\alpha$ structure of $^{12}$C(0$^+_2$) and the 2$\alpha$p one of unstable nucleus of $^9$B further expands the abilities of the above diagnostics. However, their statistical reliability is inherently lower.

\section{Development of clustering research and its current motivation} 
The list of publications on the $\alpha$-clustering is virtually endless, and several of them can be considered ``golden collections'' of physics in general. Thus, a recent review \cite{1}, including 582 references, begins with citations of nine lecture collections and reviews. Our approach has been also mentioned. The bibliography on this topic includes the proceedings of renowned international conferences, dedicated meetings, and schools. Following \cite{1,8} as well as other cited publications and freely quoting of the theoretical and experimental milestones inspired the application of the relativistic approach to $\alpha$-cluster phenomena, that will be presented below. Without substituting these sources, a summary of the main points made in them may be useful for the overview of the achievements and current research challenges in this area.

\subsection{From the $^8$Be nucleus to the Hoyle state}
Advances in chemistry and atomic physics by the beginning of the 20th century provided important analogies for the analysis of nuclear processes. Chemical bonds between atoms compose molecules possessing a colossal diversity of rotational and vibrational excitations. The emergence of these structures in nuclei is made possible by strong coupling of the quadruple of nucleons in $\alpha$-particle and the nearly bound two-neutron channel. However, at the nuclear level, physics changes radically due to the equality of the participating particles. Unlike heavy ions surrounded by electrons, the protons and neutrons have nearly equal masses, and cluster structures appear due to a delicate balance between short-range repulsive forces and blocking Pauli effects, medium-range attractions, and long-range Coulomb forces.

Based on the newly emerging field of quantum mechanics, in 1928 G. Gamow demonstrated the opportunity of particle tunneling through a potential barrier thereby explaining the $\alpha$-decay of nuclei. He also suggested that tunneling could reduce the energy required for a positively charged particle to overcome the Coulomb barrier. Nuclei were first disintegrated by protons, accelerated by J. Cockcroft and E. Walton in 1932. In accordance with conservation laws, protons with the energy up to 300 keV were captured by $^7$Li nuclei. Highly excited $^8$Be nuclei were formed, which, in turn, decayed into $\alpha$-particles, which penetrated up to 8 cm in the air (approximately 8 MeV) and were detected visually as single and paired scintillations.

\begin{figure}
\includegraphics[width=0.6\textwidth]{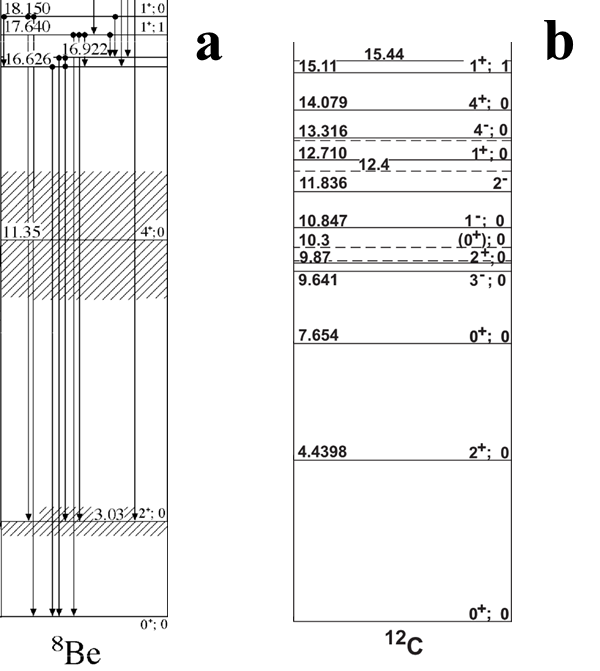}
\caption{\label{fig:2.1}Excitation spectra of the $^8$Be (a) and $^{12}$C(b) nuclei \cite{2}.}
\end{figure}

According to the established concepts, in the reaction $p$ + $^7$Li $\to$ 2$\alpha$ an intermediate nucleus $^8$Be appears in the isobaric analog state of 17.64 MeV with the width $\Gamma$ = 10.7 keV, spin-parity $J^\pi$ = 1$^+$ and isospin $T$ = 1. From the above state the electromagnetic transition occurs to the $^8$Be levels associated with the emission of nucleons (Figure~\ref{fig:2.1}(a)). When the proton energy varied from 30 keV to 18 MeV, it was observed: $\gamma$-transitions of $^8$Be$^*$(17.640) to the narrow ground state $^8$Be(0$^+$) at 92 keV ($\Gamma$ = 5.6 eV, $J^\pi$ = 0$^+$ and $T$ = 0) and the first excited state $^8$Be(2$^+$) at 3.03 MeV ($\Gamma$ = 1.5 MeV, $J^\pi$ = 2$^+$ and $T$ = 0). In addition, there is the $^8$Be(4$^+$) level at 11.35 MeV ($\Gamma$ $\approx$ 3.5 MeV, $J^\pi$ = 4$^+$, $T$ = 0). The formation of the most energetic $\alpha$-particles occurs through a doublet of levels at 16.6 MeV and 16.9 MeV with mixed isospin ($\Gamma$ = 108 and 74 keV, $J^\pi$ = 2$^+$, $T$ = 0+1). Thus, the $^8$Be excitation spectrum clearly is quantized up to the nucleon separation threshold of 17.3 MeV.

The analysis of NE exposed to cosmic rays and then in accelerators in the 1940s and 1950s yielded direct observations of target nuclear fragmentation. The resulting $\alpha$-pairs, with close ranges and smallest scattering angles, were used to identify $^8$Be(0$^+$) decays, while the $\beta$$^{-/+}-$decays of stopped $^8$Li and $^8$B fragments were taken to identify $^8$Be(2$^+$) decays in the form of hammer tracks \cite{13,14}. In subsequent studies of binary reactions with an accessible variety of nucleons, the lightest and light nuclei, the $^8$Be levels were identified indirectly \cite{2}.

The lifetime of $^8$Be(0$^+$) $\tau$ turned out to be too short to be measured by this range. Its inverse width $\Gamma$ and the energy of the $\alpha$-particles of decay were also extremely small. Extraction of the scattering parameters in elastic scattering of ions $^4$He$^+$ on a jet helium target yielded values of the binding energy of this resonance of 92.04 $\pm$ 0.05 keV and $\Gamma$ = 6.8 $\pm$ 0.25 eV \cite{53}. The value of $\tau$ is comparable with the lifetime of the $\pi^0$ meson 8.4$\times$10$^{-17}$ s. In practice, the identification of $^8$Be(0$^+$), like $\pi^0$, is determined by the instrumental resolution. The ground state parameters of the unstable $^9$B nucleus were extracted from the energy dependence of the cross section of the $^9$Be($p$,$n$)$^9$B reaction near the threshold $^8$Be(0$^+$) + $n$, where $^9$B is produced in the $^8$Be(0$^+$) + $p$ interaction. The ground state of $^9$B is 185.1 keV above the threshold $^8$Be + $p$ at $\Gamma$($^9$B) = 0.54 $\pm$ 0.21 keV \cite{54}.

The prediction, detection, and characterization of the second and first above the threshold excitations of the $^{12}$C(0$^+_2$) nucleus also deserve a reminder (Figure~\ref{fig:2.1}(b)) \cite{8}. It is named after the astrophysicist F. Hoyle who postulated its existence in the 1950s to explain the abundance of the $^{12}$C and $^{16}$O isotopes (reviewed in \cite{55}). The absence of stable isotopes with mass numbers 5 and 8 in the nucleosynthesis sequence led H. Bethe in the 1930s to hypothesize the formation of $^{12}$C in the triple $\alpha$-particle fusion. However, the probability of such a process is too low. Epic and Salpeter proposed the sequential fusion 2$\alpha$ $\to$ $^8$Be(0$^+$)$\gamma$, and Hoyle $^8$Be(0$^+$)$\alpha$ $\to$ $^{12}$C(0$^+_2$) $\to$ $^{12}$C$\gamma$ \cite{56}.

The formation energy of $^8$Be(0$^+$) is slightly higher than the average energy at the temperature of 10$^8$ K and is close to the Gamow peak at 85 keV with a width of 60 keV. Therefore, $^8$Be(0$^+$) can exist in dynamic equilibrium, continuously being formed and decaying. Playing an important role in nucleosynthesis and energy generation in stars, these processes occur in the final stage of red giant evolution, when hydrogen in the central region has been converted to helium, and gravitational compression has ensured the temperature of approximately 10$^8$ K and the density of 100 g/cm$^3$. Under these conditions, the equilibrium ratio of $^8$Be(0$^+$) to He of approximately 10$^{-10}$ is established. The width of $^8$Be(0$^+$) is limited by the Coulomb barrier, through which the alpha decay or capture must occur. At a higher decay energy, its width would be significantly larger, the lifetime would be shorter, and $^8$Be(0$^+$) would be outside the Gamow window, which would significantly affect the equilibrium content of $^8$Be(0$^+$) in helium plasma.

\begin{figure}
\includegraphics[width=0.7\textwidth]{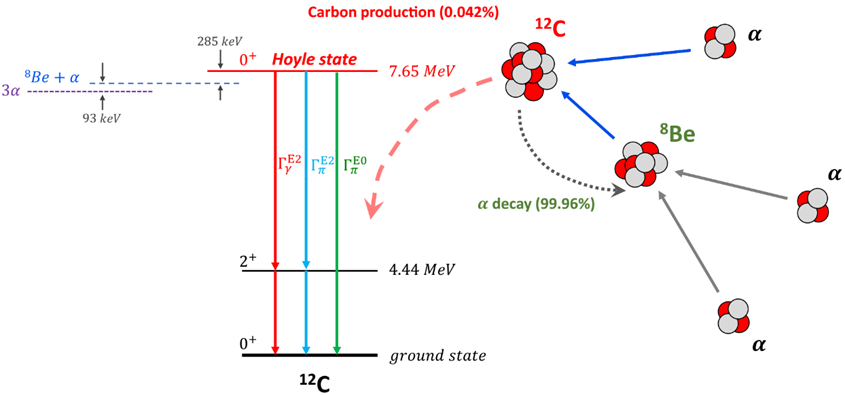}
\caption{\label{fig:2.2}Schematic representation of the triple $\alpha$-process in stars and the $^{12}$C states involved in this process. Possible electromagnetic transitions from the Hoyle state to the ground and first excited states in $^{12}$C are shown in different colors; $\pi$ is the $e^-e^+$ pair emission \cite{1}.}
\end{figure}

A detailed examination by F. Hoyle of the reaction rates and resulting abundances of $^4$He, $^{12}$C, and $^{16}$O indicated the inadequacy of this scenario and resulted in the proposal of forming the $^{12}$C via resonance in the region of the efficient interaction of $^8$Be(0$^+$) and $^4$He (Figure~\ref{fig:2.2}). The predicted value of 0.33 MeV above the 3$\alpha$-threshold was confirmed in the $^{14}$N($d$,$\alpha$) reaction of the $^{12}$C(0$^+_2$) level at 7.68 MeV with a width of less than 25 keV \cite{57,58}.
Early difficulties in detection are explained by its weak contribution to the $\alpha$-particle spectra (6\%). Thus, the synthesis of $^{12}$C occurs through the capture of a third $\alpha$-particle by $^8$Be(0$^+$). In this case, the transition to the ground state of $^{12}$C occurs with a probability of 1/2500 via 2$\gamma$-decay 0$^+$ $\to$ 2$^+$ $\to$ 0$^+$ or the production of $e^+e^-$ pairs. Being only 285 keV above the $^8$Be(0$^+$)$\alpha$ threshold, the energy of $^{12}$C(0$^+_2$) at 2.5$\times$10$^8$ K also corresponds to the peak of the Gamow window.

Being close to the critically sensitive region, $^{12}$C(0$^+_2$) provides the increase in the 3$\alpha$ reaction rate by 7–8 orders of magnitude, which is crucial for the synthesis of $^{12}$C. At the energy of several hundred keV higher, its dominance would be lost. Responsible for the generation of energy in the region of 10$^9$ K, the 3$\alpha$ process determines the luminosity of red giants. Not less important is the absence of a similar resonance in $^{16}$O in the fusion reaction $^{12}$C($\alpha$,$\gamma$)$^{16}$O. The states of $^{16}$O (7.12,1$^-$) just below the $\alpha$-decay threshold of 7.16 MeV and $^{16}$O (8.87,2$^-$) are inaccessible due to parity. This circumstance prevents rapid burnout of $^{12}$C and ensures a relative abundance of $^{12}$C/$^{16}$O of about 0.6. In this case, an alternative could be the $^{16}$O synthesis via combinations of the ground state $^{12}$C(0$^+_1$) and the excited states $^{12}$C(2$^+_1$), $^{12}$C(0$^+_2$), and $^{12}$C(3$^-$) \cite{59}. The importance of $^{12}$C(0$^+_2$) as an input to the synthesis of $^{12}$C and $^{16}$O motivates its ongoing research, as well as the search for more complex states of alpha particles and light nuclei at the binding thresholds. This topic, regularly updated, remains in the focus of nuclear theory and experiment.

\subsection{Development of clustering concepts}
Nucleons readily assemble into multi- nucleon systems, as indicated by the saturation of nuclear binding energy and density. A fundamental aspect of many-body nuclear dynamics, along with the average field of single nucleons, is their clustering. These approaches are deeply connected with unstable $\alpha$-particle states. The emerging structural complexity is reflected in the characteristics of the ground and excited states of atomic nuclei, whose description requires a unification of concepts.

Speculation that the $\alpha$-particle could serve as a structural element appeared with the formulation of the nuclear structure problem [1]. Within nuclei, $\alpha$-clusters were considered short-lived ingredients that dissolve within the nucleus after a certain period of time, after which other $\alpha$ particles are formed from the nucleons of the nucleus. An estimate of the lifetime of $\alpha$-clusters for proton or neutron exchange, significantly longer than the typical oscillation time in nuclei, assumes the presence of $\alpha$ clustering in the excited states. In the late 1930s, the resonant group method was developed, where the wave function of a nucleus was expressed as a linear combination of various cluster structures, each contributing its own weights. Note that these weights can be reflected in the probabilities of the relativistic dissociation channels of nuclei. Even-even nuclei $N$ = $Z$ were described in the model of $\alpha$ particles with connecting bonds. In the case of $^{12}$C, the $\alpha$-cluster model specifies two structures. The first, historically associated with the ground state, is an equilateral triangle, and the second is a linear arrangement (or chain). These simple representations retain a heuristic value for the excited states above the corresponding coupling thresholds.

\begin{figure}
\includegraphics[width=0.4\textwidth]{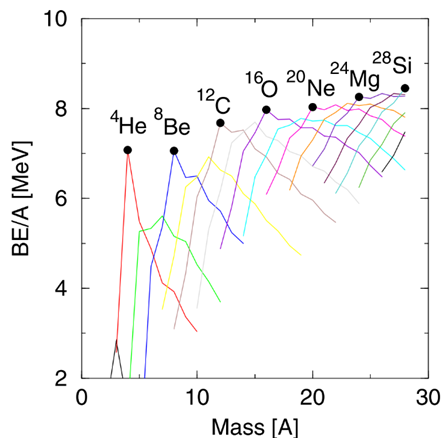}
\caption{\label{fig:2.3}Binding energy per nucleon of light nucleus systems (up to A = 28); lines connect the isotopes of each element \cite{44}.}
\end{figure}

\begin{figure}
\includegraphics[width=0.55\textwidth]{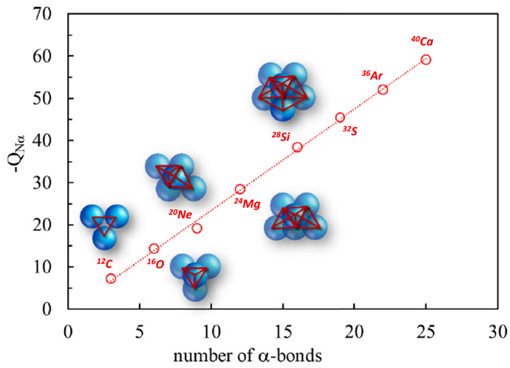}
\caption{\label{fig:2.4}Representation of the Wefelmeyer, Wheeler, and Hafstad-Teller $\alpha$-cluster model \cite{60}. The geometric arrangements of the clusters are shown according to the number of bonds between the $\alpha$-particles. The nearly linear relationship between the number of bonds and the energy required for the decay of the nucleus into $\alpha$-clusters is shown with the dotted line \cite{1}.}
\end{figure}

The binding energy of nucleons has shown that nuclei with an even and equal number of protons and neutrons (so-called $\alpha$-conjugated nuclei) are particularly stable (Figure~\ref{fig:2.3}). The number of possible $\alpha$-$\alpha$ bonds in them correlates with the binding energy values. Figure~\ref{fig:2.4} shows the dependence of the binding energy per nucleon on the energy of the first excited state for various nuclei. The linear dependence is interpreted as indicating the constancy of the $\alpha$–$\alpha$-interaction and the inertia of the $\alpha$-particle in the ground states of these nuclei. This view has not become generally accepted. On the contrary, it is believed that cluster structures diffuse into the ground states of $\alpha$-conjugated nuclei. At the same time, this dependence attests to $p$-$p$, $n$-$n$, and $n$-$p$ correlations associated with the population of common orbitals.

The shell model of non-interacting particles confined by a common mean field was proposed in analogy with the atomic structure. By incorporating spin-orbit interactions, it became possible in the late 1940s to explain magic numbers in shell fillings, magnetic moments, and single-particle excitations of medium and heavy nuclei. However, having eclipsed the first $\alpha$-particle models, it encountered difficulties in the region of light nuclei. For example, the ground state of $^6$Li with $J^\pi$ = 1$^+$ manifests itself as an $\alpha$$d$-structure with magnetic and quadrupole moments, in contradiction with the shell model. It is noteworthy that this structure and the exotically large size of $^6$Li were clearly evident in relativistic dissociation. The magnetic moment of $^9$Be was in a better agreement with the shell model than with the cluster picture of $n$$^8$Be(0$^+$). This difficulty was explained by describing the ground state of $^9$Be as a mixture of the long-lived and short-lived configurations 0.55 $^8$Be(0$^+$) + 0.45 $^8$Be(2$^+$). An indication of such a structure appeared in the relativistic fragmentation $^9$Be $\to$ 2$\alpha$ in NE \cite{5}. In the late 1950s, it was established that in the shell model it is difficult or even impossible to obtain the position of $^{12}$C(0$^+_2$) and the $J^\pi$ value of other excitations in $\alpha$-conjugated nuclei (for example, in $^{16}$O(0$^+_2$, 6.05). These conclusions, as well as emerging data on electron scattering by $\alpha$-particle levels of $^{12}$C, stimulated the development of the $\alpha$-cluster model.

\begin{figure}
\includegraphics[width=0.55\textwidth]{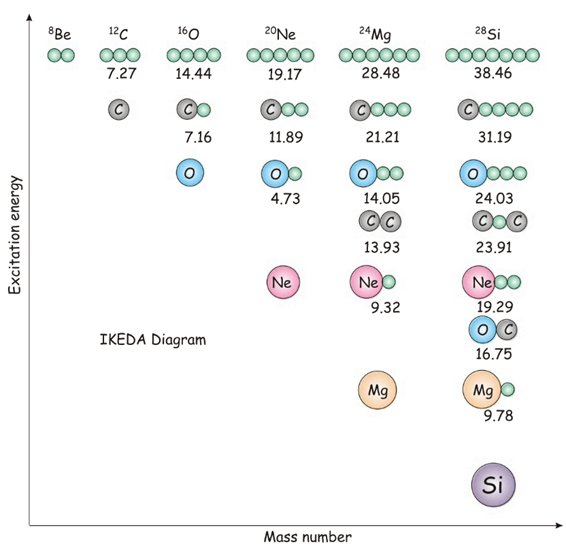}
\caption{\label{fig:2.5}Ikeda diagram \cite{44}. The diagram shows the evolution of clustering with increasing the excitation energy. The numbers indicate the excitation energies (MeV) at which cluster structures should emerge – these are the binding energies of the cluster components in the parent nucleus. An important concept conveyed by this diagram is that the cluster's degrees of freedom are released only near the cluster decay threshold. Thus, for heavy systems, the degrees of freedom N$\alpha$ appear only at the highest energy.}
\end{figure}

A new impetus was given by the prediction in 1956, radical for that time, that $^{12}$C(0$^+_2$) forms a 3$\alpha$-line. The idea that a cluster should not manifest itself in the ground state but arise as the internal energy of the nucleus increases, was recognized in the 1960s. For a nucleus to form a cluster structure, it must be energetically allowed, i.e., an energy equivalent to the mass difference between it and the clusters must be transferred to the nucleus. In other words, the cluster structure must manifest itself near the threshold and, probably, slightly below it. In the latter case, interactions between clusters are possible, which must be overcome for their complete separation. The Ikeda diagram illustrates the gradual transition from a compact ground state to the release of cluster degrees of freedom (Figure~\ref{fig:2.5}). Although the diagram shows a linear arrangement in the N$\alpha$ limit, this is not obligatory the most stable configuration. In fact, the linear structure has internal instability \cite{44}.

\subsection{Combination of cluster and shell aspects}
In the shell model, the $\alpha$-particle is a 2$p$2$n$ quartet in the 0S$_{1/2}$ orbital. The nucleon separation threshold of approximately 20 MeV, determined by the pairing of identical nucleons, results in the $\alpha$-particle's inertness in reactions and excitations. Its structure is the subject of few-nucleon physics. Due to non-central forces, the ground state wave function $J^\pi$ = 0$^+$ is represented as a mixture of three $^1S_0$, six $^3P_0$, and five $^5D_0$ orthogonal states with positive parity \cite{2}. The symmetric $S$-wave component dominates, with the $D$-wave contributing approximately 16\% and the $P$-wave contributing $\sim$1\%. In the structural hierarchy, the $\alpha$-particle is an independent object of study in the physics of few-nucleon systems like the deuteron, triton, and helion. In $\alpha$-particle-based models, quartets are assumed to form from pairs of protons and neutrons, which are associated with the zero total angular momentum. The $D$-wave may complicate this approximation.

The competition between the mean field of nucleons and their clustering leads to alternatives between shell and cluster descriptions. The transition from shells to cluster configurations was explained using the parabolic potential of a harmonic oscillator \cite{8}. In this oscillator, each nucleon moves within a linear restoring force formed by the interaction of other nucleons. The number of oscillator quanta and energy quantization are given by the solution of the Schrödinger equation in the potential where oscillations are possible along any of the Cartesian axes.

\begin{figure}
\includegraphics[width=0.4\textwidth]{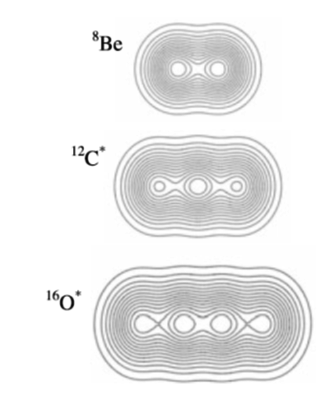}
\caption{\label{fig:2.6}Density of three configurations in the harmonic oscillator associated with the placement of $\alpha$-particles (pairs of protons and neutrons) in orbits with degeneracy 2, at deformations of 2:1, 3:1, and 4:1. The densities correspond to linear structures in the 2$\alpha$, 3$\alpha$, and 4$\alpha$ systems of $^8$Be, $^{12}$C, and $^{16}$O, respectively. In each case, the presence of $\alpha$-particles is obvious \cite{44}.}
\end{figure}

If the nucleus is deformed, for example, stretched along the z-axis, then, the potential in the x- and y-directions is reduced while maintaining the nucleus volume. It turns out that the magic numbers of deformed states can be expressed as sums of the spherical ones, localizing the related cluster structure at each deformation (Figure~\ref{fig:2.6}). At a 2:1 super deformation, cluster states manifest themselves in $^8$Be($\alpha$+$\alpha$), $^{20}$Ne($^{16}$O+$\alpha$), $^{32}$S($^{16}$O+$^{16}$O), and at a 3:1 hyperdeformation - $^{12}$C($\alpha$+$\alpha$+$\alpha$), $^{24}$Mg ($\alpha$+$^{16}$O+$\alpha$), etc. A hidden symmetry is preserved between the ground shell states and the cluster density distributions, which, in turn, form the mean field where the particles are moving. Thus, although the $\alpha$-particles themselves are not present in the nuclei, symmetries in their structure manifest themselves during deformations.

Clustering is substantiated by observable parameters. The moment of inertia of a rotating nucleus gives an idea of its deformation. The rotational levels of $^8$Be include 2$^+$(3.06) and 4$^+$(11.35). The ratio of their energy is 3.7, which is close to the expected value for the rotating nucleus (3.33). Calculations have given a distance between two alpha particles equal to twice the radius of an alpha particle. However, the rotational sequence alone does not prove either clustering or deformation. Measurements of the strengths of electromagnetic transitions allow us to verify the overlap of the structures of the initial and final states and the degree of collectivity. An important aspect of reactions involving $^{12}$C(0$^+_2$) is the mixing of cluster configurations with the ground state $^{12}$C(0$^+$1) \cite{39,45}. Studies of electromagnetic transitions between $^{12}$C(0$^+_1$) and $^{12}$C(0$^+_2$) have shown that, despite their different structures, their wave functions contain approximately 10\% impurity of each other. It is possible that during nuclear fragmentation, the deformed $^{12}$C(0$^+_2$) configuration can be populated as a component of $^{12}$C(0$^+_1$).

Another sign of clustering is the $\alpha$-decay dominance. The width of a state reveals details of its structure and decay. The greater the overlap of the initial structure with the state beyond the decay barrier, the shorter the lifetime and the larger the width. If the Coulomb and centrifugal barriers are removed, the reduced width can be compared to the Wigner limit - the widths at which $\alpha$-particles are fully formed. The width of $^8$Be(0$^+$) is close to this limit, further indicating a 2$\alpha$ structure.

\begin{figure}
\includegraphics[width=0.5\textwidth]{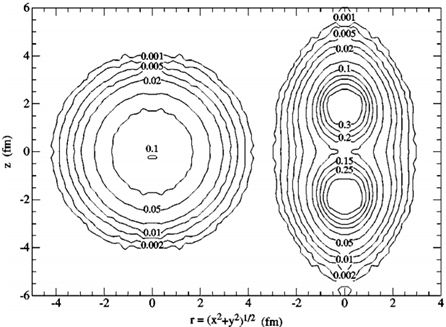}
\caption{\label{fig:2.7}Density of $^8$Be(0$^+$) calculated by the Monte Carlo method using the Green's function in the laboratory (left) and internal reference frames \cite{61}.}
\end{figure}

The de Broglie wavelength of $\alpha$-particles, $\lambda = h/\sqrt{(2M_\alpha E_\alpha)}$, estimated from the 100 keV resonance energy of $^8$Be(0$^+$), is approximately 15 Fermi. A similar estimate for the Hoyle state is approximately 20 Fermi \cite{46}. These data have indicated that the de Broglie wavelength is much larger than the distance between $\alpha$-particles, that favors the mean-field approach with Bose-state coherence. During the decay of a compound nucleus, the decay stages are usually statistically independent, and the formed $\alpha$-particles leave the nucleus one after another. However, in the coherent state, the $\alpha$-particles already exist, and their wave functions overlap. The decay can be simultaneous, preserving the phases of relative motion and lowering the Coulomb barrier. This further increases its probability for the observed unbound resonances of $^8$Be(0$^+$) and $^{12}$C(0$^+_2$), which can be interpreted as an observation of bosonic coherence.

\subsection{Influence of nucleon structure}
Contrary to the implicit assumption, $\alpha$-particles are not inert ingredients inside the nuclei. Interactions distort, polarize, and dissolve their spatial and spin structure. The nucleon degrees of freedom are fundamental. The description of the structure and excitations of light nuclei is determined by the details of the interactions and correlations between nucleons. Ab initio calculations of $^8$Be(0$^+$) \cite{61} were performed on the nucleon–nucleon interaction for all 2- and 3-particle components. The 2-particle interactions represent a parameterization of nucleon–nucleon scattering, and the 3-particle interactions are included using the parameterization of the pion exchange components. Remarkably, the nucleon degrees of freedom give the structure of $^8$Be(0$^+$) as $\alpha$-$\alpha$ pairs at 3-4 Fermi (Figure~\ref{fig:2.7}) \cite{44}. With the number of nucleons, the complexity of these calculations increases rapidly.

In the Bloch-Brink model, $\alpha$-particle quartets are composed of pairs of protons and neutrons in the 0$S$ state. The coupling of quartets in a harmonic oscillator is described by the functions:

\begin{equation}
\phi_i\left(r\right)=\ \sqrt{\frac{1}{b^3\pi^{3/2}}}\exp\left[\frac{-{(r-R_i)}^2}{2b^2}\right],
\end{equation}

where $R_i$ is a vector describing the location of the $i$-th quartet, and $b$ is a scaling parameter determining the size of the $\alpha$-particle. Since the degrees of freedom are fermionic, due to the Pauli exclusion principle, the antisymmetric wave function of the $\alpha$-ensemble is constructed as the Slater determinant:

\begin{equation}
\Phi_\alpha\left(R_1,R_2,\ldots,R_N\right)=K\mathcal{A}\prod_{i=1}^{N}{\phi_i\left(R_i\right).} 
\end{equation}

At short distances, it ensures the dissolution of $\alpha$-particles. The optimal arrangement of $\alpha$-particles is achieved by the variation optimization of the location and size of $\alpha$-particles. The corresponding configurations of the harmonic oscillator can be derived when the $\alpha$-particle separation tends to zero. The model is aimed at the states in which the $\alpha$-particle separation is such that their structure is irrelevant. This condition is achievable near the threshold, when the $\alpha$-particle can tunnel through the barrier to a weakly bound state, increasing the volume of the nucleus. $^{12}$C(0$^+_2$) is a candidate for this state, since electron scattering measurements have indicated that the volume associated with it is 3-4 times larger than that of $^{12}$C(0$^+_1$). The model obtained spatial images of $^{16}$O and $^{24}$Mg cluster systems and linear arrangements of $\alpha$-particles.

\begin{figure}
\includegraphics[width=0.55\textwidth]{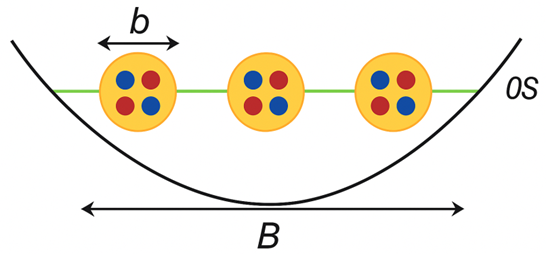}
\caption{\label{fig:2.8}Graphical representation of the THSR wave function \cite{46}. Three $\alpha$-particles are trapped in the 0$S$ state of the wide harmonic oscillator ($B$), and four nucleons of each $\alpha$-particle are trapped in the 0$S$ state of the narrow one ($b$). All nucleons are antisymmetrized.}
\end{figure}

The state described by a set of identical bosons could adopt bosonic symmetries and behave as an atomic Bose-Einstein condensate. The Bloch-Brink wave function was adapted by Tohsaki, Horiuchi, Shook, and Repke (THSR) to describe this opportunity \cite{46,47} (Figure~\ref{fig:2.8}). The THSR wave function of $\alpha$-particle triples is constructed as the antisymmetrized product of their wave functions:

\begin{equation}
\Phi_{3a}=\mathcal{A}\prod_{i=1}^{3}{\phi_{ai}\left(\vec{r_{1i}},\vec{r_{2i}},\vec{r_{3i}},\vec{r_{4i}}\right)}
\end{equation}

It refers to 12 nucleons grouped into three quartets described by $\phi_{\alpha i}$. The vectors $r_{1i}$, etc., denote the coordinates for each nucleon in the $i$-th quartet. The wave functions of each $\alpha$-particle are given by the following:

\begin{equation}
\phi_{ai}\left(\vec{r_1},\vec{r_2},\vec{r_3},\vec{r_4}\right)=e^\frac{\vec{-R^2}}{B^2}\exp{\left\{\frac{-\left[\vec{r_1}-\vec{r_2},\vec{r_1}-\vec{r_3},\ldots\right]^2}{b^2}\right\}},
\end{equation}

where $R$ is the coordinate of the quartet's center of mass. The quartet wave functions are Gaussian wave packets spatially spanned by an exponential function. As in the $\alpha$-cluster model, the parameter $b$ determines the size of the quartets, and $B$ is the size of the overall distribution. In the limit of $B$ $\to$ $\infty$, the antisymmetrization $A$ is inefficient, and the wave function is the product of wave packets of free $\alpha$-particles. Thus, although the THSR wave function is very similar to the wave function of the $\alpha$-cluster model, it possesses an additional variational degree of freedom.

\begin{figure}
\includegraphics[width=0.55\textwidth]{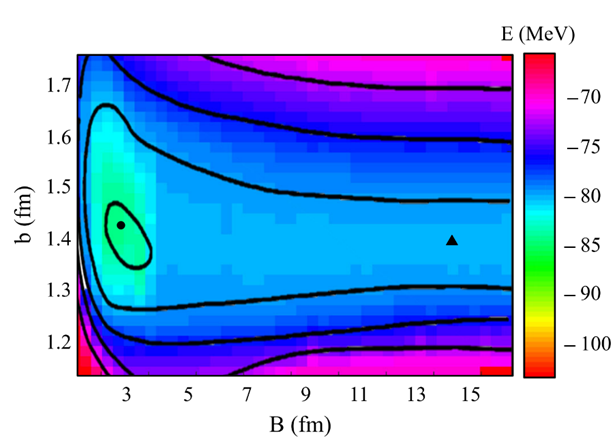}
\caption{\label{fig:2.9}Contour map of the $E_{3\alpha}$($B$, $b$) energy surface for $^{12}$C. The color map and contour lines denote binding energy \cite{48}. The dot represents a minimum on the energy surface, and the triangle marks a saddle point.}
\end{figure}
 
Possible structures are investigated by varying the parameters $b$ and $B$. The energy surface in their space is estimated as $\langle\mathrm{\Phi}_{3\alpha}|\hat{H}|\ \mathrm{\Phi}_{3\alpha}\rangle$, 
where the Hamiltonian includes the kinetic and Coulomb energies and the efficient nuclear interaction potential. Potentials reproducing the binding energy and radius of the $\alpha$-particle, and the $\alpha$-$\alpha$ scattering phase shifts, are used. The resulting surface for $^{12}$C is shown in Figure~\ref{fig:2.9}. Its minimum, indicated by the circle in Figure~\ref{fig:2.9}, corresponds to the binding energy of the ground state. The corresponding values of $b$ and $B$ at this minimum reproduce the size of the $\alpha$-particle and the compact ground state of $^{12}$C(0$^+_1$). A ridge extends from the minimum to large values of $B$, having a saddle point at $b$ $\approx$ 1.4 fm and $B$ $\approx$ 14 fm at the 3$\alpha$ energy threshold.

At the triangle point in Figure~\ref{fig:2.9}, stabilization in $^{12}$C is possible at higher B values than for $^{12}$C(0$^+_1$). Due to the energy and larger volume compared to $^{12}$C(0$^+_1$), this feature is identified as $^{12}$C(0$^+_2$). One of the major successes of this model is the reproduction of the form factor of elastic excitation by electrons of $^{12}$C(0$^+_2$) without arbitrary normalization. The agreement with these data has confirmed the spatial expansion of $^{12}$C(0$^+_2$) with the decisive influence of the internal $\alpha$-particle structure (Figure~\ref{fig:2.9}).

\begin{figure}
\includegraphics[width=0.55\textwidth]{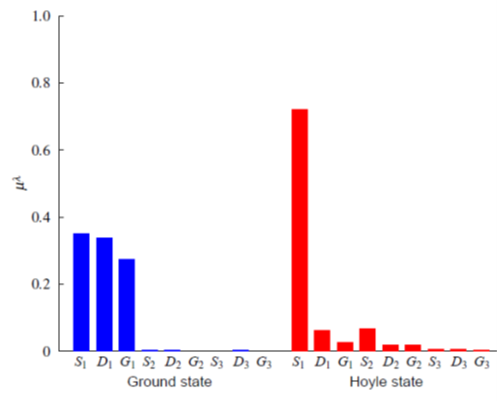}
\caption{\label{fig:2.10}Comparison of single $\alpha$-particle orbitals of the ground state of $^{12}$C and the Hoyle state \cite{47}.}
\end{figure}

The THSR approach allows one to decompose the wave function of $^{12}$C(0$^+_2$) into $\alpha$-particle orbitals. The probabilities of their occupation for $^{12}$C(0$^+_1$) and $^{12}$C(0$^+_2$) have exhibited a radical difference (Figure~\ref{fig:2.10}) \cite{47}. The predominance of the lowest 0$S$ orbital (70\%) indicates the significance of the approximation of $^{12}$C(0$^+_2$) as an ideal Bose gas. On the contrary, in accordance with the shell model, $^{12}$C(0$^+_1$) is fragmented into $s$, $d$, and $g$ levels. The THSR approach does not claim that $^{12}$C(0$^+_2$) is an $\alpha$-condensate. The contribution of orbitals other than 0$S$, estimated at 30\%, indicates a significant role of the Pauli Exclusion Principle.

\begin{figure}
\includegraphics[width=0.55\textwidth]{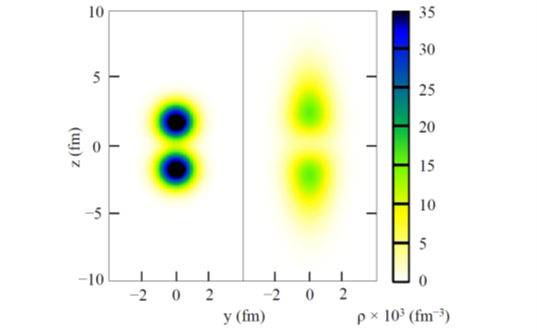}
\caption{\label{fig:2.11}Comparison of internal nucleon densities in the ground state of $^8$Be(0$^+$) calculated by using the Brink-Bloch (left) and THSR wave functions (right) \cite{48}.}
\end{figure}
 
\begin{figure}
\includegraphics[width=0.5\textwidth]{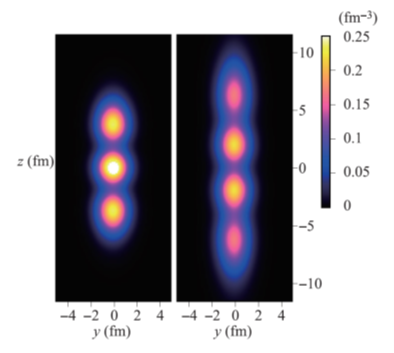}
\caption{\label{fig:2.12}Nucleon density distribution for TSHR wave function of the triplet and quartet of $\alpha$-particles \cite{62}.}
\end{figure}

A comparison of nucleon densities in the $^8$Be(0$^+$) $\alpha$-cluster and THSR models is shown in Figure~\ref{fig:2.11}. The former places $\alpha$-particles at fixed points in space, yielding a dumbbell-shaped 2$\alpha$ structure with a separation of 4 fm. The THSR model also predicts a similar dumbbell-shaped structure. However, repulsion is observed at short distances, and the tails appear at radii where the Coulomb repulsion is weakened. Figure~\ref{fig:2.12} shows the nucleon density distribution for the highly elongated TSHR wave function of the triplet and quartet of $\alpha$-particles \cite{62}.

\begin{figure}
\includegraphics[width=0.3\textwidth]{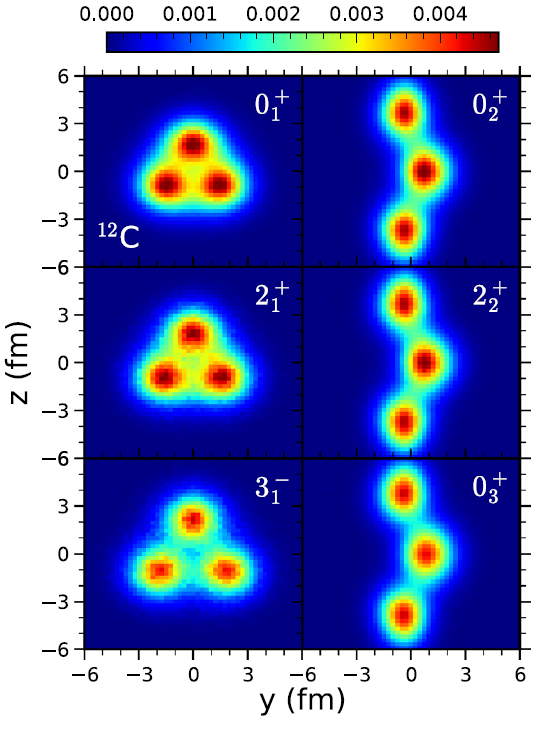}
\caption{\label{fig:2.13} Nuclear density distributions in several $^{12}$C states for a two-dimensional nuclear density projection. In each case, the orientation of the shortest mean square direction coincides with the X-axis \cite{50}.}
\end{figure}
 
A comprehensive description of the spatial structures of the $\alpha$-particle states of the $^{12}$C nucleus could provide insight into the nucleon correlations which result in nuclear binding and the opportunity of similar structures in other nuclear systems \cite{63,64}. There are two main obstacles to this. The first is the impossibility of calculating strong multiparticle correlations. The second is the impossibility of measuring the detailed spatial correlations needed to determine the internal structure of the 12-particle wave function. In efficient field theory simulations on an unlimited nuclear lattice that includes all possible multiparticle quantum correlations of protons and neutrons, $^{12}$C(0$^+_2$2) appeared along with other observed states of $^{12}$C \cite{50}. The straight chain does not appear to be the preferable configuration of $^{12}$C(0$^+_2$). The 0$^+_1$, 2$^+_1$, 3$^-$, 4$^-_1$, and 4$^+_1$ states have similar equilateral triangular shapes, as members of the 0$^+_1$ rotational band (Figure~\ref{fig:2.13}). The 0$^+_2$, 2$^+_2$, 0$^+_3$, and 4$^+_2$ states have similar obtuse-angled isosceles triangular shapes and are consistent with membership in the 0$^+_2$ rotational band (Figure~\ref{fig:2.13}). The equilateral triangle states have significant overlap with the antisymmetrized product of the mean-field shell model $^{12}$C(0$^+_1$), demonstrating duality between the shell model and cluster states. Contrary to this, the obtuse-angled isosceles triangle states have little overlap with the initial state of the shell model. The obtuse-angled shape corresponds to the orthogonality of the wave function with respect to the equilateral states. Thus, the $^{12}$C excitation spectrum contains two branches of states.

\subsection{Some experiments at low energy}
Progress in clustering experiments was made possible by forming the low-energy heavy ion beams and the introduction of silicon detectors \cite{39,65}. The energy and angle measurements of particles in the final states of nuclear reactions make it possible to determine the excitation of the resulting compound nuclei. Resonances with energies of 10–20 MeV, manifested in the reaction $^{12}$C($^{12}$C,$^{12}$C)$^{12}$C, were identified as pairs of rotating and vibrating $^{12}$C nuclei with large angular momenta. The research has expanded to more complex systems involving $^{16}$O, $^{24}$Mg, and $^{28}$Si. Sufficiently large arrays of silicon detectors enabled researchers to study breakup reactions such as $^{12}$C($^{24}$Mg, $^{12}$C$^{12}$C)$^{12}$C. The study of few $\alpha$-particle configurations carried out in the 90s became possible by reconstructing the decays of $^8$Be(0$^+$), $^{12}$C(0$^+_2$), and $^{12}$C(3$^-$). For this purpose, the strip silicon detectors, both single-ended with charge division and double-ended, were applied. This type of detector was developed in the late 80s to reconstruct the decays of charmed particles.

\begin{figure}
\includegraphics[width=0.65\textwidth]{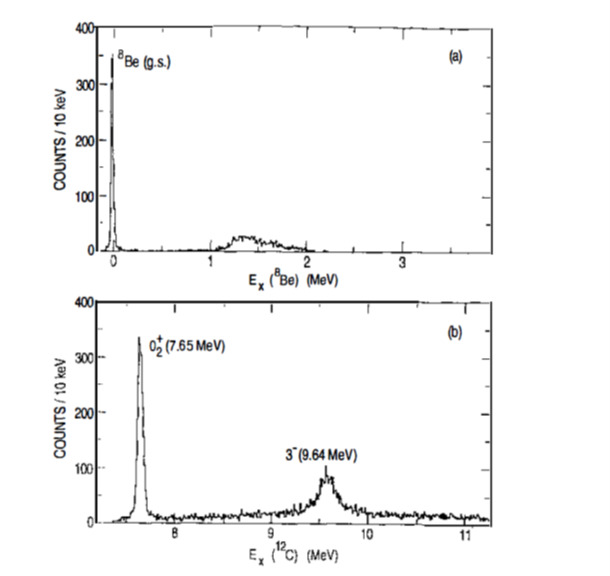}
\caption{\label{fig:2.14}Excitation spectrum for $\alpha$-particle pairs (a) and triplets (b) produced in $^{12}$C + $^{12}$C scattering at 65 MeV. The 2$\alpha$ spectrum was obtained for pairs of $\alpha$-particles from the decay of the $^{12}$C(3$^-$) level \cite{39}.}
\end{figure}

The main spectrum of 3$\alpha$-excitations of $^{12}$C begins with the $^{12}$C(3$^-$) level at 9.64 MeV, which has a width of $\Gamma$ = 46 keV (Figure~\ref{fig:2.1}). The proposed structure of $^{12}$C(3$^-$) corresponds to rotation around the axis passing through the center of a triangle where each $\alpha$-particle carries a unit of angular momentum \cite{44}. The 2 MeV gap between $^{12}$C(3$^-$) and $^{12}$C(0$^+_2$) allows one to identify parallelly their 3$\alpha$-decays. At 10-20 MeV/nucleon, the reactions $^{12}$C($^{12}$C,2$\alpha$)$^{16}$O and $^{12}$C($^{12}$C,3$\alpha$)$^{12}$C were studied. The decays of $^8$Be(0$^+$) were reconstructed with a resolution of 40 keV, and $^{12}$C(0$^+_2$) and $^{12}$C(3$^-$) having a width of 46 keV - 70 keV, correspondingly (Figure~\ref{fig:2.14}).

The above provided to search for resonances decaying into $^{12}$C(0$^+_2$) and $^{12}$C(3$^-$) \cite{39}. The reconstructed excitation energy reveals the production of an alpha particle from the decay of $^8$Be(0$^+$) or the decay of $^{12}$C levels. The kinetic energy and scattering angle of the decaying nucleus are used to calculate the energy balance $Q$ of the two-body reaction. Identification of the final state of inelastic scattering in each event is possible. As a result, the $Q$ distributions of the $^{12}$C($^{12}$C,$^{12}$C(0$^+_2$,3$^-$))$^{12}$C reactions showed many narrow excitations in the $^{12}$C$^*$ pairs 0$^+_2$ + 0$^+_1$, 0$^+_2$ + 2$^+_1$, 0$^+_2$ + 0$^+_2$, 0$^+_2$ + 3$^-$, 0$^+_1$ + 3$^-$, 2$^+_1$ + 3$^-$, 0$^+_2$ + 3$^-$, 3$^-$ + 3$^-$. A peak at 32.5 MeV in the c.m.s. (about 5 MeV above the mass threshold) with a width of about 5 MeV was found in the excitation function of this reaction. The initial assumption that a 6$\alpha$-chain is formed in the deformed compound nucleus $^{24}$Mg has not been confirmed. Due to its difference from the ground state of $^{24}$Mg, the formation of the 2$^{12}$C(0$^+_2$) pair with a higher angular momentum is more promising. Later, the experimental studies of cluster structures and unstable states using silicon detectors were shifted to beams of light radioactive isotopes (recently \cite{51}).

State-of-art experiments searching for 4$\alpha$BEC have continued a similar application of strip silicon detectors. They are based on compact spectrometers providing significant solid angle coverage, using beams of light nuclei at several tens of MeV per nucleon \cite{66,67,68,69,70,71,72}. Silicon detectors having the best energy resolution are placed in vacuum volumes near ultrathin targets. Identification of unstable nuclei and states is carried out on the energy and angular correlations in the detected ensembles of $\alpha$ particles.

The experiment with full registration of $\alpha$-particle fragments of the projectile in the reaction $^{40}$Ca(25 MeV/nucleon) + $^{12}$C indicated the increase in the contribution of $^8$Be to a 6$\alpha$ multiplicity \cite{66}. This fact contradicts to the model, which has predicted a decrease (Table 2 in \cite{66}). A search for decays of the state $^{16}$O(0$^+_6$,15.1 MeV) $^{20}$Ne(12 MeV/nucleon) + $^4$He \cite{67} and $^{16}$O(160, 280, 400 MeV) + $^{12}$C \cite{69} was carried out. Recently, the data on $^{16}$O(45 MeV) + $^{12}$C $\to$ 4$\alpha$ in full kinematics \cite{71} have been analyzed for all possible configurations. The excitation function was reconstructed directly from 4$\alpha$, as well as for special decay channels such as $^{12}$C(0$^+_2$)$\alpha$, $^{12}$C(3$^-$)$\alpha$, and 2$^8$Be. However, the search for the 15.1 MeV state remains unsuccessful in all cases \cite{70}. Coincidence measurements of the 386 MeV $\alpha$-particles scattered by 0$^\circ$ in the $^{20}$Ne($\alpha$,$\alpha$’)5$\alpha$ reaction were carried out \cite{72}. The newly observed states at 23.6, 21.8, and 21.2 MeV in $^{20}$Ne are claimed to be strongly associated with the 4$\alpha$BEC candidate and are themselves $\alpha$BEC candidates.

Although the status of $\alpha$BEC observations remains uncertain \cite{69}, it has been established that $^{12}$C(0$^+_2$) is produced by fragmentation of more than just $^{12}$C. This fact indicates that $^{12}$C(0$^+_2$), like $^8$Be(0$^+$), is independent of the parent nucleus. Similar universality should be exhibited by $\alpha$BEC candidates. It turns out that experiments searching for 4$\alpha$BEC states have reached their practical statistical limit. Focusing on peripheral collisions of heavier nuclei, having higher-energy, is required. It is also required to combine the data obtained over the broadest possible energy range and, on this basis, confirm the universality of $\alpha$BEC, a relativistic invariant representation of unstable states.

Light nuclei near the stability boundaries have very few bound states. Most of their states are in the continuum, being unstable to particle emission. For nuclei beyond the stability boundaries, all the states are in the continuum, including the ground states. Along with radioactive nuclei, the instability of these nuclei does not reduce their value to expand nuclear astrophysics scenarios. They can also play an intermediate role in inverse fusion processes, since their decay products are He and H stable isotopes. Studying the continuum spectroscopy of these light nuclei requests to reconstruct the invariant masses of the final ensembles. In this regard, it is important to note the research headed by R. Charity with strip silicon detectors in the beams of light radioactive nuclei at the University of Michigan Superconducting Cyclotron and the Texas A\&M University Cyclotron (review \cite{68}). For example, in the $^9$C + $^9$Be reaction at 70 MeV per nucleon, the invariant mass method identified 42 resonances in the continuum that decay into output channels containing charged particles. These resonances represent continuum states in the nuclei $^{5,6,7}$Li, $^{6,7,8}$Be, $^{7,8,9}$B, and $^{8,9,10}$C. Only half of these resonances were associated with binary output channels, while the rest were associated with 3, 4 and 5-particle decays. We highlight some observations regarding the dynamics of the emergence of the unstable states.

It is proposed to consider the fragmentation of a projectile nucleus as a two-stage process. The time scale of the initial instantaneous step was estimated from the fragment accelerations after the decay in the target's Coulomb field. It turned out to be the same interaction time of the projectile nucleus with the target. Resonances can arise during this initial step and decay in the second step, forming delayed protons and residual products. To maintain the time separation between the two steps, so that the resonance decay is independent of its mode of origin, the resonance lifetime must be significantly longer than the interaction time of the projectile with the target. For the reactions considered in this work, this time is estimated as the time spent by the projectile to traverse the diameter of the target nucleus, i.e., 5 $\times$ 10$^{-23}$ s. A resonance with a decay width of 2.5 MeV has a lifetime of 2.4 $\times$ 10$^{-22}$ s, which meets this requirement. Thus, resonances of this width and the narrower ones can be studied in the reactions considered in this work. It is possible that several mechanisms could provide the yield channels of projectile fragmentation under consideration. Sometimes there are processes where one or two projectile nucleons strongly interact with the target and are knocked out of the projectile, leaving a bound or resonant state. However, the question arises when considering resonances which are much lighter than the projectile: are other projectile nucleons not accounted for in the resonance removed during the knockout process? An alternative scenario is that the projectile undergoes multifragmentation, producing nucleons, possibly some light clusters, and a resonance.

Although the presented arguments have implications for the levels beyond both the proton and neutron stability boundaries due to detection efficiency the attention is focused on proton-rich states. The knock-on and double-neutron knock-on reactions can be viewed as pure methods to search for proton resonances beyond the stability boundary. The first step, the direct knock-on process, produces no protons, and, thus, only delayed protons from the second step (resonant decay), are detected. However, there is a limit to how far one can probe beyond the proton stability boundary, restricting such "pure" reactions to those having no ``dirty'' prompt protons from the knock-on reaction. Furthermore, in the invariant mass method, such pure processes cannot be completely separated from ``dirty'' processes which produce this ``dirty'' prompt protons in addition to "pure" delayed protons from the resonance decay. This is true even for the knock-on and double-neutron knock-on reactions. For example, the yield of $p$ + ($Z$ projectile - 1, $A$ projectile - 2) events may have a contribution from the delayed protons produced by resonance decay after the 1$n$ knock-out reaction, or from the events where both the proton and neutron are rapidly removed from the projectile in the first step. In the invariant mass spectrum, the resonance peak will be located against the background of the latter events. Three-neutron knock-out should, in principle, allow exploration even further, beyond the experimental boundary, but the very small yields expected for this reaction may be overwhelmed by such a background. Since invariant mass backgrounds will always be present in these fragmentation reactions, it is important to consider whether fragmentation of particles with higher atomic numbers could give a similar resonance with more intense yields and better peak-to-background ratios.

\begin{figure}
\includegraphics[width=1.0\textwidth]{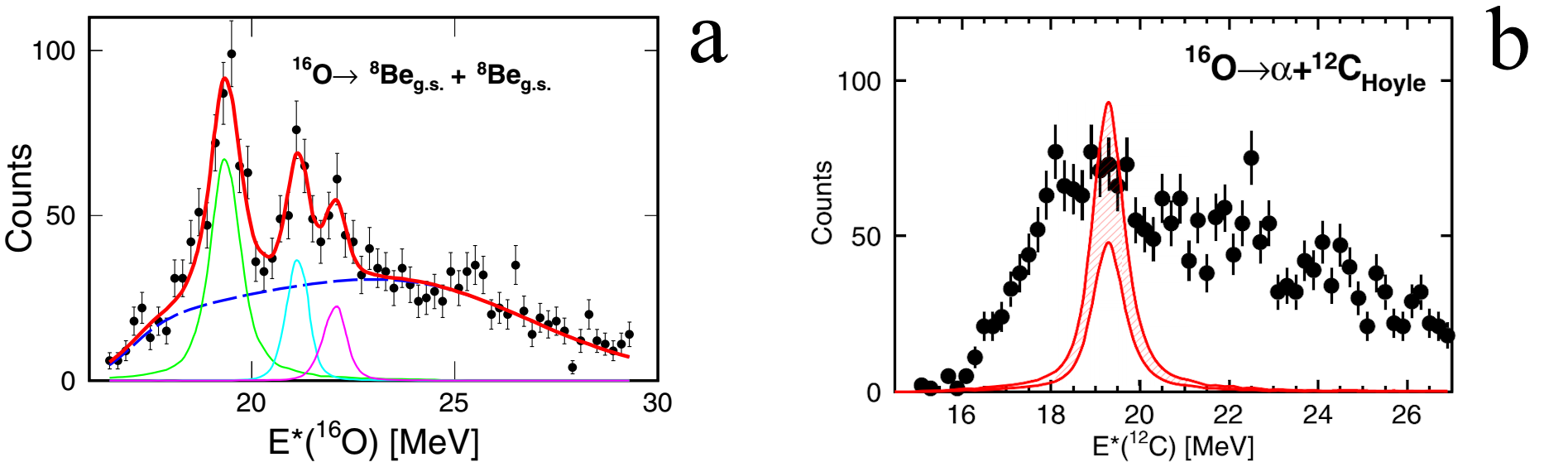}
\caption{\label{fig:2.15}(a) Excitation energy spectrum of 2$^8$Be(0$^+$) \cite{73}. (b) Excitation energy spectrum of $^{12}$C(0$^+_2$)$\alpha$ \cite{73}.}
\end{figure}
 
The $^{10}$C isotope has only one bound excited state, 2$^+_1$, above which most levels decay to 2$\alpha$2$p$. When these $^{10}$C states are formed from the decay of the incident $^{13}$O particle, there is an opportunity of producing two prompt protons in addition to the two protons from the resonance decay. Consequently, there are potentially many ways to generate background signals beneath the $^{10}$C resonance peaks. The problem is simplified by selecting successive decays of these $^{10}$C states. The invariant mass distribution of two 2$\alpha$$p$ ensembles has shown a strong narrow peak just above the threshold, which is located against a negligible background. By isolating this peak, the events with an intermediate $^9$B fragment can be clearly distinguished and the proton from the $^{10}$C resonance decay to $^9$B can be identified. Thus, these events can be taken as $^9$B$p$. The $^{10}$C excitation energy distribution contains three narrow peaks and regions determined by the background from prompt protons.

A large number of 4$\alpha$ events have been detected with a mixed $^{15}$O/$^{17}$Ne beam \cite{73}. Their production requires the transfer of neutrons, whose packaging into $\alpha$-particles also serves as a source of higher excitation energy. Although the invariant mass distribution has no peaks, these events can be produced in various scenarios. One interesting case is the fission of $^{16}$O into 2$^8$Be(0$^+$) pair. The events have been selected where the relative energy between one pair of $\alpha$-particles is consistent with the decay of $^8$Be(0$^+$), and similarly for the remaining pair. The relative energy distribution has a very sharp peak for $\alpha$-pairs from $^8$Be(0$^+$) decay, and there is virtually no background beneath it. The excitation energy spectrum for such events contains a large peak at 19.26 MeV and a broader structure at $\sim$21 MeV (Figure~\ref{fig:2.15}(b)). One can isolate these 4$\alpha$-events by selecting those where three of the four $\alpha$-particles have the invariant mass of $^{12}$C(0$^+_2$) (Figure~\ref{fig:2.15}(a)).

It is worth noting that the representation of data in terms of the invariant mass variable, denoted as the excitation energy $E^*$, was adopted at approximately the same time and independently in the BECQUEREL experiment. The term "invariant mass $Q$" is used there, with a note indicating which mass constant is subtracted for ease of presentation. The concept is the same, but in our case, the mass is calculated assuming conservation of the initial velocity of the projectile by relativistic fragments. A comparison of the data reveals several similarities, which may be based on a common mechanism to form the unstable states. It is possible to say that the unstable states may not simply result from shaking the parent nuclei, from which they precipitate like ingredients. Another way is as follows: they may form during the final-state interaction of clusters and nucleons with low relative $4$-momenta (velocities) because of their coalescence. The invariant approach allows one to project low-energy observations into the relativistic region. In addition, they can be included as fundamental elements of multiple fragmentation events of heavy nuclei that is not feasible at low energy.

\section{Search for the Hoyle state in the dissociation of light nuclei} 
\subsection{$^8$Be and $^9$B in relativistic dissociation of $^9$Be and $^{10}$C} 
\begin{figure}
\includegraphics[width=1.0\textwidth]{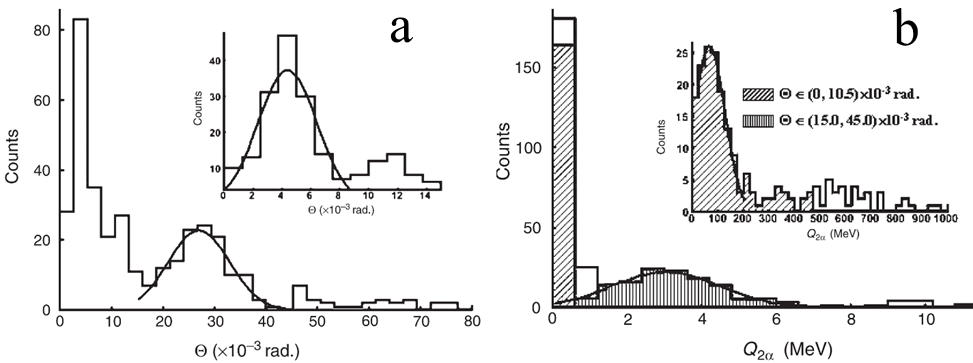}
\caption{\label{fig:3.1}Distribution of $\alpha$-pairs in the fragmentation $^9$Be $\to$ 2$\alpha$ at 1.2 GeV per nucleon over the opening angle $\Theta$ (a) and the invariant mass $Q_{2\alpha}$ (b) \cite{3,74}.}
\end{figure}

$^9$Be is the simplest object to test the approach based only on the $\Theta$ angle in relativistic $\alpha$-pairs (Figure~\ref{fig:3.1}(a)) \cite{3,74}. The invariant mass distribution of $\alpha$-pairs $Q_{2\alpha}$ $^9$Be $\to$ 2$\alpha$ at 1.2 GeV per nucleon has indicated that in 80\% of cases, the ``long-lived'' $^8$Be(0$^+$) and ``short-lived'' $^8$Be(2$^+$) are formed with close probabilities (Figure~\ref{fig:3.1}(b)). This conclusion is consistent with the use of the superposition (0.56$^8$Be(0$^+$)+0.44$^8$Be(2$^+$))n in calculating the magnetic moment of $^9$Be.

\begin{figure}
\includegraphics[width=0.6\textwidth]{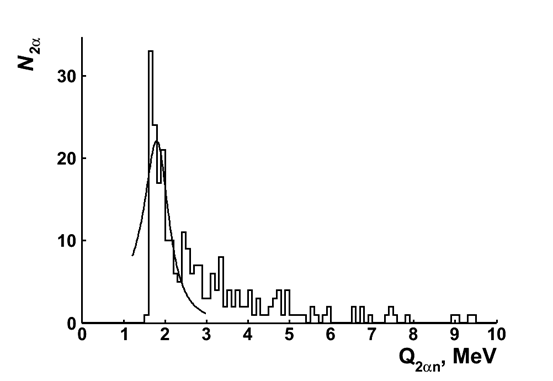}
\caption{\label{fig:3.2}Distribution over invariant masses of triplets consisting of pairs of $\alpha$-particles $Q_{2\alpha}$ $<$ 0.25 MeV and neutrons in dissociation $^9$Be $\to$ 2$\alpha$ at 2 GeV/$c$ per nucleon; the curve - Breit-Wigner distribution \cite{6}.}
\end{figure}

To date, the number of measured stars $^9$Be $\to$ 2$\alpha$ has reached 712 \cite{6}. The average value of the total transverse momentum in the dissociation $^9$Be $\to$ 2$\alpha$ is about 10 MeV/$c$ per nucleon, which is several times less than the Fermi momentum of nucleon motion (100-200 MeV/$c$), which manifests itself in the emission of neutrons. Neglecting the total transverse momentum, one can estimate the transverse momentum carried away by a neutron and then, – the invariant masses $Q_{2\alpha n}$. Under the condition $Q_{2\alpha}$ $<$ 0.25 MeV in the presence of $^8$Be(0$^+$), a peak appears in the distribution $Q_{2\alpha n}$ (Figure~\ref{fig:3.2}) near the $^8$Be(0$^+$)$p$ threshold equal to 1.665 MeV. Approximation of this distribution by the Breit-Wigner function yields a resonance energy 1.80 $\pm$ 0.01 MeV at width $\Gamma$= 732 keV determined by the experimental resolution and the approximations made. These parameters are consistent with approximately 100 decays of the level $^9$Be$^*$(1.684 MeV, $\Gamma$ = 217 keV, $J^\pi$ = 2$^+$) up to $Q_{2\alpha n}$ $<$ 2 MeV. Their contribution to the $^9$Be $\to$ $^8$Be(0$^+$) channel is 33 $\pm$ 4\% and 14 $\pm$ 2\% to the total statistics. Under the condition 0.25 $<$ $Q_{2\alpha}$ $<$ 0.85 MeV, i.e., with a veto on $^8$Be(0$^+$) in the region of 2 $<$ $Q_{2\alpha n}$ $<$ 3 MeV, where the signal of $^9$Be(2.43 MeV) is expected, no more than 25 events or no more than 4\% of the dissociation events $^9$Be $\to$ 2$\alpha$, correspondingly.

\begin{figure}
\includegraphics[width=1.0\textwidth]{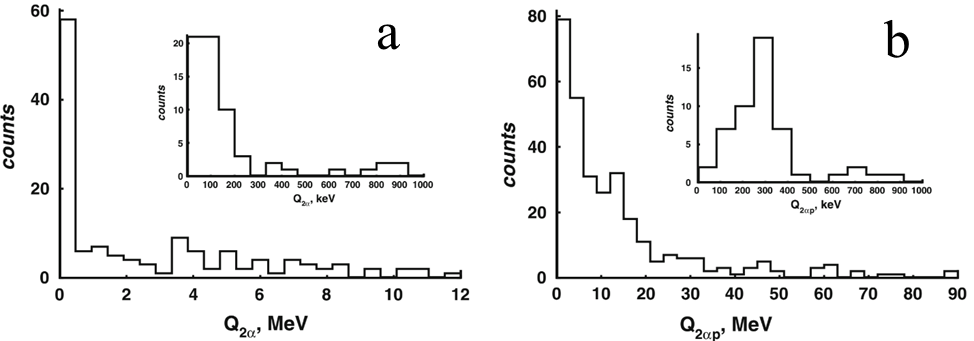}
\caption{\label{fig:3.3}Distribution in the coherent dissociation events of $^{10}$C $\to$ 2$\alpha$2$p$ at 1.2 GeV per nucleon over invariant masses $Q_{2\alpha}$ (a) and $Q_{2\alpha p}$ (b) \cite{3,75}.}
\end{figure}

The $^{10}$C nucleus is the only stable super-Boromean-type system where the removal, of at least, one cluster or nucleon results in an unbound state. The threshold for $^{10}$C $\to$ 2$\alpha$2$p$ is 3.73 MeV, for $^9$B$p$ it is 4.01 MeV and for $^8$Be(0$^+$)2$p$ it is 3.82 MeV. The coherent dissociation of $^{10}$C is dominated by the 2He + 2H channel (82\%). The $Q_{2\alpha}$ and $Q_{2\alpha p}$ distributions (Figure~\ref{fig:3.3}) of 184 ``white'' $^{10}$C $\to$ 2$\alpha$2$p$ stars indicate a (30 $\pm$ 4)\% contribution from $^9$B decays, which exclusively involve $^8$Be(0$^+$) \cite{3,75}.

\begin{figure}
\includegraphics[width=0.5\textwidth]{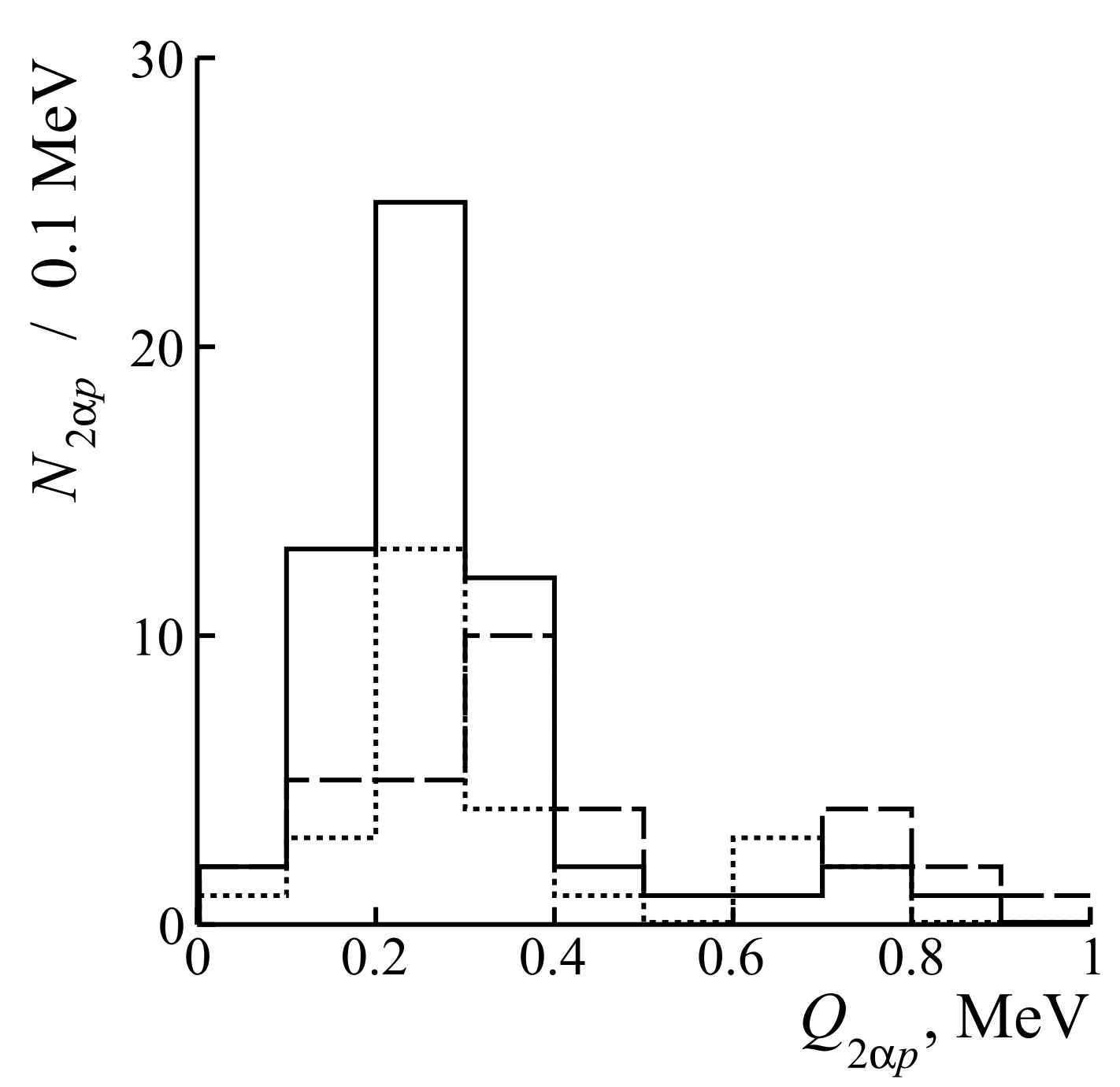}
\caption{\label{fig:3.4}Distribution of the number of 2$\alpha$$p$ triples $N_{2\alpha p}$ over invariant mass $Q_{2\alpha p}$ ($<$ 1 MeV) in the events of coherent dissociation $^{10}$C $\to$ 2He2H (solid line) and dissociation $^{11}$C $\to$ 2He2H (dots) and $^{10}$B $\to$ 2HeH (dashed line) \cite{5}.}
\end{figure}

The identification of $^8$Be(0$^+$) and $^9$B simplified due to the composition of $^{10}$C allows one to address their contribution to the dissociation of $^{10}$B and $^{11}$C. Angular measurements were performed in 318 $^{10}$B $\to$ 2He + H dissociation events at 1.0 GeV per nucleon, among which 20 $^9$B $\to$ $^8$Be(0$^+$)$p$ decays were identified, satisfying the condition $Q_{2\alpha p}$($^9$B) $<$ 0.5 MeV (Figure~\ref{fig:3.4}). Similarly, 22 $^9$B decays were found in 154 $^{11}$C $\to$ 2He + 2H events at 1.2 GeV per nucleon (Figure~\ref{fig:3.4}). Thus, the universal condition $Q_{2\alpha p}$($^9$B) was established in the dissociation of $^{10}$C, $^{10}$B, and $^{11}$C. Moreover, when identifying the decays $^9$B $\to$ $^8$Be$p$, the criterion $Q_{2\alpha}$($^8$Be) $<$ 200 keV was confirmed under the purest conditions.

\subsection{Dissociation of $^{12}$C} 
The development of reconstruction based on the invariant masses of relativistic decays of $^8$Be and $^9$B has shown the prospect of searching for $^{12}$C(0$^+_2$) decays which should manifest themselves as the narrowest $\alpha$-triples. The starting material was 200 $\mu$m NE layers on 2 mm glass, measuring 9 by 12 cm$^2$, produced by Slavich company. In 2016–2017, they were longitudinally exposed in the biomedical beam of $^{12}$C nuclei of the Institute of High Energy Physics (Protvino) at 450 MeV (1 momentum GeV/$c$) per nucleon with a flux of 2000-4500 cm$^{-2}$ of 2\% uniformity. The search for 3$\alpha$ events was carried out along the bands transverse to the beam direction. 86 $^{12}$C $\to$ 3$\alpha$ events were measured, including 36 ``white'' stars \cite{76}. The first step involved a reanalysis of the measurements of 72 ``white'' $^{12}$C $\to$ 3$\alpha$ stars by G.M. Chernov's group (Tashkent) \cite{77} as well as 114 ``white'' stars by A.Sh. Gaitinov's group (Alma-Ata) in NE layers exposed to $^{12}$C nuclei at 4.5 GeV/$c$ per nucleon at the JINR Synchrophasotron \cite{76}. The $^{12}$C(0$^+_2$) problem was not addressed at that time. 

In the distribution of the invariant mass of $\alpha$-triplets, a peak was observed at $Q_{3\alpha}$ $<$ 1 MeV where $^{12}$C(0$^+_2$) decays may be reflected. For events at 3.65 GeV per nucleon, $\langle Q_{3\alpha} \rangle$ (RMS) = 441 $\pm$ 34 (190) keV, and at 450 MeV per nucleon 346 $\pm$ 28 (85) keV. It was concluded that $^{12}$C(0$^+_2$) decays are observed with a probability of about 10-15\%. Angular measurements allow us to draw conclusions about the dynamics of $^{12}$C(0$^+_2$) occurrence based on the distribution of $\alpha$-particle triples over their total transverse momentum $P_{Tsum}$. Its mean value $\langle P_{Tsum} \rangle$ (RMS) is 190 $\pm$ 19 (118) MeV/$c$ which corresponds to the nuclear diffraction mechanism. In the case of electromagnetic dissociation on Ag and Br nuclei which are NE part the expected limitation $P_{Tsum}$ $<$ 100 MeV/$c$. It is surprising that such a ``fragile'' formation as $^{12}$C(0$^+_2$) can arise in relativistic collisions as a whole ensemble ``rebounding'' with a transverse momentum characteristic of strong interactions, not the electromagnetic ones. The $Q_{3\alpha}$ distribution contains the $^{12}$C(3$^-$) peak between 2 and 4 MeV. 

At present the emission angles are measured in 510 $^{12}$C $\to$ 3$\alpha$ events with a momentum of 4.5 GeV/$c$ per nucleon \cite{6}. Such a significant level of statistics is achieved due to a targeted search while transverse scanning. Its part corresponds to a proportional set of 164 coherent dissociations and 160 3$\alpha$-dissociations accompanied with target nuclear fragments. Analysis of the relative yields of $^8$Be(0$^+$), $^{12}$C(0$^+_2$) and $^{12}$C(3$^-$), and the distributions over the transferred transverse momentum of these two types of dissociation in this sample have not revealed differences. The total statistics are discussed below. The number of $Q_{2\alpha}$($^8$Be) decays $<$ 200 keV in it is 221.

\begin{figure}
\includegraphics[width=0.7\textwidth]{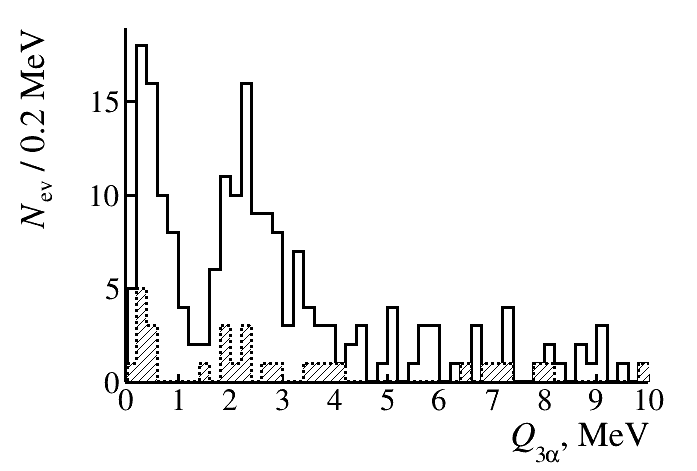}
\caption{\label{fig:3.5}Combined distributions of the invariant mass $Q_{3\alpha}$ of $\alpha$-triples with the presence of an $\alpha$-pair $Q_{2\alpha}$($^8$Be) $<$ 200 keV in the dissociation $^{12}$C $\to$ 3$\alpha$ at 3.65 GeV per nucleon (solid) \cite{6} and 450 MeV per nucleon (dots, shaded) \cite{76}.}
\end{figure}

At $Q_{2\alpha}$($^8$Be) $<$ 200 keV two peaks are observed in the distribution over $Q_{3\alpha}$ (Figure~\ref{fig:3.5}). The first with the average value $Q_{3\alpha}$(RMS) = 417 $\pm$ 27 (165) keV corresponds to $^{12}$C(0$^+_2$), and the second one - with the parameters of the Rayleigh distribution $Q_{2\alpha}$($\sigma$) = 2.4 $\pm$ 0.1 MeV – $^{12}$C(3$^-$). Since the peaks are well separated for $^{12}$C(0$^+_2$) decays, the soft condition $Q_{3\alpha}$ $<$ 1 MeV is adopted. The rapid decrease in the contribution $Q_{3\alpha}$ $>$ 4 MeV determines the upper limit of $^{12}$C(3$^-$). The contributions of $^8$Be(0$^+$), $^{12}$C(0$^+_2$) and $^{12}$C(3$^-$) are estimated as 43 $\pm$ 4, 11 $\pm$ 2, 19 $\pm$ 2\%, respectively. The contribution of $^{12}$C(0$^+_2$) to the decays of $^8$Be(0$^+$) is 26 $\pm$ 4\%, and $^{12}$C(3$^-$) is 44 $\pm$ 6\%, and their ratio is 0.6 $\pm$ 0.1. 

Consistent with the accepted concepts, the joint identification of the decays of $^8$Be(0$^+$), $^{12}$C(0$^+_2$) and $^{12}$C(3$^-$) in the cone of relativistic fragmentation of the $^{12}$C nucleus only at the angles of emission of $\alpha$-particles serves as a strong argument in favor of the approach used and indicates the opportunities of its application to heavier nuclei. The initial identification of $^{12}$C(0$^+_2$) at 450 MeV per nucleon (Figure~\ref{fig:3.5}) brings this opportunity to the energy range where such electron experiments become feasible.

\subsection{Dissociation of $^{16}$O}
\begin{figure}
\includegraphics[width=0.7\textwidth]{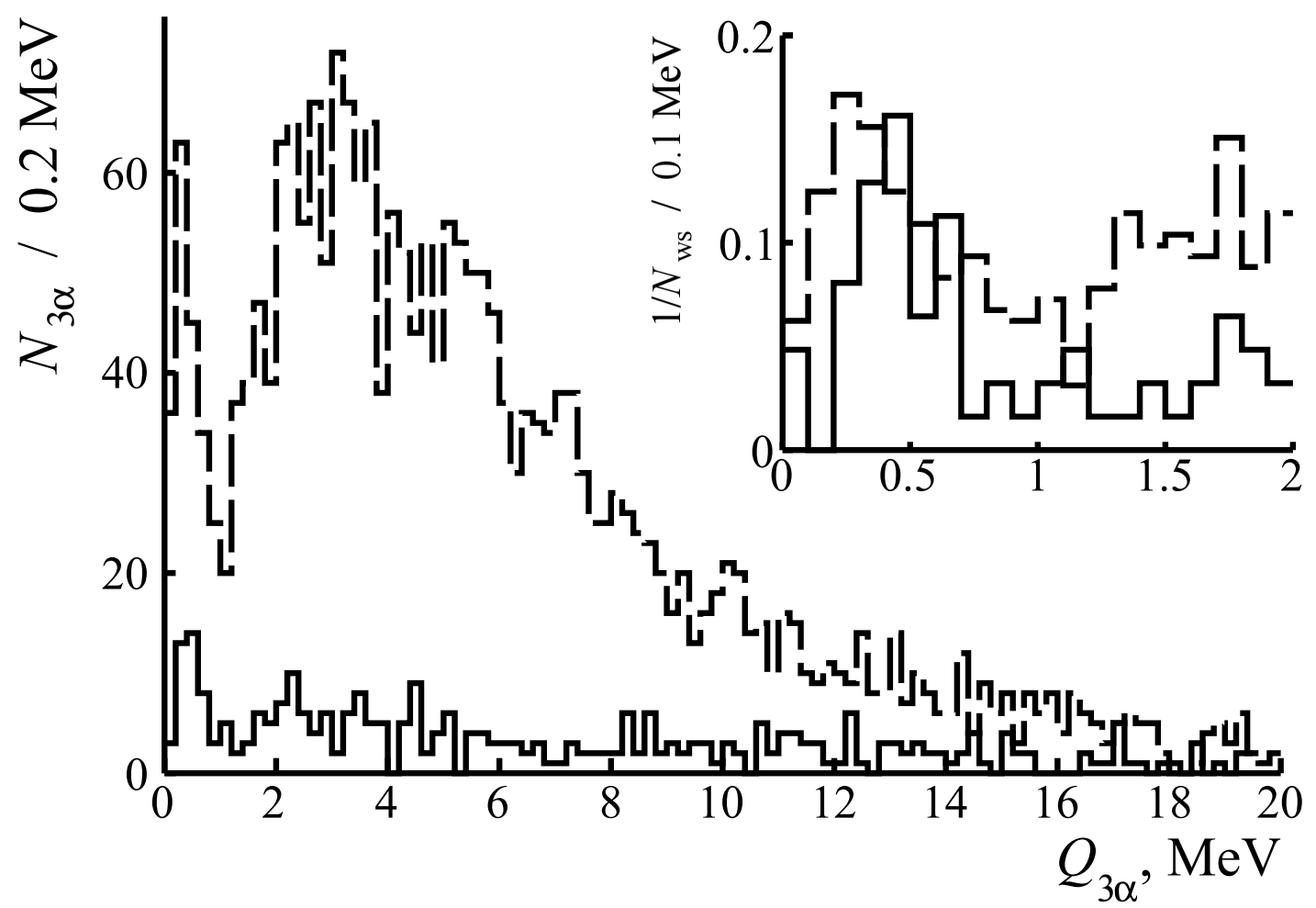}
\caption{\label{fig:3.6}Distribution of number of 3$\alpha$-triples $N_{3\alpha}$ over invariant mass $Q_{3\alpha}$ in 316 ``white'' stars $^{12}$C $\to$ 3$\alpha$ (solid) and 641 ``white'' stars $^{16}$O $\to$ 4$\alpha$ (dashed) at 3.65 A GeV; in the inset: the enlarged part $Q_{3\alpha}$ $<$ 2 MeV normalized to numbers of ``white'' stars Nws \cite{79}.}
\end{figure}

Decays of $^{12}$C(0$^+_2$) can arise in the dissociation of $^{16}$O $\to$ $^{12}$C$^*$ ($\to$ 3$\alpha$) + $\alpha$. Their identification by the invariant mass of 3$\alpha$-triples $Q_{3\alpha}$ is possible taking 641 ``white'' stars $^{16}$O $\to$ 4$\alpha$ at 4.5 GeV/$c$ per nucleon measured in the groups headed by G.M. Chernov and A.Sh. Gaitinov \cite{78}. Their distribution over $Q_{3\alpha}$ has shown similarity with the same $Q_{3\alpha}$ distribution (in terms of measurement conditions) of 316 ``white'' stars $^{12}$C $\to$ 3$\alpha$ of the statistics described above (Figure~\ref{fig:3.6}). In both cases, peaks are observed in the region of $Q_{3\alpha}$ $<$ 0.7 MeV with mean values (RMS) 417 $\pm$ 27 (165) keV for $^{12}$C and 349 $\pm$ 14 (174) keV for $^{16}$O. Thus, the contribution of $^{12}$C(0$^+_2$) in the case $^{12}$C $\to$ 3$\alpha$ is 11 $\pm$ 3\%, and of $^{16}$O $\to$ 4$\alpha$ - 22 $\pm$ 2\% \cite{79}. The increase in 3$\alpha$ combinations in $^{16}$O $\to$ 4$\alpha$ results in a noticeable increase in the $^{12}$C(0$^+_2$) contribution.

\begin{figure}
\includegraphics[width=0.7\textwidth]{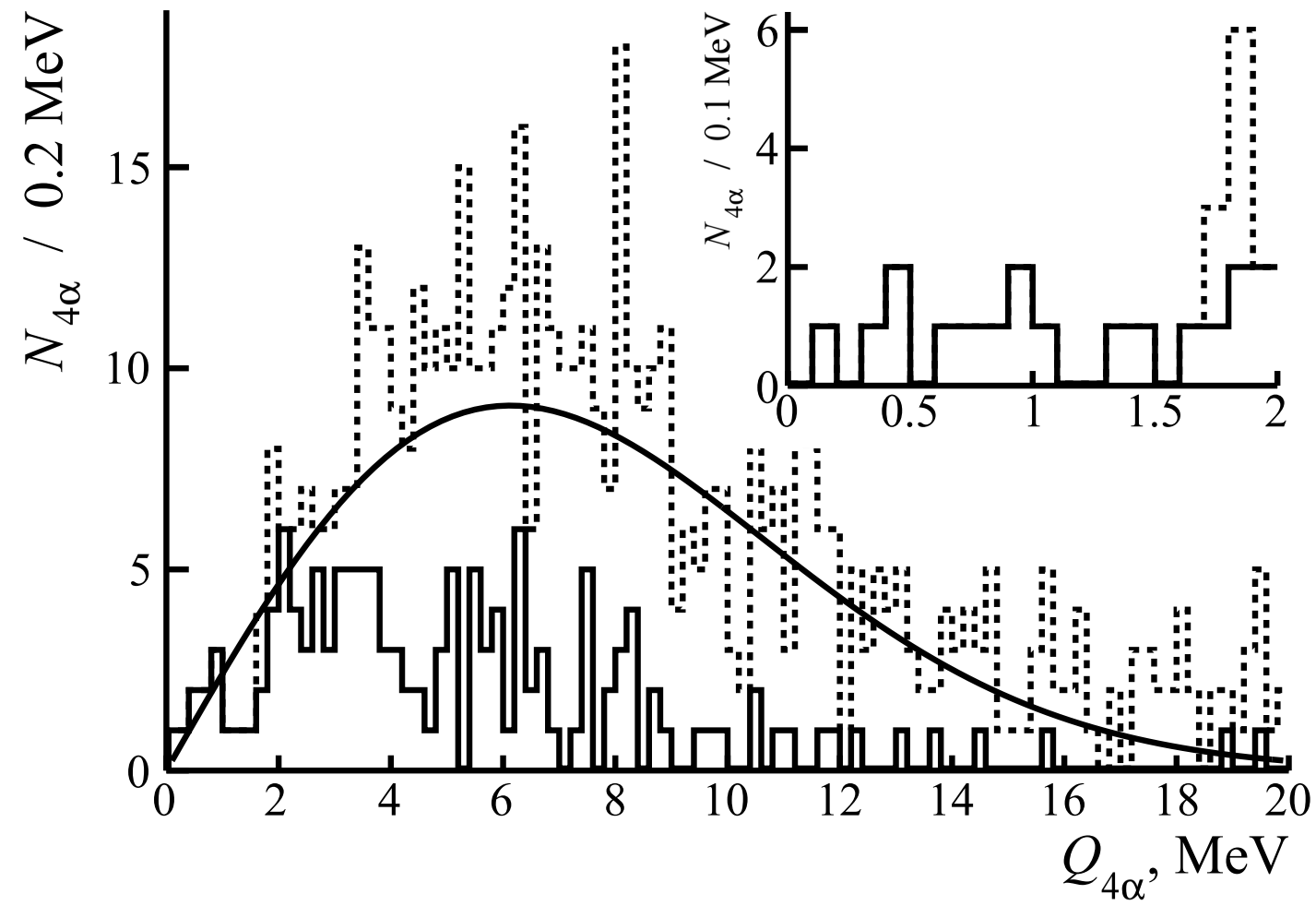}
\caption{\label{fig:3.7}Distributions over invariant mass $Q_{4\alpha}$ in 641 ``white'' stars $^{16}$O $\to$ 4$\alpha$ at 3.65 GeV per nucleon of all 4$\alpha$-quartets (dots) and $^{12}$C(0$^+_2$)$\alpha$ events (solid line); smooth line - Rayleigh distribution; the inset: the enlarged part $Q_{3\alpha}$ $<$ 2 MeV \cite{5}.}
\end{figure}

\begin{figure}
\includegraphics[width=0.7\textwidth]{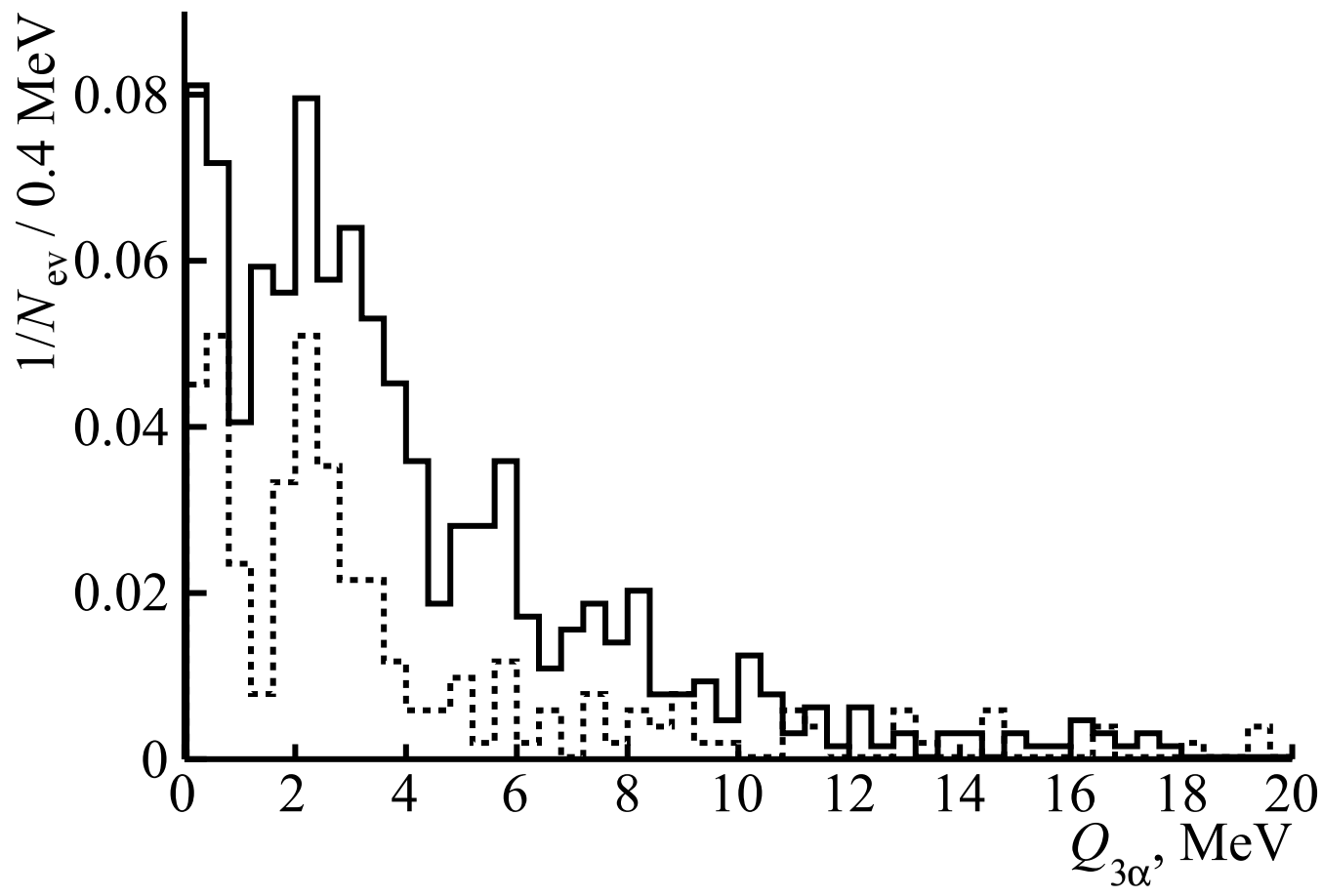}
\caption{\label{fig:3.8}Distribution over $Q_{3\alpha}$ in the events $^{12}$C $\to$ $^8$Be(0$^+$)$\alpha$ (dots) and $^{16}$O $\to$ $^8$Be(0$^+$)2$\alpha$ (solid); normalized to the number of events \cite{7}.}
\end{figure}
 
$^{12}$C(0$^+_2$) can appear in $\alpha$-decays of $^{16}$O(0$^+_6$) \cite{5}. The distribution of ``white'' stars $^{16}$O $\to$ 4$\alpha$ over the invariant mass of 4$\alpha$-quartets $Q_{4\alpha}$ (Figure~\ref{fig:3.7}) indicates 9 events $Q_{4\alpha}$ $<$ 1 MeV with the average value of 624 $\pm$ 84 (252) keV. Then the estimate of the contribution of $^{16}$O(0$^+_6$) $\to$ $\alpha$ + $^{12}$C(0$^+_2$) is 1.4 $\pm$ 0.5\% when normalized to the entire statistics and 7 $\pm$ 2\% when normalized to the number of $^{12}$C(0$^+_2$) in it. It can be concluded that the above dissociation dominates in the formation of $^{12}$C(0$^+_2$), and the search for its 4$\alpha$ ``precursor'' is possible but requires significantly more statistics accessible in electronic experiments at several hundred MeV per nucleon.

The condition for the presence of $^8$Be(0$^+$) leads to the signal of $^{12}$C(3$^-$) in the $Q_{3\alpha}$ distribution (Figure~\ref{fig:3.8}) \cite{7}. However, to estimate the signal contribution, it is desirable to weaken the combinatorial background. First of all, the already identified events can serve as a source, including 139 $^{12}$C(0$^+_2$)$\alpha$ and 36 2$^8$Be(0$^+$). After their removal, 196 events remain in the range adopted for $^{12}$C(3$^-$), with the average value (RMS) of 2.48 $\pm$ 0.06 MeV (1.0). In 105 from them there are single $\alpha$-triplets, and 91 – have the double ones with the statistical weight of 0.5. The next contribution can be given by the events $^{16}$O $\to$ $^8$Be(0$^+$)$^8$Be(2$^+$) whose direct identification is impossible. Assuming the equality of $^8$Be(0$^+$) and $^8$Be(2$^+$) in the cases of $^{12}$C and $^{16}$O and their independent formation, the contribution of $^8$Be(0$^+$)$^8$Be(2$^+$) can be estimated as approximately equal to the contribution of 2$^8$Be(0$^+$). It is subtracted from the number of $^{12}$C(3$^-$) candidates. Taking this remark into account, it can be stated that $^{12}$C(3$^-$) decays have been identified in the dissociation of $^{16}$O $\to$ 4$\alpha$. Then the contribution of the $^{12}$C(0$^+_2$)$\alpha$ channel was 23 $\pm$ 2\%, $^{12}$C(3$^-$)$\alpha$ – 32 $\pm$ 2\%, 2$^8$Be(0$^+$) – 6 $\pm$ 1\%, and the ratio of the $^{12}$C(3$^-$)$\alpha$ and $^{12}$C(0$^+_2$)$\alpha$ channels was equal to 1.4 $\pm$ 0.1.

There are theoretical observations that due to the extremely small energy gap of $^{16}$O(0$^+_6$) $\to$ $^{12}$C(0$^+_2$)$\alpha$, equal to only 296 keV the width of this decay is also extremely small amounting to about 10$^{-9}$ MeV \cite{46}. It is proposed to consider the excitation of $^{16}$O(0$^+_6$) in the model of a $^{12}$C(0$^+_2$)$\alpha$ container where the conversion to the ground state $^{12}$C(0$^+_2$)$\alpha$ $\to$ $^{12}$C(0$^+_1$)$\alpha$ occurs with the subsequent decay having the estimated width of 0.2 MeV. In the search for the 5$\alpha$-condensate, the decay channel $^{16}$O(0$^+_6$)$\alpha$ $\to$ $^{12}$C(0$^+_1$)$\alpha$$\alpha$ with internal conversion $^{12}$C(0$^+_2$)$\alpha$ $\to$ $^{12}$C(0$^+_1$)$\alpha$ was proposed \cite{49}.

The search for the decay $^{16}$O(0$^+_6$) $\to$ $^{12}$C(0$^+_1$)$\alpha$ is important in the context of nuclear astrophysical synthesis of the isotope $^{12}$C. It can serve as an alternative to the fusion of $^8$Be(0$^+$)$\alpha$ $\to$ $^{12}$C(0$^+_2$). In the case of $^{16}$O(0$^+_6$) $\to$ $^{12}$C(0$^+_1$)$\alpha$, one of the $\alpha$-particles in the quartet serves as a kind of catalyst, eliminating the need of electromagnetic transition. The coexistence of the decays $^{12}$C(0$^+_2$)$\alpha$ and $^{12}$C(0$^+_1$)$\alpha$ within the 165 keV width of $^{16}$O(0$^+_6$) cannot be ruled out. Below we present the first results of the analysis of binary dissociation $^{16}$O $\to$ $^{12}$C(0$^+_1$)$\alpha$.

\begin{figure}
\includegraphics[width=1.0\textwidth]{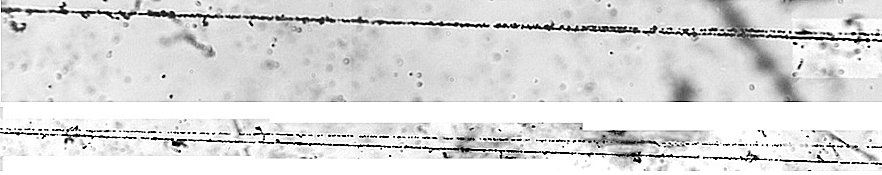}
\caption{\label{fig:3.9}Macrophotograph of the coherent dissociation event $^{16}$O $\to$ C + He at 3.65 GeV per nucleon. In the upper photograph on the left, the primary track of $^{16}$O is visible, accompanied by short traces of $\delta$-electrons. At a displacement of 1 mm in the direction of the fragment jet, their traces are distinguishable (lower photo).}
\end{figure}
 
\begin{figure}
\includegraphics[width=1.0\textwidth]{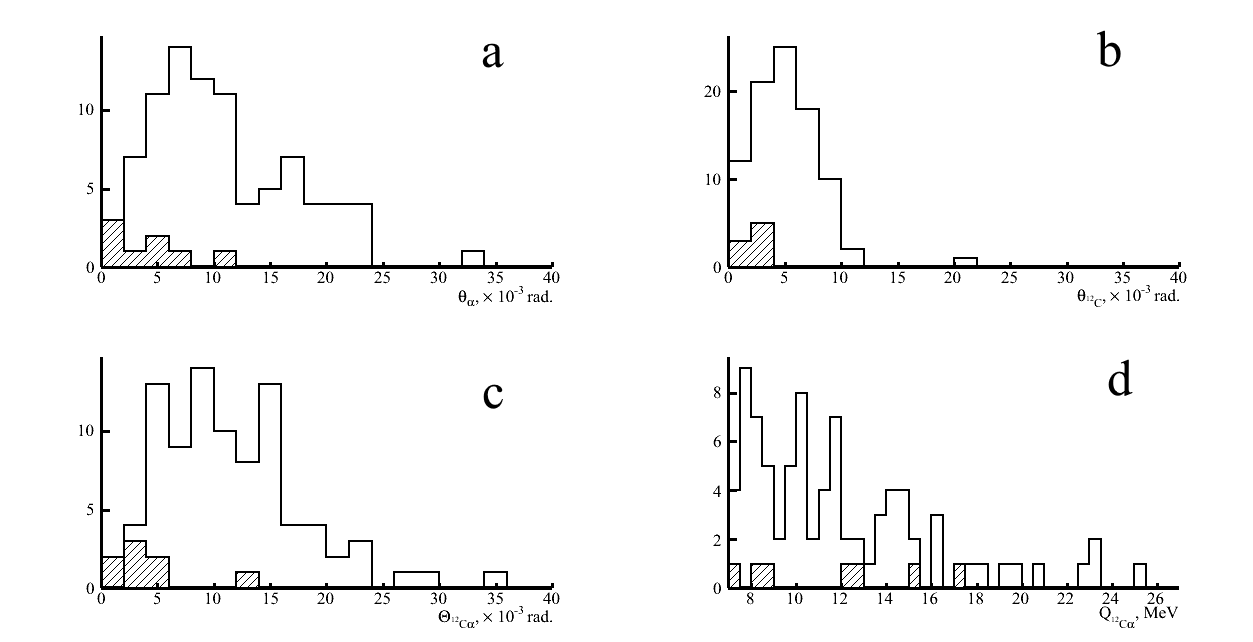}
\caption{\label{fig:3.10}The first measurements of events $^{16}$O $\to$ $^{12}$C$\alpha$ at 4.5 and 15 GeV/$c$ per nucleon (shaded): emission angles $\theta_\alpha$ (a) and $\theta_{^{12}\mathrm{C}}$ (b), opening angles $\Theta_{^{12}\mathrm{C}\alpha}$ between $^{12}$C and $\alpha$ fragments (c) and the invariant masses $Q_{^{12}\mathrm{C}\alpha}$ \cite{7}.}
\end{figure}

Recently, an accelerated search for C + He events was initiated by transversely scanning NE layers exposed to $^{16}$O nuclei at 3.65 GeV per nucleon at the JINR Synchrophasotron in the 1980s (example in Figure~\ref{fig:3.9}) \cite{7}. The directions in pairs of converging tracks of He and a heavier fragment, are measured and traced to the interaction vertex. By the present moment 67 events have been measured, including 22 ``white'' stars. The distributions of the polar emission angles $\theta_{^{12}\mathrm{C}}$ and $\theta_\alpha$ (Figure~\ref{fig:3.10}(a) and ~\ref{fig:3.10}(b)) were obtained from coordinate measurements on the secondary tracks. They are described by the Rayleigh distribution with parameters $\sigma_{\theta^{12}\mathrm{C}}$ = (3.9 $\pm$ 0.3) mrad and $\sigma_{\theta\alpha}$ = (8.4 $\pm$ 0.8) mrad, and 9.4 $\pm$ 0.9 mrad – for the opening angles $\Theta_{^{12}\mathrm{C}\alpha}$ (Figure~\ref{fig:3.10}(c)).

These measurements assuming conservation of momentum per nucleon and the $^{12}$C + $\alpha$ correspondence yield the invariant mass distribution of 89 pairs shown in Fig. 3.10d. It exhibits a $Q_{^{12}\mathrm{C}\alpha}$ grouping near 8.1 (25), 10.0 (17), 11.8 (15), and 14.3 (14) MeV with an RMS of about 0.5 MeV. The $Q_{^{12}\mathrm{C}\alpha}$ values of the 2$^{\mathrm{nd}}$ and 3$^{\mathrm{rd}}$ groups correspond to $\alpha$-unstable levels starting from $^{16}$O$^*$(8.9) up to the $^{15}$N$p$ threshold of 12.1 MeV \cite{2}. The 1$^{\mathrm{st}}$ group turns out to be unexpected, especially since it is absent from the dissociation on protons considered below. The reason for its appearance here may be the bremsstrahlung of some $^{12}$C fragments in the field of NE heavy nuclei. While this secondary effect appears when calculating invariant masses, it may be absent in spectroscopy.

Eight events, including two ``white'' stars, with the average value $\langle Q_{^{12}\mathrm{C}\alpha} \rangle$ (RMS) = 15.1 $\pm$ 0.2 (0.6) MeV, are candidates for the decay of $^{16}$O(0$^+_6$). This decay variant has the potential to be confirmed while statistics accumulating. Moreover, it stimulates the analysis of emulsions exposed to $^{16}$O nuclei at 15 GeV per nucleon at AGS BNL in the 1990s. This case offers methodological advantages associated with greater beam compactness, a smaller contribution of accompanying $\alpha$-particles, and the fragmentation cone three times smaller. The invariant representation of the data for both irradiations will allow independent comparison of the conclusions and verification of the universality of $^{16}$O(0$^+_6$) production. Measurements of the first events at 15 GeV per nucleon are added to Figure~\ref{fig:3.10} (shaded).

\subsection{Dissociation of $^{16}$O on protons}
As noted in the Introduction, measurements of the emission angles and total momenta of $P_{fr}$ fragments in 11,000 interactions of $^{16}$O nuclei at the momentum of $P_0$ = 3.25 GeV/$c$ per nucleon are available \cite{26}. To determine the charges of fragments heavier than He, the condition that the sum of the charges of secondary particles in the event is equal to 9, is sufficient. The $P_{fr}$/$P_0$ ratios allow one to verify the effect of isotopic identification of fragments \cite{27}.

\begin{figure}
\includegraphics[width=0.7\textwidth]{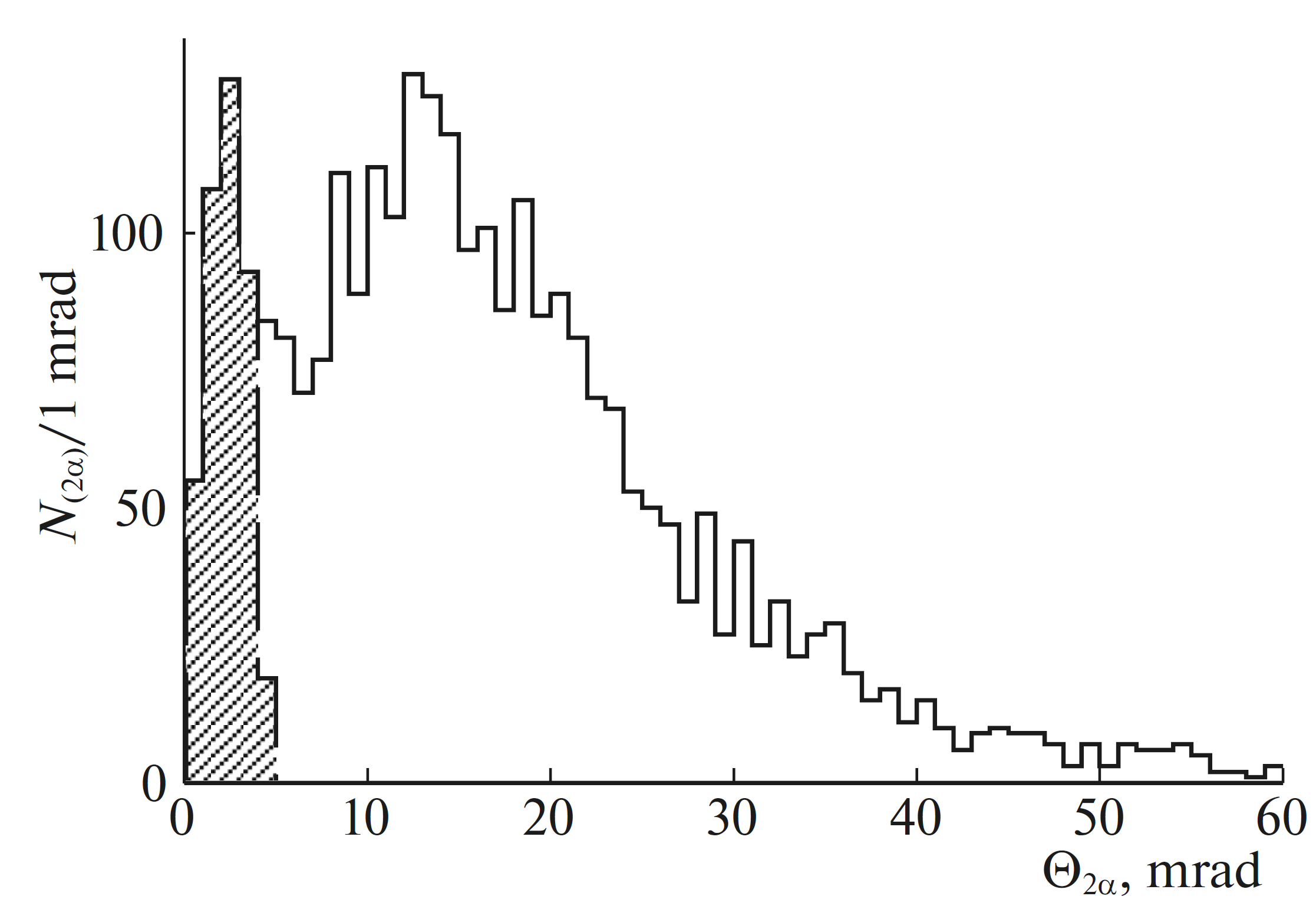}
\caption{\label{fig:3.11}Distribution by the expansion angle $\Theta_{2\alpha}$ of 2$\alpha$-pair combinations for all statistics (solid line) and with the condition $Q_{2\alpha}$($^8$Be) $\leq$ 0.2 MeV (dashed line) in the fragmentation of $^{16}$O nuclei at 3.25 GeV/$c$ per nucleon on protons \cite{27}.}
\end{figure}

\begin{figure}
\includegraphics[width=1.0\textwidth]{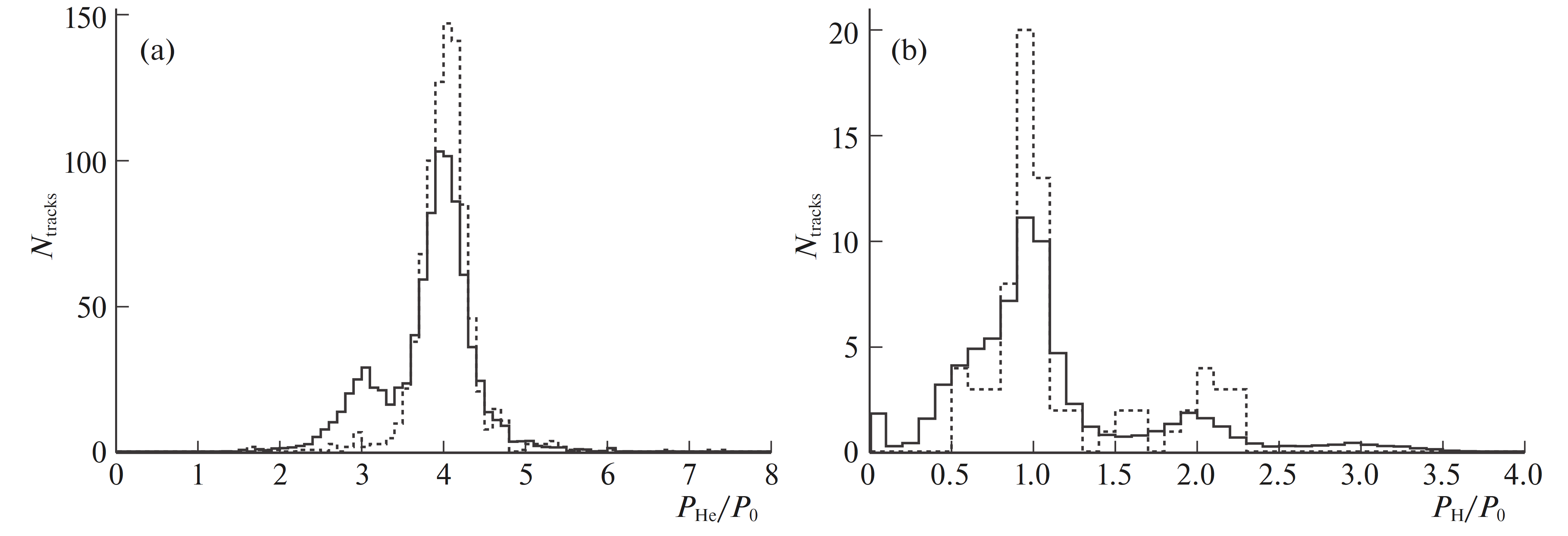}
\caption{\label{fig:3.12}Distribution of relativistic fragments of H (a) and He (b) over the ratios of their measured momenta $P_\mathrm{H}$ and $P_\mathrm{He}$ to the initial momentum per nucleon $P_0$ (solid line) in the fragmentation of $^{16}$O nuclei at 3.25 GeV/$c$ per nucleon on protons; the dotted line indicates samples with the conditions $Q_{2\alpha}$($^8$Be) $\leq$ 0.2 MeV and $Q_{2\alpha p}$($^9$B) $\leq$ 0.5 MeV normalized to the main distributions \cite{27}.}
\end{figure}

The VPK-100 events also exhibit a peak in the initial part of the 2$\alpha$-pair opening angle distribution $\Theta_{2\alpha}$, which corresponds to $^8$Be(0$^+$) decays (Figure~\ref{fig:3.11}). The $^8$Be signal disappears when calculating $Q_{2\alpha}$ due to insufficiently accurately measured $P_{\mathrm{He}}$ momenta. The situation is corrected by fixing the momenta, as in the NE case. The $P_{\mathrm{He}}$ and $P_{\mathrm{H}}$ values normalized to the initial momentum $P_0$ (per nucleon) identify the He and H isotopes. According to Figure~\ref{fig:3.12}, the condition $Q_{2\alpha}$($^8$Be) $\leq$ 0.2 MeV eliminates the $^3$He contribution, and the proton contribution is equal to 90\% among the H fragments.

\begin{figure}
\includegraphics[width=1.0\textwidth]{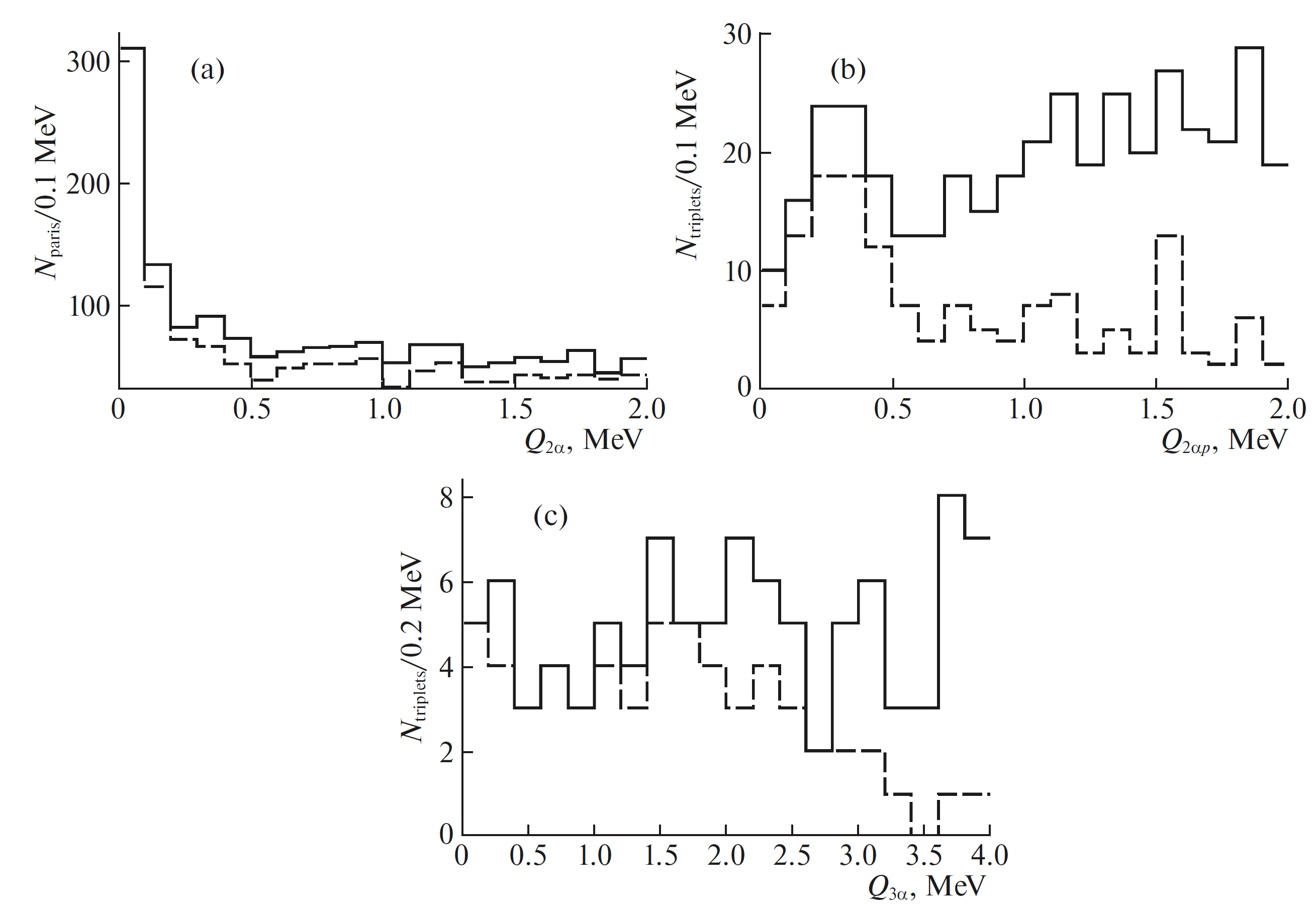}
\caption{\label{fig:3.13}Distribution of fragmentation events of $^{16}$O nuclei at 3.25 GeV/$c$ per nucleon on protons over $Q_{2\alpha}$ (a), $Q_{2\alpha p}$ (b) and $Q_{3\alpha}$ (c); distributions with conditions on $^4$He, protons and $^8$Be (dashed line) are added \cite{27}.}
\end{figure}

Figure~\ref{fig:3.13} shows the invariant mass distributions of all 2$\alpha$-pairs $Q_{2\alpha}$, 2$\alpha p$-triples $Q_{2\alpha p}$, and 3$\alpha$-triples $Q_{3\alpha}$ calculated by using the angles determined in VPK-100. Distributions with selection of $^4$He (3.5 $\leq$ $P_{\mathrm{He}}$/$P_0$ $\leq$ 4.5), protons (0.5 $\leq$ $P_\mathrm{H}$/$P_0$ $\leq$ 1.5), and $^8$Be ($Q_{2\alpha}$($^8$Be) $\leq$ 0.2 MeV), are added. The fixed-momentum variant depends only on the fragment emission angles and shows peaks for $^8$Be and $^9$B. A small number of $^{12}$C(0$^+_2$) candidates are presented.

\begin{figure}
\includegraphics[width=1.0\textwidth]{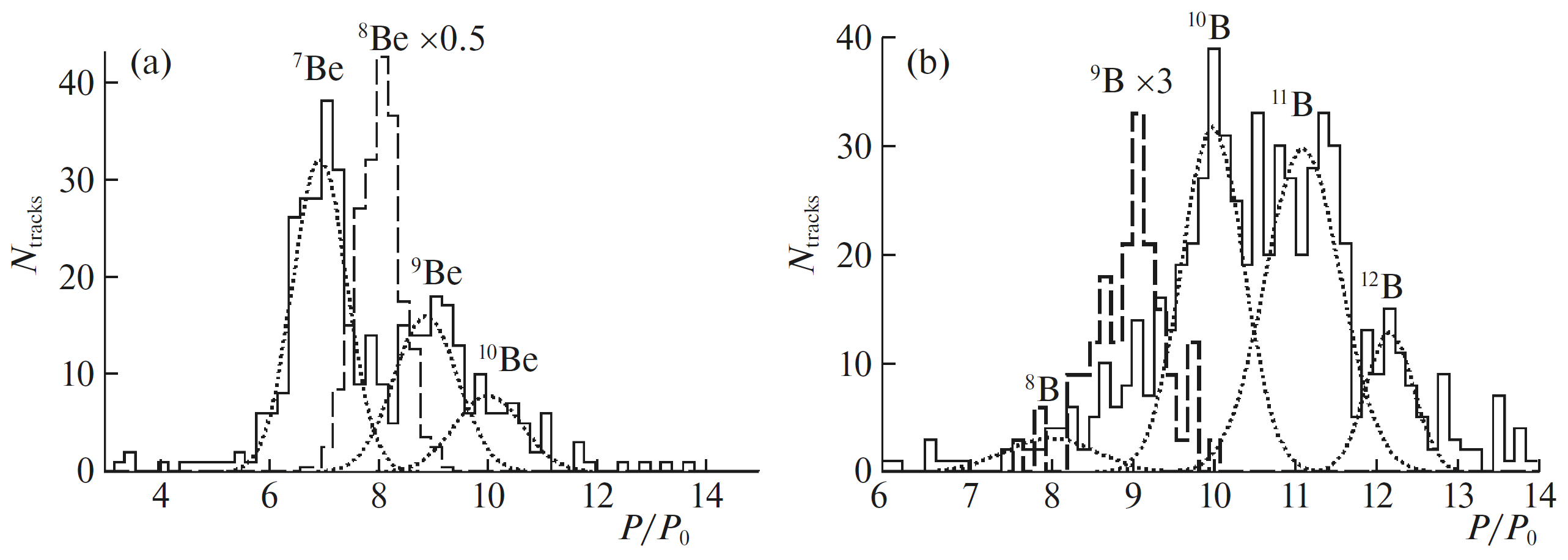}
\caption{\label{fig:3.14}Distribution of relativistic fragments of (a) Be and (b) B over the ratios of their measured momenta $P_\mathrm{H}$ and $P_\mathrm{He}$ to the initial momentum per nucleon $P_0$ (solid line) in fragmentation of $^{16}$O nuclei at 3.25 GeV/$c$ per nucleon on protons. Dots show the approximations by the sums of Gaussians; data on $^8$Be and $^9$B decays are superimposed by dashed lines \cite{27}.}
\end{figure}
  
The analysis of pulses in the magnetic field allows us to compare the ratio of contributions to the $^{16}$O + $p$ fragmentation of the stable and unstable isotopes of Be and B under identical observation conditions. Figure~\ref{fig:3.14} illustrates the distributions of these fragments over the ratio $P_\mathrm{Be(B)}$/$P_0$. The data on the $^8$Be and $^9$B decays are presented with factors of 0.5 and 3. Parameterization with Gaussians allows one to isolate peaks with half-widths approximately equal to 0.5 and to estimate the isotope statistics. Superposition of the distributions for the total momenta of 2$\alpha$-pairs $P_{2\alpha}$/$P_0$ at $Q_{2\alpha}$($^8$Be) $\leq$ 0.2 MeV and 2$\alpha p$-triplets $P_{2\alpha p}$/$P_0$ at $Q_{2\alpha p}$($^9$B) $\leq$ 0.5 MeV demonstrates them in the ranges corresponding to $^8$Be and $^9$B. The statistics of the $^{7-10}$Be (196, 345, 92, and 46) and $^{8-12}$B (33, 60, 226, 257, and 70) fragments allow their comparison. The ratio of $^9$B to $^9$Be (0.7 $\pm$0.1) confirms the $^9$B identification.

\begin{figure}
\includegraphics[width=0.7\textwidth]{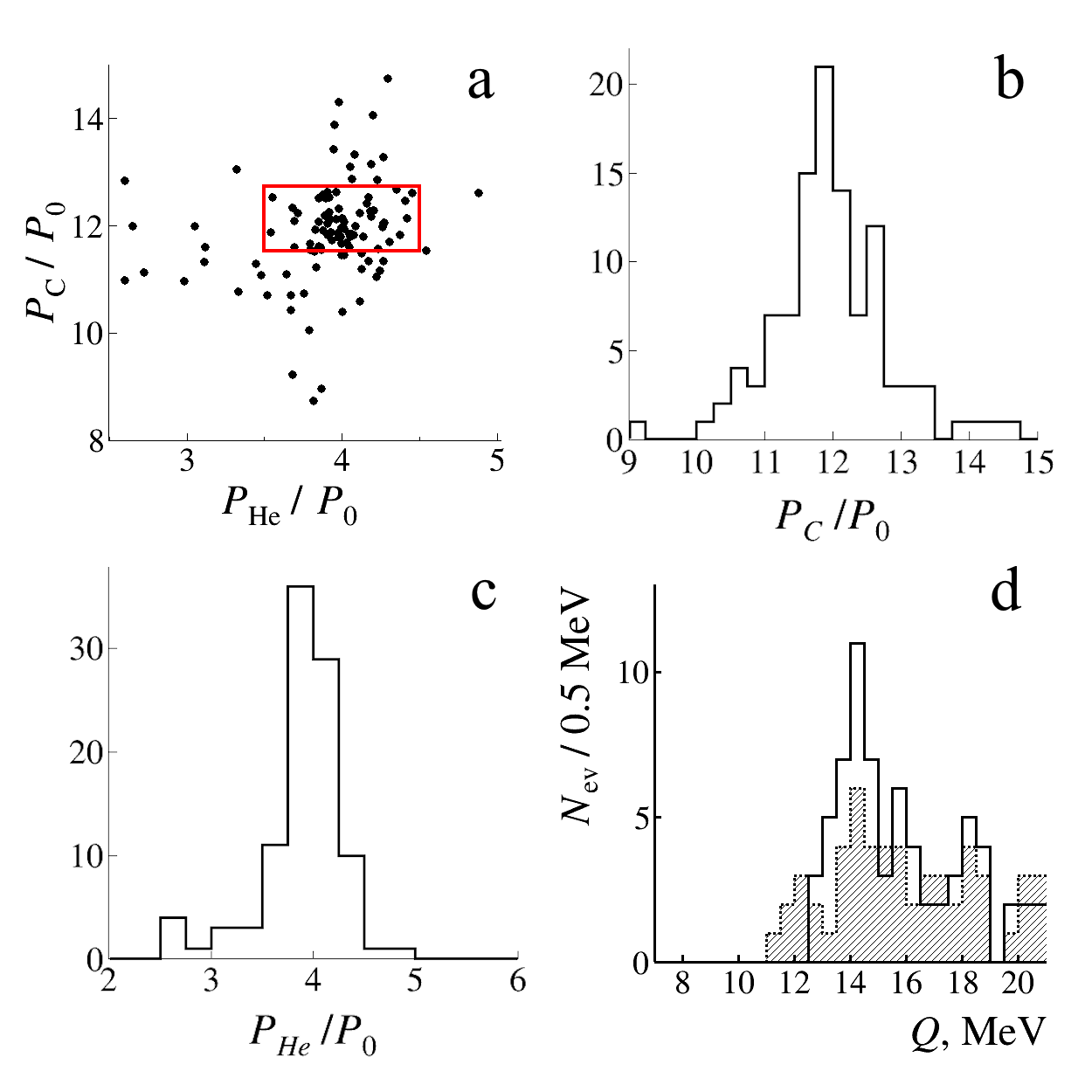}
\caption{\label{fig:3.15}Distribution of 3-prong events C + He (+$p$) in a hydrogen bubble chamber over $P_\mathrm{He}$/$P_0$ and $P_\mathrm{C}$/$P_0$ (a), projections (b and c) and invariant mass $Q_{^{12}\mathrm{C}\alpha}$ (hatched) and $Q_{^{15}\mathrm{N}p}$ (d) \cite{27}.}
\end{figure}

The hypothesis of dominance of the $^{12}$C(0$^+_1$)$\alpha$ pair (7.16 MeV above the $^{16}$O threshold) in the C + He topology was tested on 214 3-prong C + He (+$p$) events. Figure~\ref{fig:3.15}(a) shows the distribution over $P_\mathrm{He}$/$P_0$ and $P_\mathrm{C}$/$P_0$. The contribution of $^{13}$C + $^3$He events with a threshold of 22.8 MeV is negligible. In the $P_\mathrm{C}$/$P_0$ projection, signals of lighter C isotopes do not appear (Figure~\ref{fig:3.15}(b)), and in the $P_\mathrm{He}$/$P_0$ projection there is no $^3$He (Figure~\ref{fig:3.15}(c)) whose formation would correspond to the channels with thresholds above 30 MeV. Excitation of $^{12}$C(2$^+_1$) can contribute to broadening of the $P_\mathrm{C}$/$P_0$ distribution. Thus, the $^{12}$C$\alpha$ channel dominates that allows one to neglect the others. Figure~\ref{fig:3.15}(d) shows the invariant mass distribution of $Q_{^{12}\mathrm{C}\alpha}$ for the events highlighted in Figure~\ref{fig:3.15}(a). It includes the events both near the threshold resonances and in the $^{16}$O(0$^+_6$) region. We have added the distribution of $^{15}$N$p$ (12.1 MeV above the $^{16}$O threshold) over invariant mass $Q_{^{15}\mathrm{N}p}$ in 3-prong $^{15}$N$p$(+$p$) events selected in the same way (Figure~\ref{fig:3.15}(d)).

\subsection{Dissociation of $^{14}$N} 
\begin{figure}
\includegraphics[width=1.0\textwidth]{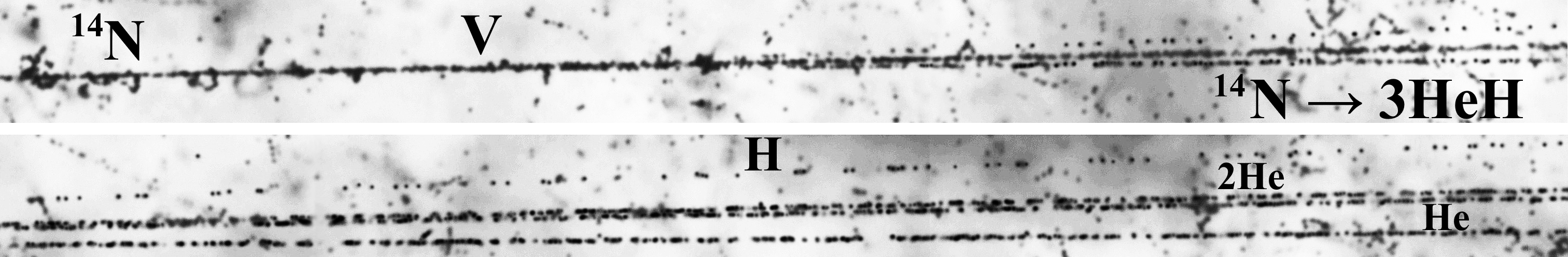}
\caption{\label{fig:3.16}Macrophotograph of the coherent dissociation $^{14}$N $\to$ $^8$Be(0$^+$)$\alpha$ + H event at of 2 GeV per nucleon.}
\end{figure}

Dissociation of the $^{14}$N nucleus can serve as a common source of $^9$B and $^{12}$C(0$^+_2$) \cite{80}. The material for this study was NE layers exposed to $^{14}$N nuclei at 2 GeV per nucleon at the JINR Nuclotron in 2004. The charge topology of $^{14}$N dissociation was previously established, and the contribution of $^8$Be was estimated at 25-30\% (example in Figure~\ref{fig:3.16}) \cite{81}. The role of $^9$B and $^{12}$C(0$^+_2$) remains an important issue. To date, the statistics of measured events for the leading dissociation channel $^{14}$N $\to$ 3$\alpha$ + H has been brought to 128, including 29 ``white'' stars, and 54 of $^{14}$N $\to$ 3$\alpha$ with target fragments.

\begin{figure}
\includegraphics[width=0.7\textwidth]{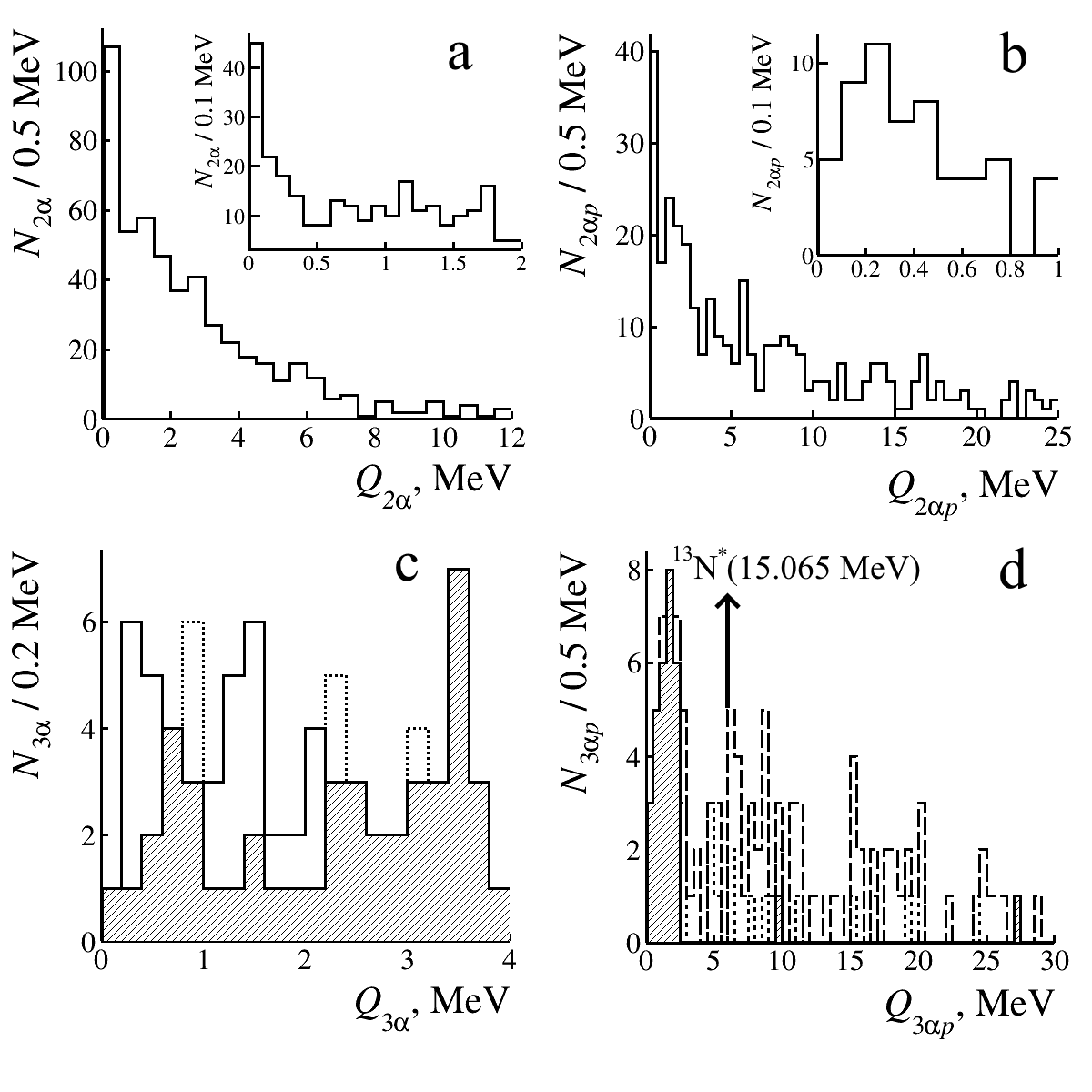}
\caption{\label{fig:3.17}Distribution of invariant masses $Q$ for the dissociation $^{14}$N $\to$ 3$\alpha$ (+ H) at 2 GeV per nucleon; a - $Q_{2\alpha}$; b - $Q_{2\alpha p}$; c - $Q_{3\alpha}$ (dots - all events, solid line with $^8$Be(0$^+$), hatching - $^9$B vetoed); d - $Q_{3\alpha p}$ (dashed line - all events, dots with $^8$Be(0$^+$), hatching - with $^9$B or $^{12}$C(0$^+_2$)) \cite{7}.}
\end{figure}

Figure~\ref{fig:3.17}(a) shows the $Q_{2\alpha}$ distribution of $\alpha$-particle pairs. The average value of $\langle Q_{2\alpha}\rangle$ in 62 events containing $\alpha$-pairs with opening angles $\Theta_{2\alpha}$ $<$ 6 mrad is 114 $\pm$ 10 keV at RMS 92 keV. For 62 $\alpha$-pairs (53 events) in the region of $Q_{2\alpha}$ $<$ 0.2 MeV, the $\langle Q_{2\alpha}\rangle$ is 76 $\pm$ 7 keV at RMS 61 keV. The contribution of $^8$Be(0$^+$) events satisfying this condition to the dissociation $^{14}$N $\to$ 3$\alpha$ + H, is 41 $\pm$ 6\%.

Figure~\ref{fig:3.17}(b) shows the $Q_{2\alpha p}$ distribution of 2$\alpha p$ triplets. In the region $Q_{2\alpha p}$ $<$ 0.5 MeV, adopted for $^{12}$C(0$^+_2$), there are 30 events with $\langle Q_{2\alpha p}\rangle$ = 263 $\pm$ 20 keV at RMS 127 keV. They allow one to estimate the contribution of $^9$B as 23 $\pm$ 4\% and 53\% of $^8$Be(0$^+$) decays are attributed to $^9$B decays. In the case of dissociation of $^{10}$B nuclei, such a contribution was equal to 39\% of the total number of events, and for ``white'' stars it is equal to 50\%.

Figure~\ref{fig:3.17}(c) shows the $Q_{3\alpha}$ distribution of 3$\alpha$ triplets. With the $Q_{3\alpha}$ $<$ 0.7 MeV cutoff adopted for $^{12}$C(0$^+_2$), 13 events with $\langle Q_{3\alpha}\rangle$ = 431 $\pm$ 35 keV at 125 keV RMS or 10\% of the statistics of this channel are distinguished. Although these parameters correspond to the expected $^{12}$C(0$^+_2$) signal, these events coincide approximately half with the events with $^9$B candidates. The events of the 3He channel without H in the fragmentation cone are not discussed in the context of $^{12}$C(0$^+_2$) due to their smaller statistics and the possible contribution of the excitation decays $^9$Be(1.67) $\to$ 2$\alpha$($n$) \cite{6}.

Figure~\ref{fig:3.17}(d) illustrates the $Q_{3\alpha p}$ distribution of 3$\alpha p$ quadruples, for which this problem is not significant. It shows the expected position of the isobar analog state $^{13}$N$^*$(15.065). When it decays to $^9$B$\alpha$, $^{12}$C(0$^+_2$)p or $^{12}$C(3$^-$)$p$, $^8$Be(0$^+$) should be present. However, introducing the condition on the presence of $^8$Be(0$^+$) in the event closes this opportunity at the given level of statistics. The condition on $^9$B or $^{12}$C(0$^+_2$) leads to a radical compression of this distribution with $\langle Q_{3\alpha p}\rangle$ equal to 2.5 $\pm$ 0.1 MeV at RMS 0.6 MeV.

Thus, the contributions of $^9$B and $^{12}$C(0$^+_2$) to the leading channel of 3HeH dissociation of the $^{14}$N nucleus are estimated at 23\% and 10\%. However, there is an overlap. It is unclear whether the resolution limit has been reached or there is a more fundamental overlap of $^9$B and $^{12}$C(0$^+_2$). The joint condition on $^9$B or $^{12}$C(0$^+_2$) indicates that the $\alpha$-particles and protons in the events with $^9$B or $^{12}$C(0$^+_2$) but not attributed to decays, turn out to be close to $Q_{3\alpha p}$.

\section{Search for enhancement of unstable states with $\alpha$-particle multiplicity}
\begin{figure}
\includegraphics[width=0.7\textwidth]{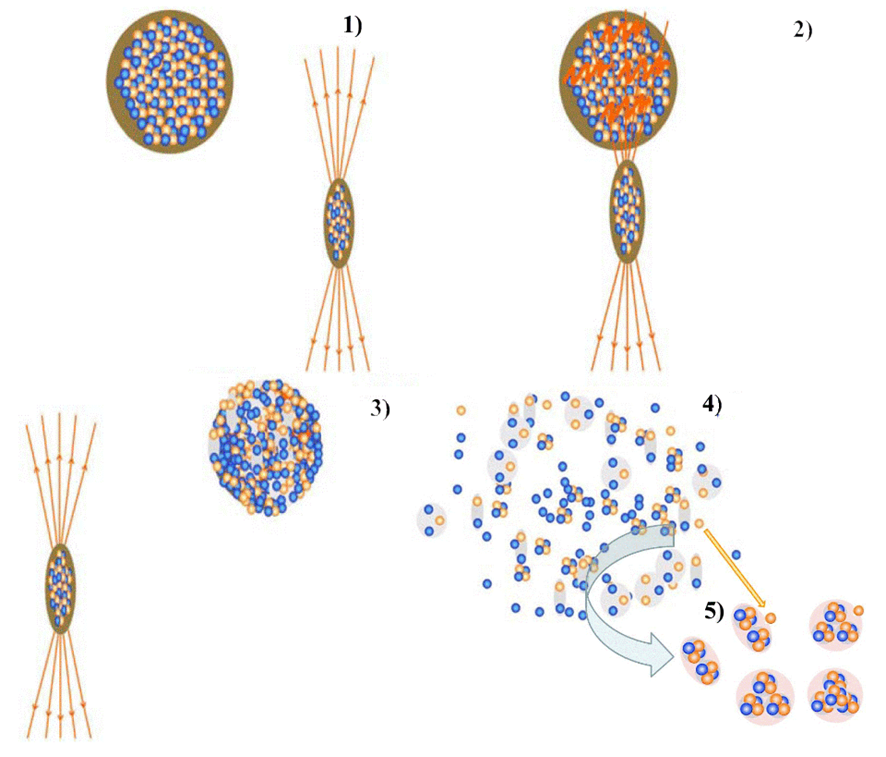}
\caption{\label{fig:4.1}Scenario of the formation of unstable states \cite{27}: approach of nuclei (1), transfer of excitation to the nucleus under study (2), transition to a system containing real lightest nuclei and nucleons (3), this system decay (4), sticking together and picking up some of the fragments into unstable states (5).}
\end{figure}

The enhancement of $^8$Be(0$^+$) and $^{12}$C(0$^+_2$) with increasing $\alpha$-particle multiplicity $n_\alpha$ in the dissociation $^{12}$C $\to$ 3$\alpha$ and $^{16}$O $\to$ 4$\alpha$ may indicate the mechanism of their generation. It seems unlikely that having exotic sizes they virtually arise inside or at the periphery of the parent nucleus, manifesting them during prompt fragmentation. Such a scenario would lead to a suppression of their yields with increasing $n_\alpha$. An alternative is generation in a brief fusion of the resulting $\alpha$-particles in the reaction $\alpha$ + $\alpha$ $\to$ $^8$Be(0$^+$) + $\gamma$ (or recoil particles) followed by the possible pickup of accompanying $\alpha$-particles and nucleons. The above secondary interaction may lead to increasing in the yield of $^8$Be(0$^+$) with increasing $n_\alpha$, followed by $^9$B and $^{12}$C(0$^+_2$) (Figure~\ref{fig:4.1}). This effect deserves verification in cases of the heavier nuclei whose achievable number $n_\alpha$ increases rapidly with the mass number.

\subsection{Analysis of available data}
This hypothesis was tested using the preserved measurements of interactions of the relativistic nuclei $^{16}$O, $^{22}$Ne, $^{28}$Si, and $^{197}$Au in nuclear neutron emission, which were obtained by scanning along primary tracks (without sampling) \cite{4,27}. These data were obtained by participants in the emulsion collaboration at the JINR Synchrophasotron in the 1980s and the EMU collaboration at the AGS (BNL) and SPS (CERN) synchrotrons in the 1990s \cite{82,83,84,85}. Being detailed and systematic, these data retain their uniqueness in terms of relativistic fragmentation. The conclusion about the limiting fragmentation regime over the widest possible range of nuclei and initial energy values expressed in the invariance of the charge composition of the fragments and the scale-invariant behavior of their spectra, retains its fundamental value. This archive was saved and transferred to N.G. Peresadko (FIAN) for the BECQUEREL experiment in the late 2010s, since it was the data on the ``white'' stars $^{12}$C and $^{16}$O used above. However, the correlations within fragment ensembles have not been studied. 

\begin{figure}
\includegraphics[width=0.6\textwidth]{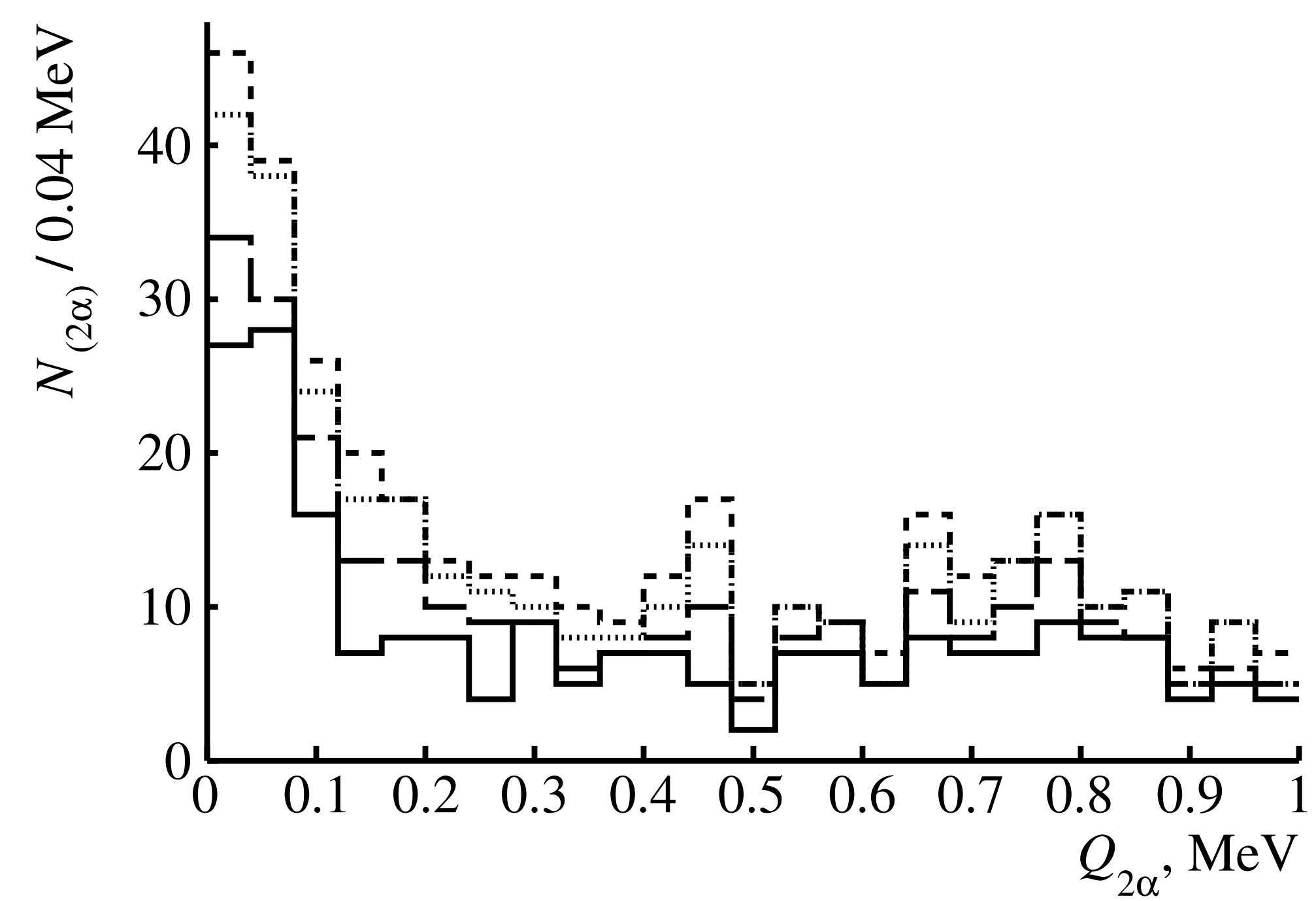}
\caption{\label{fig:4.2}Distribution of 2$\alpha$-pairs $N_{(2\alpha)}$ over in the $Q_{2\alpha}$ invariant mass range $<$ 1 MeV in the fragmentation of $^{16}$O nuclei at 3.65 GeV (solid line); for 15 (long dashed line), 60 (dots) and 200 (short, dashed line) GeV per nucleon; the data are added successively \cite{4}.}
\end{figure}

\begin{figure}
\includegraphics[width=0.6\textwidth]{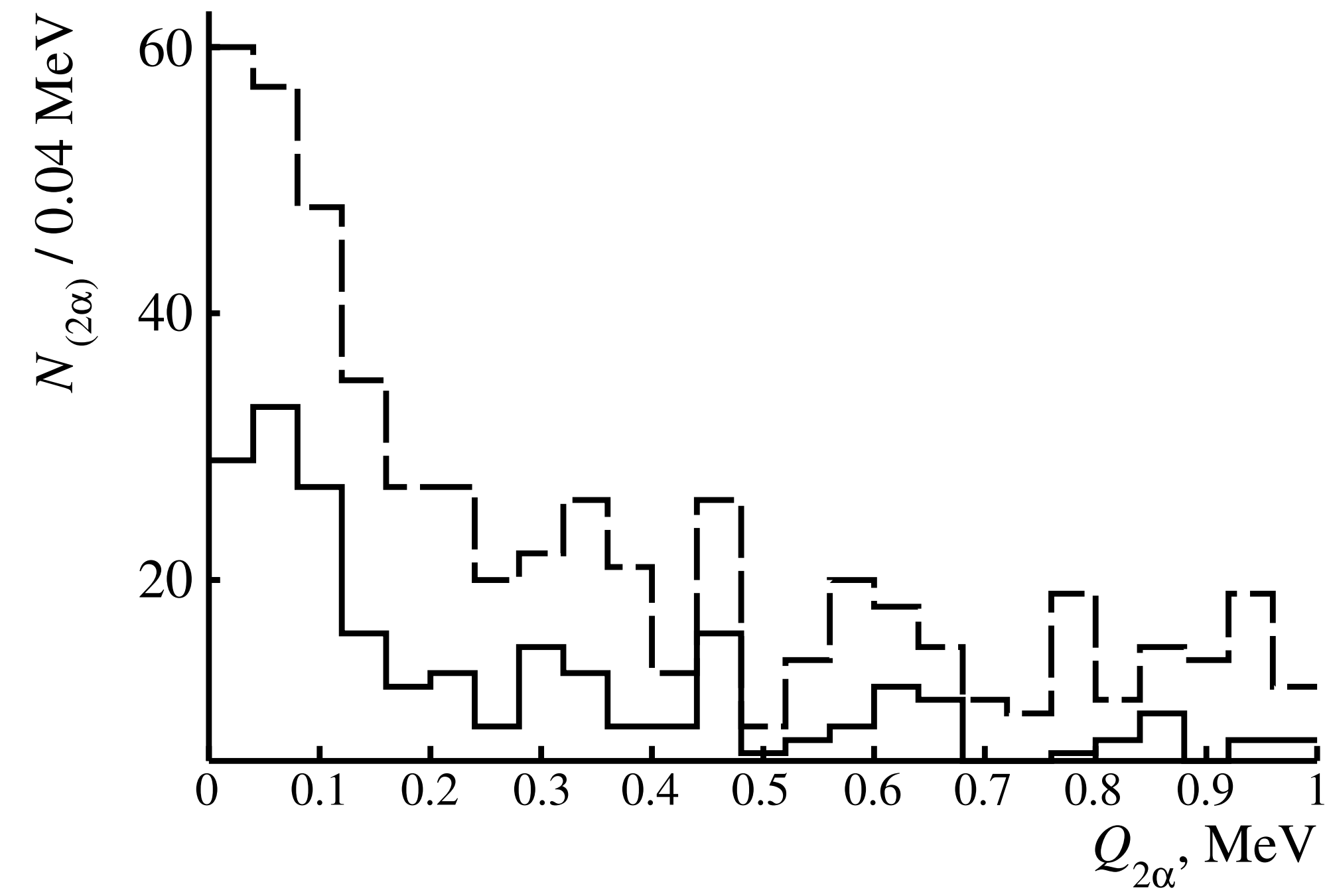}
\caption{\label{fig:4.3}Distribution of 2$\alpha$-pairs $N_{(2\alpha)}$ over $Q_{2\alpha}$ $<$ 1 MeV in the fragmentation of $^{22}$Ne at 3.22 (solid line) and $^{28}$Si at 14.6 GeV per nucleon (added by the dotted line) \cite{4}.}
\end{figure}

There are measurements of the $^{16}$O nucleus interactions found in primary track tracing, including 2823 at 3.65, 689 at 14.6, 885 at 60 and 801 at 200 GeV per nucleon. The distributions of all $\alpha$-pair combinations $N_{(2\alpha)}$ for these interactions with respect to the invariant mass $Q_{2\alpha}$ $\leq$ 1 MeV are summarized in Figure~\ref{fig:4.2}. At the beginning of the distribution, the concentration of $\alpha$-pairs is observed, and the condition $Q_{2\alpha}$($^8$Be) $\leq$ 0.2 MeV is taken to select $^8$Be(0$^+$) decays. The similarity of the $Q_{2\alpha}$ distributions in the covered energy range allows one to summarize the statistics for reducing statistical errors. The increase in $N_{n\alpha}$($^8$Be)/$N_{n\alpha}$ (\%) with $n_\alpha$ is observed: 2 (8 $\pm$ 1), 3 (23 $\pm$ 3), and 4 (46 $\pm$ 14).

4308 $^{22}$Ne events at 3.22 and 1093 $^{28}$Si events at 15 GeV per nucleon allow one to further extend the $n_\alpha$ range. In both cases, no change in the condition $Q_{2\alpha}$($^8$Be) $\leq$ 0.2 MeV is required (Figure~\ref{fig:4.3}). The $n_\alpha$ $\geq$ 3 statistics for $^{28}$Si was tripled by transverse scanning. The growth of the ratio $N_{n\alpha}$($^8$Be)/$N_{n\alpha}$ (\%) continued: for $^{22}$Ne $n_\alpha$ = 2 (6 $\pm$ 1), 3 (19 $\pm$ 3), 4 (31 $\pm$ 6) and $^{28}$Si 2 (3 $\pm$ 2), 3 (13 $\pm$ 5), 4 (32 $\pm$ 6), 5 (38 $\pm$ 11).

\begin{figure}
\includegraphics[width=0.7\textwidth]{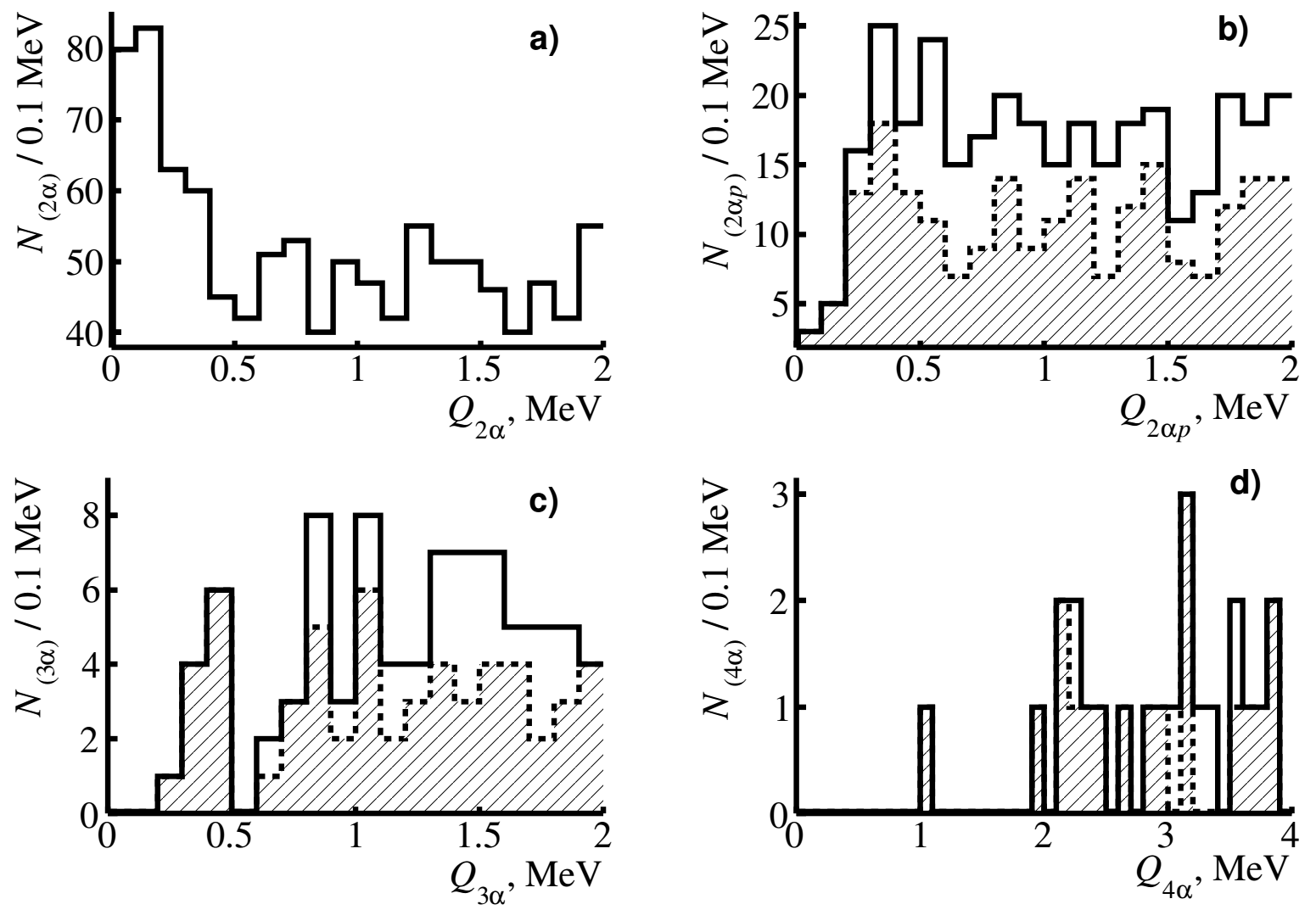}
\caption{\label{fig:4.4} Distributions of invariant masses $Q$ of 2$\alpha$-pairs (a), as well as 2$\alpha p$-triples (b), 3$\alpha$-triples (c) and 4$\alpha$-quadruples (d) under the conditions $Q_{2\alpha}$ $\leq$ 0.4 MeV (solid) and $Q_{2\alpha}$ $\leq$ 0.2 MeV (shaded) in the fragmentation of $^{197}$Au nuclei at 10.7 GeV per nucleon \cite{4}.}
\end{figure}

Figure~\ref{fig:4.4}(a) illustrates the distribution of 2$\alpha$-pairs at small $Q_{2\alpha}$ values in 1316 interactions of $^{197}$Au nuclei at 10.7 GeV per nucleon. The complication of the measurements requires softening of the selection $Q_{2\alpha}$($^8$Be) $\leq$ 0.4 MeV to save an efficiency. To check the correlation $N_{n\alpha}$($^8$Be) and minimize the background in $N_{n\alpha}$($^{12}$C(0$^+_2$)), the condition $Q_{2\alpha}$($^8$Be) $\leq$ 0.2 MeV is also used. The distributions of 2$\alpha p$-triples (b), 3$\alpha$-triples (c), and 4$\alpha$-quartets (d), which, according to these conditions, contain at least one $^8$Be(0$^+$) candidate, are also included in Figure~\ref{fig:4.4}. Distributions (c) and (d) contain 2$\alpha p$ and 3$\alpha$ triplets satisfying the conditions $Q_{2\alpha p}$($^9$B) $\leq$ 0.5 MeV and $Q_{3\alpha}$($^{12}$C(0$^+_2$)) $\leq$ 0.7 MeV, respectively. Their statistics and contributions are presented in Table 4.1, the condition $Q_{2\alpha}$($^8$Be) $\leq$ 0.4 MeV is taken into account. The channels $n_\alpha$ $\geq$ 11 are summed to reduce statistical errors. The ratio $N_{n\alpha}$($^8$Be)/$N_{n\alpha}$ (\%) shows a rapid increase to the values of the order of 0.5 when multiplicity $n_\alpha$ = 10 is reached. This trend is preserved when the selection condition is tightened to $Q_{2\alpha}$($^8$Be) $\leq$ 0.2 MeV, despite the decrease of statistics

\begin{table}
\caption{Statistics of the events involving at least one $^8$Be(0$^+$), $^{12}$C(0$^+_2$), or $^9$B candidate, or at least two $^8$Be(0$^+$) candidates provided $Q_{2\alpha}$($^8$Be) $\leq$ 0.4 MeV, among the $N_{n\alpha}$ $^{197}$Au fragmentation events with multiplicity $n_\alpha$. The summary statistics for the channels with $n_\alpha$ $\geq$ 11 are shown in italics \cite{4}.}
\label{tab:1}
\begin{tabular}{|c|c|c|c|c|}
\toprule
$n_\alpha$ & 
\makecell{$N_{n\alpha}$($^8$Be)/$N_{n\alpha}$ \\ (\% $N_{n\alpha}$)} &
\makecell{$N_{n\alpha}$($^9$B) \\ (\% $N_{n\alpha}$($^8$Be)} &	
\makecell{$N_{n\alpha}$($^{12}$C(0$^+_2$) \\ (\% $N_{n\alpha}$($^8$Be))} &	\makecell{$N_{n\alpha}$(2$^8$Be) \\ (\% $N_{n\alpha}$($^8$Be))} \\
\colrule
2 & 3/133 (2 $\pm$ 1) & -	& - & - \\
3 &	14/162 (9 $\pm$ 3) &	1 (7)	& - & 	- \\
4 & 25/161 (16 $\pm$ 4) &	7 (28 $\pm$ 12) &	2 (8 $\pm$ 6) &	- \\
5 &	23/135 (17 $\pm$ 4) &	5 (22 $\pm$ 11) &	- &	1 (4) \\
6 &	31/101 (31 $\pm$ 7) &	9 (29 $\pm$ 11) &	2 (6 $\pm$ 4) &	- \\
7 &	31/90 (34 $\pm$ 7) &	6 (19 $\pm$ 9) &	2 (6 $\pm$ 4) &	3 (10 $\pm$ 6) \\
8 &	32/71 (45 $\pm$ 10) &	8 (25 $\pm$ 10) &	2 (6 $\pm$ 4) &	2 (7 $\pm$ 5) \\
9 &	29/54 (54 $\pm$ 13) &	9 (31 $\pm$ 12) &	3 (10 $\pm$ 6) &	5(17 $\pm$ 8) \\
10 &	22/39 (56 $\pm$ 15) &	4 (18 $\pm$ 10) &	- &	5(23 $\pm$ 12) \\ \hline
11 &	\makecell{10/15 (67 $\pm$ 27) \\ \textit{19/30 (63 $\pm$ 19)}} &	\makecell{3 (30 $\pm$ 20) \\ \textit{7 (37 $\pm$ 16)}} & \makecell{1 (10) \\
\textit{2(11 $\pm$ 8)}} &	\makecell{2(20 $\pm$ 16) \\ \textit{6 (32 $\pm$ 15)}} \\ \hline
12 &	226 2/5 &	56 1 &	13 - &	22 1 \\
13 &	2/4	& 1 &	- &	1 \\
14 &	3/3 &	1 &	- &	1 \\
15 &	1/1 &	- &	- &	- \\
16 &	1/2 &	1 &	1 &	1 \\
\botrule
\end{tabular}
\label{tab1}
\end{table}

\begin{figure}
\includegraphics[width=0.7\textwidth]{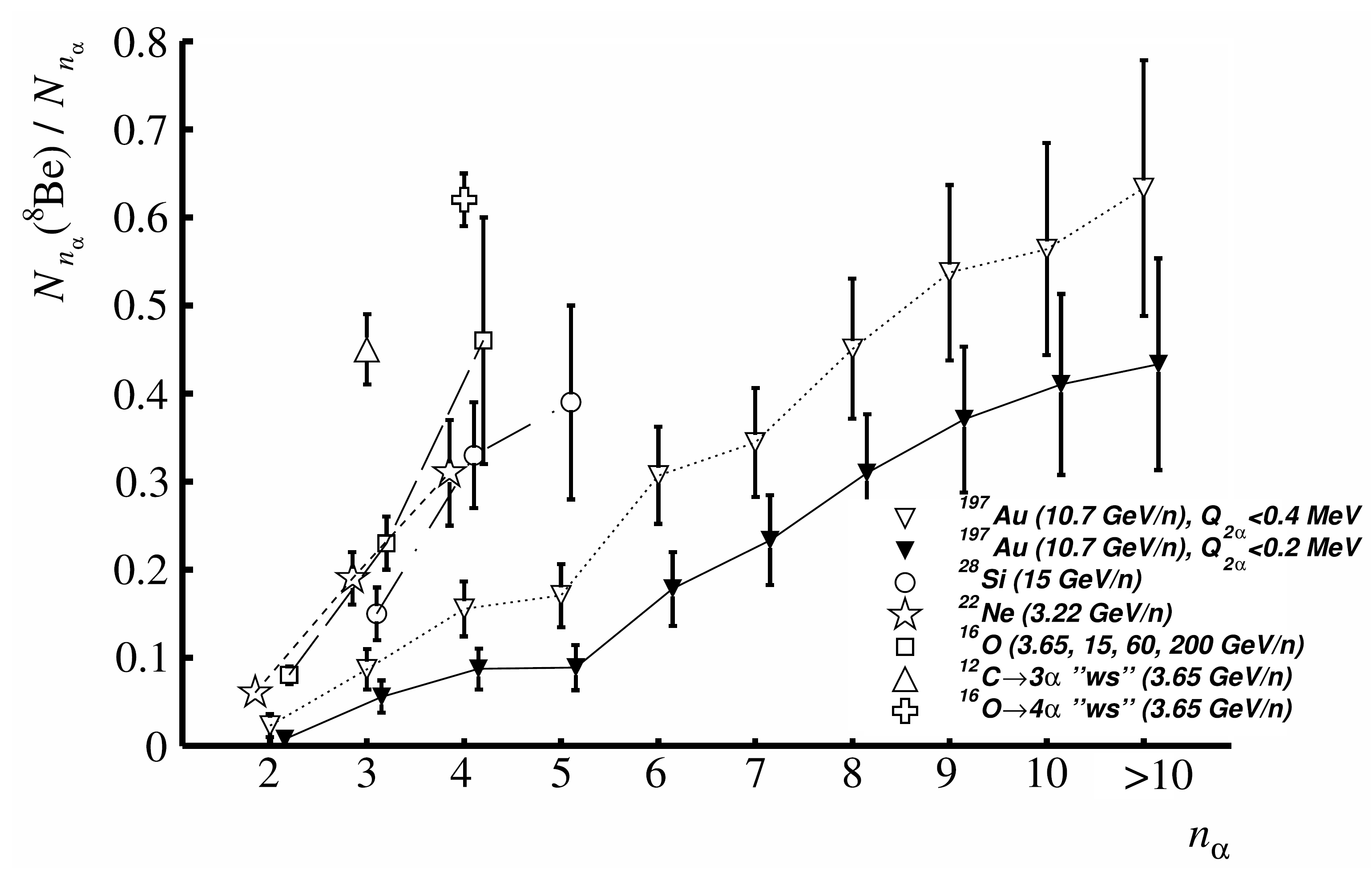}
\caption{\label{fig:4.5} Dependence of relative contribution of $N_{n\alpha}$($^8$Be) decays to the statistics of $N_{n\alpha}$ events over the multiplicity of $\alpha$-particles $n_\alpha$ in relativistic fragmentation of C, O, Ne, Si, and Au nuclei. The ``white'' stars $^{12}$C $\to$ 3$\alpha$ and $^{16}$O $\to$ 4$\alpha$ are marked (WS); for convenience, the points are somewhat displaced around the values of $n_\alpha$ and connected with the lines \cite{4}.}
\end{figure}

The ratios of $N_{n\alpha}$($^9$B), $N_{n\alpha}$($^{12}$C(0$^+_2$)) and $N_{n\alpha}$(2$^8$Be) to $N_{n\alpha}$($^8$Be) do not show any noticeable change with $n_\alpha$ (Table~\ref{tab:1}) indicating their increase to $N_{n\alpha}$. However, the statistical errors allow one to identify only the trend. The summary statistics of $N_{n\alpha}$($^9$B), $N_{n\alpha}$($^{12}$C(0$^+_2$)) and $N_{n\alpha}$(2$^8$Be) normalized to $N_{n\alpha}$($^9$B), $N_{n\alpha}$($^{12}$C(0$^+_2$)) and $N_{n\alpha}$(2$^8$Be) is equal to 25 $\pm$ 4\%, 6 $\pm$ 2\% and 10 $\pm$ 2\%, respectively. The $Q_{4\alpha}$ distribution indicates the presence of near-threshold 4$\alpha$-quartets, where the decays $^{12}$C(0$^+_2$) and 2$^8$Be are reconstructed under the condition: $Q_{2\alpha}$($^8$Be) $\leq$ 0.2 MeV (Figure~\ref{fig:4.4}(d)). The investigation of this problem requires a different level of $n_\alpha$-ensemble statistics, which is fundamentally accessible in the search for transverse events.

The ratios $N_{n\alpha}$($^8$Be)/$N_{n\alpha}$ (\%) in $n_\alpha$ fragmentation of relativistic nuclei $^{16}$O, $^{22}$Ne, $^{28}$Si, and $^{197}$Au in NE are summarized in Figure~\ref{fig:4.5}. They clearly indicate a rapid increase in the contribution of $^8$Be(0$^+$) with increasing the number of relativistic $\alpha$-particles. Being proportional to $^8$Be(0$^+$) the contributions of $^9$B and $^{12}$C(0$^+_2$) also increase with $n_\alpha$. In the case of $^{197}$Au, the growth trends are traced up to relativistic 10 $\alpha$-particles in the event.

\subsection{New measurements in Si and Kr exposures} 
A targeted search for $n_\alpha$ $>$ 2 events was performed in the 2020s in the BECQUEREL experiment by transversely scanning layers of nuclear cores longitudinally irradiated in the 1980s with $^{28}$Si nuclei at 15 GeV per nucleon at the JINR Synchrophasotron. The number of the found events was 439, including 3$\alpha$ (245), 4$\alpha$ (143), 5$\alpha$ (45), and 6$\alpha$ (6), four times exceeding the EMU collaboration statistics for $n_\alpha$ $>$ 2 at 15 GeV per nucleon. The emission angles of relativistic $\alpha$ particles were measured with particular care.

\begin{figure}
\includegraphics[width=1.0\textwidth]{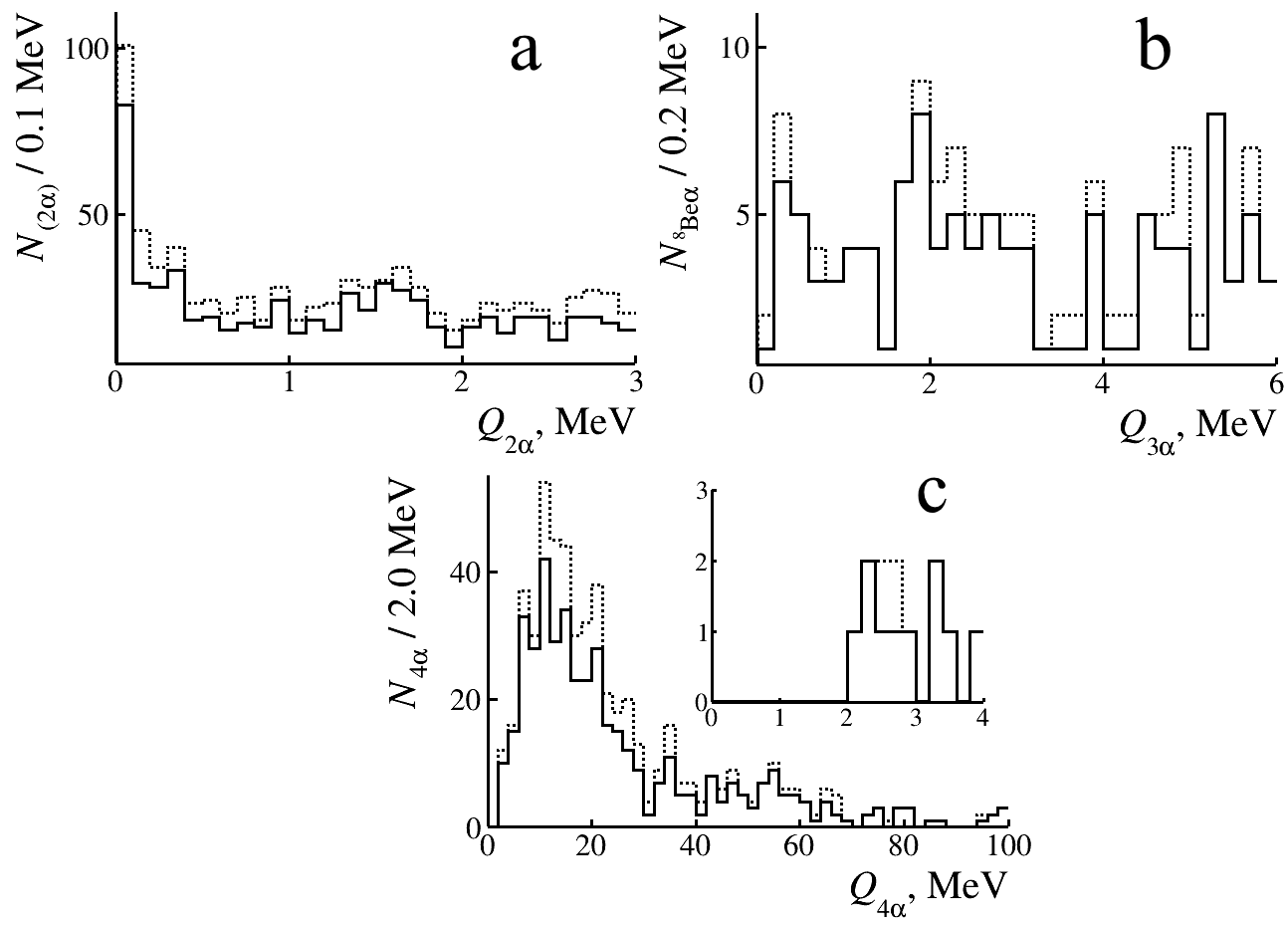}
\caption{\label{fig:4.6} Distributions of invariant masses $Q$ of 2$\alpha$-pairs (a), 3$\alpha$-triples (c) and 4$\alpha$-quartets (d) in the fragmentation of $^{28}$Si nuclei at 15 GeV per nucleon; the dotted line indicates measurements of events $n_\alpha$ $>$ 2 at 15 GeV per nucleon by the EMU collaboration.}
\end{figure}
 
Figure~\ref{fig:4.6} shows the onsets of the invariant mass distributions of all $\alpha$-pairs $Q_{2\alpha}$ (a), $\alpha$-triples $Q_{3\alpha}$ (b) and $\alpha$-quartets $Q_{4\alpha}$ (c). 101 events with $^8$Be(0$^+$) decay $Q_{2\alpha}$ $<$ 0.2 MeV having $\langle Q_{2\alpha}\rangle$ (RMS) = 79 $\pm$ 4 (55) keV were identified, including 3$\alpha$ (37), 4$\alpha$ (44), 5$\alpha$ (16), and 6$\alpha$ (4). Among them we have identified 15 $^{12}$C(0$^+_2$) decays $Q_{3\alpha}$ $<$ 0.7 MeV with $\langle Q_{3\alpha}\rangle$ = 416 $\pm$ 39 (172) keV, including 3$\alpha$ (4), 4$\alpha$ (5), 5$\alpha$ (5), and 6$\alpha$ (1). The ratio of $^{12}$C(0$^+_2$) to $^8$Be(0$^+$) is 0.15 $\pm$ 0.04. There are no candidates for the decay $^{16}$O(0$^+_6$) $\to$ 4$\alpha$.

The NE layers of nuclear were exposed to $^{84}$Kr nuclei at 950 MeV per nucleon at the SIS-18 (GSI) synchrotron in the 1990s. In 875 stars identified over primary tracks, the multiplicities and directions of secondary traces were measured \cite{86}. The statistics of $n_\alpha$ events were as follows: 174 for 2$\alpha$, 117 for 3$\alpha$, 69 for 4$\alpha$, 54 for 5$\alpha$, 27 for 6$\alpha$, 19 for 7$\alpha$, 12 for 8$\alpha$, 2 for 9$\alpha$, and 1 for 13$\alpha$. Due to the uncertainty introduced by deceleration, the analysis was performed on the NE initial part. In principle, the energy decrease can be calculated and taken while calculating the invariant masses. Thus, the covered energy region and the visible NE area can be increased. 

\begin{figure}
\includegraphics[width=0.6\textwidth]{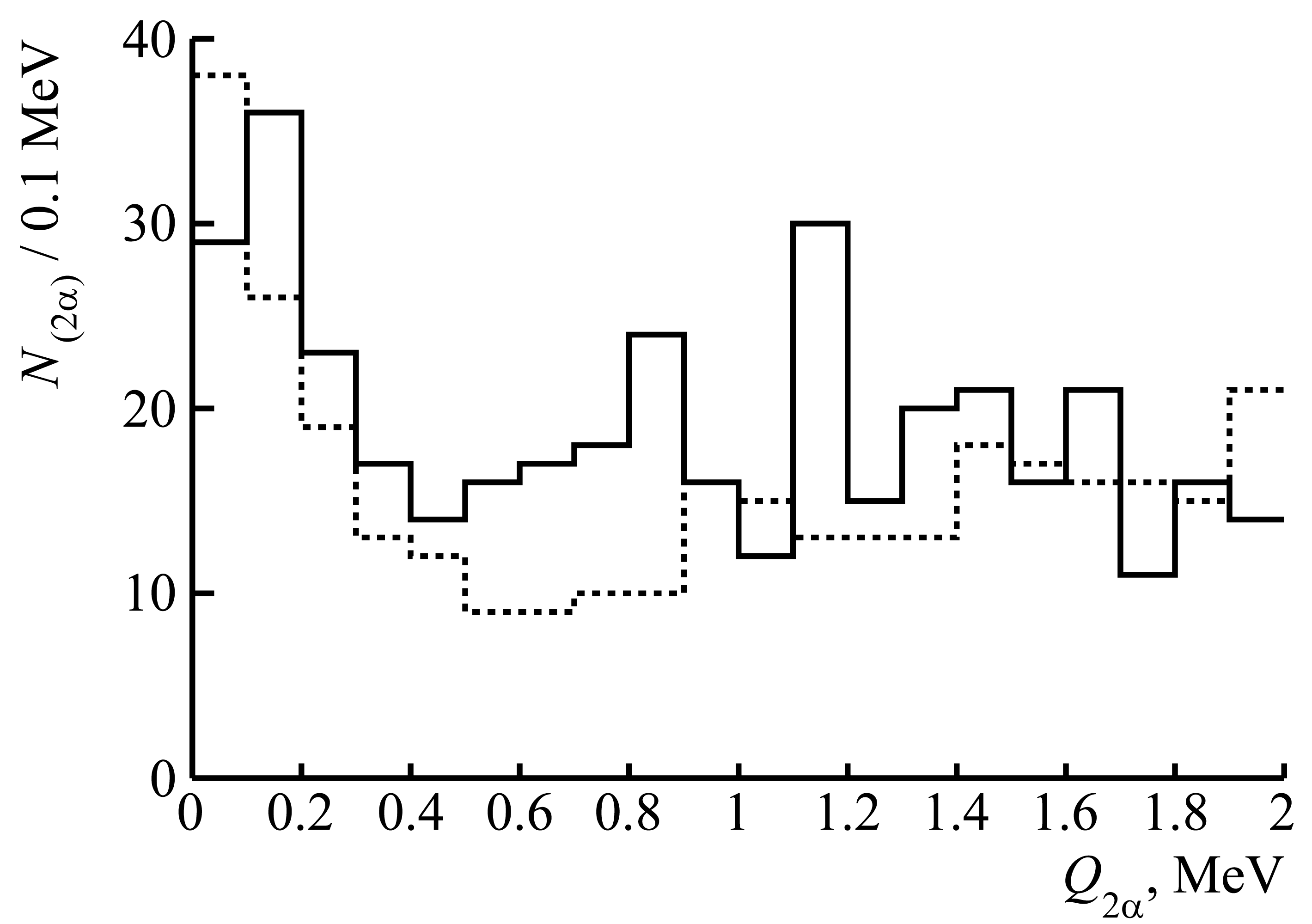}
\caption{\label{fig:4.7} Distribution of all combinations of $\alpha$-particle pairs produced in the fragmentation of $^{84}$Kr nuclei at 600-950 MeV per nucleon over invariant mass $Q_{2\alpha}$ $<$ 2 MeV (a) according to recent measurements \cite{87} and earlier data \cite{86} without taking into account the stopping at the average value of 875 MeV per nucleon (dotted curve).}
\end{figure}

\begin{figure}
\includegraphics[width=0.7\textwidth]{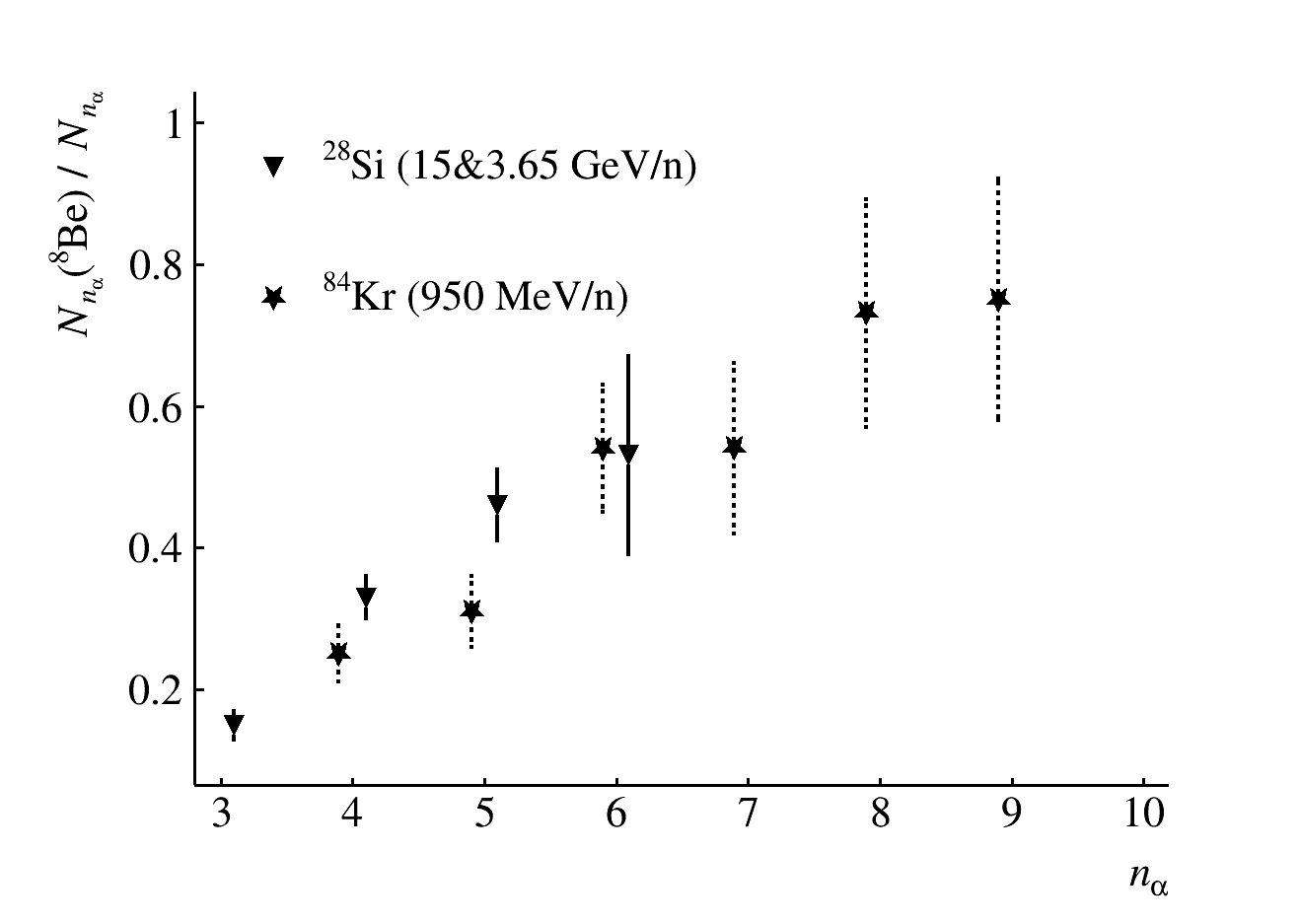}
\caption{\label{fig:4.8}  Dependence of the ratio $N_{n\alpha}$($^8$Be)/$N_{n\alpha}$ in fragmentation in NE of $^{28}$Si nuclei at 3.65 and $^{84}$Kr below 1 GeV per nucleon on the number of $\alpha$-particles $n_\alpha$.}
\end{figure}
 
The BECQUEREL experiment has recently measured another 278 stars with $n_\alpha$ $>$ 2, found by means of transverse scanning: 3$\alpha$ - 56, 4$\alpha$ - 73, 5$\alpha$ - 69, 6$\alpha$ - 34, 7$\alpha$ - 16, 8$\alpha$ - 18, 9$\alpha$ - 6, $>$10$\alpha$ - 9 \cite{87}. Their coordinates were used to calculate $Q_{(2-4)\alpha}$ for event-by-event correction. Remaining uniform up to 6 cm, the energy loss is about 9 MeV/mm, and the range before stopping is about 8 cm. In the previous data, the energy was assumed to be equal to 875 MeV per nucleon. In addition, the momentum of the fragments was taken with a factor of 0.8. Being unimportant to select $Q_{2\alpha}$($^8$Be) $<$ 0.4 MeV, this correction further allows one to preserve the selection condition of $Q_{3\alpha}$($^{12}$C(0$^+_2$)) $<$ 0.7 MeV, focusing on the grouping of events.

Figure~\ref{fig:4.7} shows the $Q_{2\alpha}$ distribution of 173 stars with $n_\alpha$ $>$ 3. For the best selection of $^8$Be(0$^+$) decays, the emission angles in this sample were determined from the averaged values after five-fold measurements of the coordinates of five points in the trace of each $\alpha$-particle at a distance of up to 500 $\mu$m from the vertex. This distribution coincides with the $Q_{2\alpha}$ distribution in 184 stars with $n_\alpha$ $>$ 3 from the previous measurements. The ratio $N_{n\alpha}$($^8$Be)/$N_{n\alpha}$ for both samples has confirmed the universality of the $^8$Be(0$^+$) growth with $n_\alpha$ for one more nucleus and at the lowest energy. The new data are included in Figure~\ref{fig:4.8}. New measurements have identified 12 decays of 2$^8$Be(0$^+$) and 9 of $^{12}$C(0$^+_2$) as well as a 4$\alpha$-quartet ($n_\alpha$ = 6) having an isolated value of $Q_{4\alpha}$ = 0.6 MeV, corresponding to the both: 2$^8$Be and $^{12}$C(0$^+_2$). The energy of the Kr nucleus is 700 MeV per nucleon.

Thus, measurements of the fragmentation of relativistic nuclei from hundreds of MeV per nucleon to tens of GeV per nucleon over a wide range of the parent nucleus mass numbers have enabled the identification of the relativistic decays of $^8$Be(0$^+$), $^9$B, and $^{12}$C(0$^+_2$). The contributions of $^8$Be(0$^+$) and the proportional contributions of $^9$B and $^{12}$C(0$^+_2$) rapidly increase while increasing $\alpha$ particle number (Figures~\ref{fig:4.5} and ~\ref{fig:4.8}). The observed universality allows one to develop a theory of nuclear fragmentation which takes into account the secondary interactions characteristic of low-energy nuclear physics, extending to nuclear astrophysics.

\subsection{New opportunities in Xe exposure at NICA} 
Using an Olympus BX63 intelligent microscope, secondary stars produced by neutrons in the fragmentation cone were studied under $^{84}$Kr irradiation \cite{88}. For the planar component of neutron transverse momenta, estimated from the angles of the observed secondary stars, the Rayleigh distribution parameter was (35 $\pm$ 7) MeV/$c$, corresponding to the energy in the parent nucleus frame of approximately 1.3 MeV. Neutron multiplicity can be estimated from the probability of star formation, and the effects of neutron ``coat'' can be studied. Information on the relative yields of neutrons, as well as the deuterons and tritons that bind them, is of both fundamental and applied importance.

However, moving into the energy below the limiting fragmentation degrades identifying the unstable states. Also, going beyond the working range of H and He isotope identification based on $P\beta c$ values becomes a limitation. The expansion of the fragmentation cone with decreasing energy worsens the capabilities of such measurements. Using this method would allow one to estimate the ratios of the lightest isotopes and their energy from emission angles, thereby providing the estimation of the density of rarefied matter which appears in the relativistic fragmentation of heavy nuclei. The Olympus BX63 intelligent microscope makes it possible to begin solving this extremely complex problem by simultaneously measuring any traces in a semi-automated mode.

\begin{figure}
\includegraphics[width=1.0\textwidth]{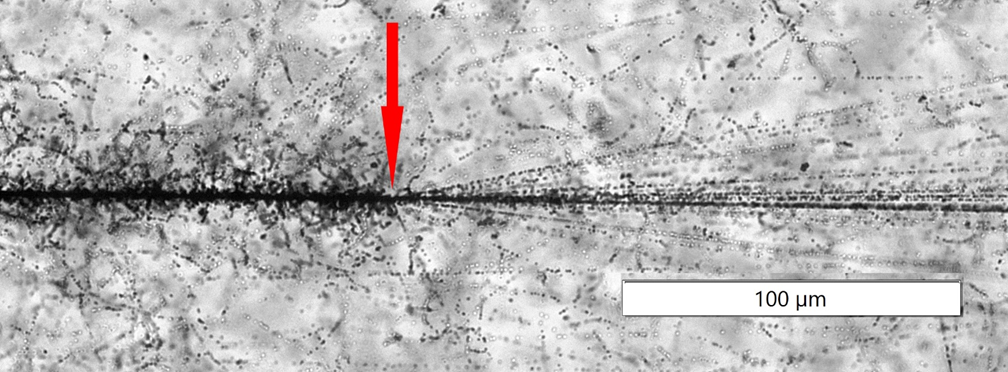}
\caption{\label{fig:4.9}Macrophotograph of the peripheral interaction of the Xe nucleus at 3.8 GeV per nucleon showing the formation of fragments in a cone of up to 20$^\circ$. The primary trace is accompanied with a ``cloud'' of $\delta$-electrons.}
\end{figure}

Recently, the opportunity of applying the proven approaches was revealed by an NE exposure to $^{124}$Xe nuclei at 3.8 GeV per nucleon at NICA/Nuclotron \cite{89}. A CR-39 solid-state track detector (SSTD) was used to determine the beam intensity, position, and profile. The exposed NE layers were developed at LHEP, and the SSTD detectors were processed at the Flerov Laboratory of Nuclear Reactions. The microscope software allows one to carry out direct counting of nuclei passing through the SSTD and determine the beam parameters. The incoming traces of the Xe nucleus and multiple fragments are visible in the macrophotograph taken with Olympus BX63 microscope at 40-fold magnification (Figure~\ref{fig:4.9}). Thus, computer microscopy has made it possible to combine these methods under the conditions where the use of electron microscopy is difficult.

\section{Conclusion}
Since the early 2000s, the NE method has been used in the BECQUEREL experiment to study the nuclear clustering in light isotopes, including radioactive ones. In the relativistic dissociation of $^9$Be, $^{10}$C, $^{10}$B, and $^{11}$C, decays of $^8$Be(0$^+$) and $^9$B were identified on the invariant masses of $\alpha$-pairs down to 0.2 MeV and 2$\alpha p$-triples down to 0.5 MeV determined from emission angles in the approximation of conservation of the initial momentum per nucleon. This experience has indicated the opportunity to identify the $^{12}$C(0$^+_2$) decays in the $^{12}$C $\to$ 3$\alpha$ dissociation in the mass range down to 0.7 MeV missed in earlier studies. Since the decay energies of these three states are significantly lower than the nearest excitations with the same nucleon composition, their selection by the upper limit on the corresponding invariant mass has proved to be sufficient.

Over the past decade, the focus has been concentrated on searching for the dissociation of the accessible variety of nuclei for the decays $^{12}$C(0$^+_2$) and its state of the assumed 4$\alpha$ analog $^{16}$O(0$^+_6$) which should be accompanied by decays of $^8$Be(0$^+$). This search was motivated by the concept of $\alpha$-particle Bose-Einstein condensate whose these three states are assumed to be forms of this condensate. Focusing on the relativistic decays of $^8$Be(0$^+$) and $^{12}$C(0$^+_2$), which produce He fragments with extremely small scattering angles, has given the search for condensate states not only methodological clarity but also visibility.

The search for $^{12}$C(0$^+_2$) decays was performed using the statistics of 590 $^{12}$C $\to$ 3$\alpha$ events at 450 MeV and 3.65 GeV per nucleon from earlier measurements and measurements doubled in this experiment. It has been established that in the $^{12}$C $\to$ $^8$Be(0$^+$)$\alpha$ channel which accounts for 43\% of the statistics there is a contribution of 26\% from $^{12}$C(0$^+_2$) and additional 45\% from the subsequent excitation $^{12}$C(3$^-$). The simultaneous identification of $^{12}$C(0$^+_2$) and $^{12}$C(3$^-$) justifies the applied approach being critical to understand the dissociation of $^{12}$C.

The $^{12}$C(0$^+_2$) peak was also established from early measurements of 648 events of coherent dissociation $^{16}$O $\to$ 4$\alpha$ at 3.65 GeV per nucleon. Thus, $^{12}$C(0$^+_2$) manifests itself not only as an excitation of $^{12}$C, but also as an analogue of the universally formed nucleus $^8$Be(0$^+$). The contribution of $^{12}$C(0$^+_2$) was 22\%. It has demonstrated the increase with the number of $\alpha$-particles compared to the coherent dissociation of $^{12}$C where the $^{12}$C(0$^+_2$) is equal to 11\%. The contribution of $^{12}$C(3$^-$) to the dissociation $^{16}$O $\to$ $^8$Be(0$^+$)2$\alpha$ equal to 32\% is also enhanced in comparison to the case of $^{12}$C similarly to $^{12}$C(0$^+_2$).

The contribution of $^{16}$O(0$^+_6$) candidates to the $^{12}$C(0$^+_2$)$\alpha$ channel does not exceed 7\%. The $^{16}$O(0$^+_6$) $\to$ $^{12}$C(0$^+_1$)$\alpha$ channel proposed as an alternative is currently being measured. The search for the $^{16}$O(0$^+_6$) $\to$ $^{12}$C(0$^+_1$)$\alpha$ decay is important as a variant of the $^{12}$C isotope synthesis where one of the $\alpha$-particles in the quartet serves as a catalyst.

The joint search for $^9$B and $^{12}$C(0$^+_2$) decays in the leading dissociation channels $^{14}$N $\to$ 3He(+H) has been performed. The contribution of $^9$B decays turned out to be 23\%. Although 10\% of the statistics are consistent with $^{12}$C(0$^+_2$) decays and the invariant mass distributions of $\alpha$-triples do not allow one to estimate the contribution of $^{12}$C(0$^+_2$) being half-overlapped by the $^9$B$\alpha$ channel.

The weakening of $^{12}$C(0$^+_2$) in the partially fermionic dissociation of $^{14}$N compared to the $\alpha$-particle dissociation of $^{12}$C and $^{16}$O may indicate the competition between unbound excitations of lighter odd nuclei. This circumstance emphasizes the efficiency of $^{12}$C(0$^+_2$) generation in purely bosonic dissociation channels of $^{12}$C and $^{16}$O.

The increase in the number of $\alpha$-particles in heavy nuclei can enhance the generation of boson states while limiting the fermionic ones. The review of measurements of the relativistic fragmentation of $^{16}$O, $^{22}$Ne, $^{28}$Si, $^{84}$Kr, and $^{197}$Au nuclei has revealed a strong correlation between $^8$Be(0$^+$) production and alpha particle multiplicity. $^9$B and $^{12}$C(0$^+_2$) follow the same trend. This conclusion is inconsistent with model concepts, which, on the contrary, predict a suppression of $^8$Be(0$^+$) while increasing $\alpha$-particle multiplicity.

Identification of the Hoyle state needs to study the diluted, cold nuclear matter in the temperature and density range of red giants. The ratios of the identified $^{1,2,3}$H and $^{3,4}$He isotopes can characterize the emerging matter holistically. In this regard, NE layers exposed to heavy nuclei at several GeV per nucleon at the NICA accelerator complex will enable the application of the developed approaches to the holistic analysis of ensembles of H and He isotopes and heavier fragments of unprecedented complexity.

It is justified to hope that advances in image analysis will enable a completely new dimension in using the unique characteristics of nuclear emulsions in nuclear structure studies, low-energy nuclear physics, and applied research. These challenges require to develop the automated microscopy. However, this development is sure to be based on the classical nuclear emulsion method whose foundations were laid out eight decades ago in cosmic ray physics.

\section{Acknowledgments}
While still unresolved, the condensate problem proved to be a kind of signpost leading to inspiring findings The shift toward this topic was made possible by the influence and support of colleagues. In the 2000s, the results of the BECQUEREL experiment were presented at the EXON conferences organized by Professor Yuri Penionzhkevich (Dubna), who passed away in August 2025. At the EXON-2009 conference, he organized our discussion with Professor Wolfram von Oerzen (Berlin), who was interested in multiple alpha particle generations and the observability of  $^8$Be forks among them for condensate searches. Having included this possibility in his review, he recommended our paper to the founder of this research, Professor Peter Schuck (Orsay), at the 2nd Workshop on Nuclear Clustering in 2010. At this workshop, Professor Gerd Röpke (Rostock) asked about the possibility of identifying decays of the Hoyle state, which was answered affirmatively, although the nature of the process was unclear at the time. The transition to practice became possible with new nuclear emulsion samples, the production of which was initiated by Natalia Polukhina (Moscow). Thanks to the active support of Vladimir Pikalov (Protvino), they were irradiated in a carbon beam in 2016. Our initial results continued our collaboration with Sergei Kharlamov (Moscow) and Natalia Peresadko (Troitsk), using existing measurements and accumulating new ones. The data they provided from their emulsion collaborations were analyzed during the 2020-2021 lockdown period. Our presentation at the 2019 cluster meeting in Trento was made possible thanks to the support of Prof. David Blaschke (Wroclaw). Thus, this review was made possible thanks to the scientific solidarity of these colleagues and the colleagues whose names are among the authors of the cited papers. We hope we have not disappointed them.
Research on the BECQUEREL experiment, which began two decades ago, would not have been possible without the constant support of our research leader at JINR, Professor Aleksandr Malakhov. We express our sincere gratitude to Irina Zarubina, who proofread this extensive text, and to Svetlana Chubakova, who edited its translation.

\end{document}